%% file: main.tex
\def\year{2019}\relax
\documentclass[letterpaper]{article} 
\usepackage{aaai19}  
\usepackage{times}  
\usepackage{helvet}  
\usepackage{courier}  
\usepackage{url}  
\usepackage{graphicx}  
\newif\ifdraft
\draftfalse

\usepackage{amsmath}
\usepackage{mwe,tikz}\usepackage[percent]{overpic}
\usepackage[capitalize]{cleveref}
\crefname{subsection}{subsection}{subsections}
\usepackage{color}
\usepackage{pgfplots}
\pgfplotsset{compat=newest}
\pgfplotsset{every axis legend/.append style={%
		cells={anchor=west}}
}
\pgfplotsset{every y tick label/.append style={font=\footnotesize}}
\pgfplotsset{every x tick label/.append style={font=\footnotesize}}
\pgfplotsset{every axis x label/.append style={font=\footnotesize}}
\pgfplotsset{every axis y label/.append style={font=\footnotesize}}
\pgfplotsset{every axis legend/.append style={font=\footnotesize}}
\pgfplotsset{every axis title/.append style={font=\footnotesize}}
\usepgfplotslibrary{polar}
\usetikzlibrary{arrows}
\tikzset{>=stealth'}

\definecolor{ra_1}{rgb}{1.0, 1.0, 1.0}
\definecolor{ra_2}{rgb}{0.0, 1.0, 1.0}
\definecolor{ra_3}{rgb}{0.5647058823529412, 0.9333333333333333, 0.5647058823529412}
\definecolor{ra_4}{rgb}{0.11764705882352941, 0.5647058823529412, 1.0}
\definecolor{ra_5}{rgb}{0.0, 1.0, 0.0}
\definecolor{ra_6}{rgb}{0.0, 0.0, 1.0}
\definecolor{ra_7}{rgb}{0.13333333333333333, 0.5450980392156862, 0.13333333333333333}
\definecolor{ra_8}{rgb}{0.0, 0.0, 0.5019607843137255}
\definecolor{ra_9}{rgb}{0.0, 0.39215686274509803, 0.0}
\usepgfplotslibrary{groupplots}

\usepackage{tikz}
\usepackage{tabu}
\usepackage{changepage}
\usepackage{amsmath}
\usepackage{booktabs}
\usepackage{soul}
\usepackage[utf8]{inputenc}
\usepackage{comment}
\usepackage{flushend}

\usepackage[per-mode=symbol,detect-all,binary-units=true]{siunitx}
\let\DeclareUSUnit\DeclareSIUnit
\let\US\SI
\let\us\si
\DeclareUSUnit\inch{in}
\DeclareUSUnit\foot{ft}

\newcommand{\shivam}[1]{\ifdraft{\color{blue}[#1 -- Shivam]}\fi}

\newcommand{\kyle}[1]{{\ifdraft\color{olive}[#1 -- Kyle]}\fi}

\newcommand*\samethanks[1][\value{footnote}]{\footnotemark[#1]}

\usepackage{xspace}
\newcommand\system{VerticalCAS\xspace}
\newcommand{\citet}[1]{\citeauthor{#1} \shortcite{#1}}

\title{Verifying Aircraft Collision Avoidance Neural Networks\\ Through Linear Approximations of Safe Regions}

\author{Kyle D. Julian\thanks{Equal Contribution} \and Mykel J. Kochenderfer\\ Stanford University, Stanford, CA 94305
\AND Shivam Sharma\samethanks \and Jean-Baptiste Jeannin \\ University of Michigan, Ann Arbor, MI 48109
}
\begin{document}

\maketitle

\begin{abstract}
    The next generation of aircraft collision avoidance systems frame the problem as a Markov decision process and use dynamic programming to optimize the alerting logic. The resulting system uses a large lookup table to determine advisories given to pilots, but these tables can grow very large. To enable the system to operate on limited hardware, prior work investigated compressing the table using a deep neural network. However, ensuring that the neural network reliably issues safe advisories is important for certification. This work defines linearized regions where each advisory can be safely provided, allowing Reluplex, a neural network verification tool, to check if unsafe advisories are ever issued. A notional collision avoidance policy is generated and used to train a neural network representation. The neural networks are checked for unsafe advisories, resulting in the discovery of thousands of unsafe counterexamples.
\end{abstract}

\section{Introduction}

Over the last decade, neural network representations have become popular in decision making systems for a variety of domains. Neural networks are state-of-the-art for image recognition systems~\cite{simonyan2014very,he2016deep} and can learn to play games at super-human levels~\cite{mnih2015human,silver2016mastering}. In these domains, a mistake by the neural network may have minor consequences; however, neural networks can also be used in safety-critical systems where a failure could be catastrophic. For example, neural networks have been used to steer autonomous cars given images~\cite{bojarski2016end} and guide unmanned aircraft to waypoints~\cite{julian2017neural}. If a neural network steers an autonomous car off the road or directs an aircraft into an obstacle, the result could be be expensive or lead to loss of life. In order for neural networks to be used for such applications, confidence must be established in their safe operation.

In the last few years, new research has resulted in tools to verify safety properties of neural networks. One tool, Reluplex, uses a Satisfiability Modulo Theories solver and extends the simplex method for neural networks with rectified linear unit (ReLU) activation functions to determine whether any input in a specified input region produces outputs with a desired property~\cite{katz2017reluplex}. Another approach defines neural network verification as a reachability problem that can be solved using a mixed integer linear program formulation~\cite{lomuscio2017approach}. Furthermore, a tool known as AI2 uses an overapproximation of the neural network to quickly verify safety properties of neural networks~\cite{gehr2018ai}. These tools enable network properties to be rapidly verified, but more work is needed to develop properties that will ensure safe operation of neural network systems.

This work focuses on the verification of neural networks used for aircraft collision avoidance. We created a highly simplified aircraft collision avoidance policy that uses vertical maneuvers using value iteration~\cite{egorov2017pomdps}. This policy, which we call \system, is loosely based on an early prototype of the next generation airborne collision avoidance system for commercial aircraft, ACAS~Xa~\cite{kochenderfer2015decision}. Although \system is not the ACAS~Xa system that will be flown on real aircraft, \system serves as a simple and open-source collision avoidance policy that can be used in the development of desirable properties. These properties should also hold for other vertical collision avoidance systems. After generating a collision avoidance policy, a neural network is trained to represent the original discrete policy~\cite{julian2016policy}.

Previous work has developed equations to verify the safety of the tabular collision avoidance policy by defining ``safeable'' regions for each advisory~\cite{jeannin2017formally}. In order to verify these properties for the neural network representation, the equations are linearized to enable the use of linear program solvers used by Reluplex. This paper describes the linearization process, introduces a new variable $\tau$, the time to loss of horizontal separation, and describes the formulation and verification of ``safeable'' regions for neural networks.




\section{\system}
\begin{figure}
    \begin{center}
    {
    \input{variables.tex}
    }
    \caption{Three input variables of \system neural network and ownship puck defined by $r_p$ and $h_p$ on ownship centered coordinate frame.}
    \label{fig:VerticalCAS_var}
    \end{center}
\end{figure}
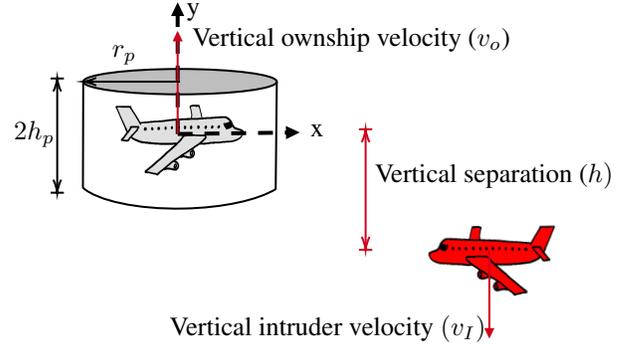

The \system collision avoidance system used throughout this paper is inspired by ACAS~Xa, which frames aircraft collision avoidance as a Markov decision process (MDP)~\cite{kochenderfer2015decision}. The ACAS~Xa system is the successor to the current Traffic alert and Collision Avoidance System (TCAS) and provides pilots with advisories to change their vertical rate to prevent a possible near mid-air collision (NMAC). A NMAC is defined as an \textit{intruder} aircraft coming inside the ownship \textit{puck} which is described in \cref{fig:VerticalCAS_var} as a region $h_p=\US{100}{\foot}$ above and below, and $r_p=\US{500}{\foot}$ radially around the ownship aircraft (the aircraft where the collision avoidance system is installed). 

\system has 5 inputs which describe the system's state:
\begin{enumerate}
    \item $h$ (ft): Altitude of intruder relative to ownship
    \item $v_O$ (ft/s): ownship vertical climb rate
    \item $v_I$ (ft/s): intruder vertical climb rate
    \item $a_\text{prev}$: previous advisory
    \item $\tau$ (sec): time to loss of horizontal separation
\end{enumerate}
The first 3 inputs are spatial and velocity quantities that are described in \cref{fig:VerticalCAS_var}. Relative altitude $h$ varies from \US{-8000}{\foot} to \US{8000}{\foot}, and the aircraft climb rates vary from \US{-100}{\foot\per\second} to \US{100}{\foot\per\second}.

Previous advisory ($a_\text{prev}$) dictates which advisories \system can issue given the most recent advisory. This restricts the network from issuing conflicting advisories like strong ascend or descend advisories immediately after a clear of conflict advisory which can be confusing to pilots. 

Time to loss of horizontal separation ($\tau$) is the time till the horizontal separation between the intruder and ownship is less than $r_p$. A more explicit definition of $\tau$ is
\begin{equation}
    \tau = \frac{r-r_p}{r_v}
\end{equation}
where $r$ is the horizontal separation between the ownship and intruder, and $r_v$ is the relative horizontal velocity between the two aircraft. 

\subsection{Markov Decision Process Policy}
\system is computed using local approximation value iteration as implemented by the Julia package called POMDPs.jl~\cite{egorov2017pomdps}. The states, dynamics, rewards, and advisories reflect an early prototype of the ACAS~Xa system described by \citet{kochenderfer2015decision}. Each state $s \in \mathcal{S}$ represents a discrete encounter geometry between the ownship and intruder aircraft and has five dimensions which are the inputs outlined above.

{\centering
\input{Table_advisories.tex}
}

The system issues a new advisory $a$ every $\epsilon$ seconds, and there are nine possible advisories as described in \cref{tab:advisories}, where $g$ is Earth's sea-level gravitational acceleration~\cite{jeannin2017formally}. Each vertical advisory is defined by a target velocity $v_{lo}$ and sign $w$. If $w=1$, the ownship can assume a velocity in the range $[v_{lo}, +\infty)$, and if $w=-1$ the ownship can assume a velocity in the range $(-\infty, v_{lo}]$. In addition to the advisory $(w,v_{lo})$, the ownship has to accelerate at least $a_{lo}$ until it is in the acceptable velocity range defined by the issued advisory.

The transition model $T(s,a,s')$ and reward model $r(s,a)$ used for vertical collision avoidance are explained in previous work~\cite{kochenderfer2015decision}. Local approximation value iteration is used to compute the state-action values, $Q(s,a)$, such that the finite-horizon Bellman equation holds for all states and actions:
\begin{equation}
    Q(s,a) = r(s,a) + \sum_{s'}\max_{a'}T(s,a,s')Q(s',a')
\end{equation}
Because $s'$ might not be exactly one of the discrete states $s \in S$, multilinear interpolation is used to compute $Q(s',a')$. After computing the $Q$ values using local approximation value iteration, the advisory associated with the highest $Q$ value for a given state is the best advisory and is issued by the system. In addition, because the computed policy tends to advise Clear of Conflict in cases where an NMAC is imminent and unavoidable, the advisory at time $\tau=6$ is used in situations where $\tau<6$.

\subsection{Neural Network Representation}
\begin{figure}
    \centering
    \input{PolicyPlots_Example.tex}
    \caption{Example policy plots for a climbing ownship and level-flying intruder using the MDP table (top) and neural network (bottom)}
    \label{fig:PolicyPlots_Example}
\end{figure}
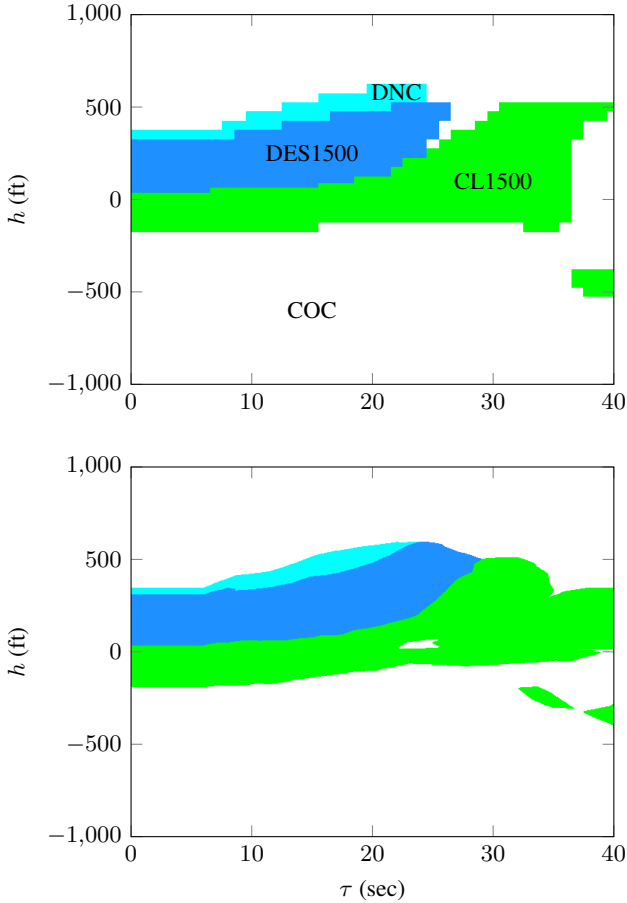

Storing the MDP policy with fine resolution in a table format can require large amounts of storage space, which may prevent implementation on limited avionics hardware. One approach to compressing the policy representation approximates the policy using a neural network through the use of supervised learning and an asymmetric loss function, which encourages the neural network to simultaneously approximate the $Q$-values and highest scoring advisory~\cite{julian2016policy}. One network was trained for each previous advisory $a_\text{prev}$, resulting in nine fully connected neural networks using six hidden layers of 45 hidden units each. Each hidden layer uses rectified linear unit (ReLU) activation, which is defined as $\text{ReLU}(x) = \max(0,x)$~\cite{ReLU}. Each network uses the remaining four state variables as inputs and outputs a value associated with each possible advisory. Each neural network was trained for 200 epochs using AdaMax optimization~\cite{Adam} implemented in Keras~\cite{keras} with the Theano backend~\cite{theano}, which requires an hour to train on an NVidia Titan X GPU.

\Cref{fig:PolicyPlots_Example} plots the advisory the system would give to the ownship if the intruder were at each location in the plot. In this scenario, the ownship is climbing while the intruder is maintaining a constant altitude. If the intruder is approaching the ownship from above, the system alerts the ownship to stop climbing (DNC) or descend (DES1500) in order to prevent an NMAC. If the intruder is a little below the ownship, the system advises the pilot to continue climbing (CL1500). In other locations, a collision is not imminent and the system alerts clear-of-conflict (COC). The neural network representation is a smooth approximation of the original table policy. Although the network appears to represent the table well, verification is needed to ensure that the neural network alerts safely at all times.

\section{Safe Regions}
\begin{figure}[t]
    \begin{center}
    \input{SafeRegion1.tex}
    \caption{Intruder in the safe region for ownship advisory CL1500 and $v_O<0$}
    \label{fig:SafeRegion1}
    \end{center}
\end{figure}
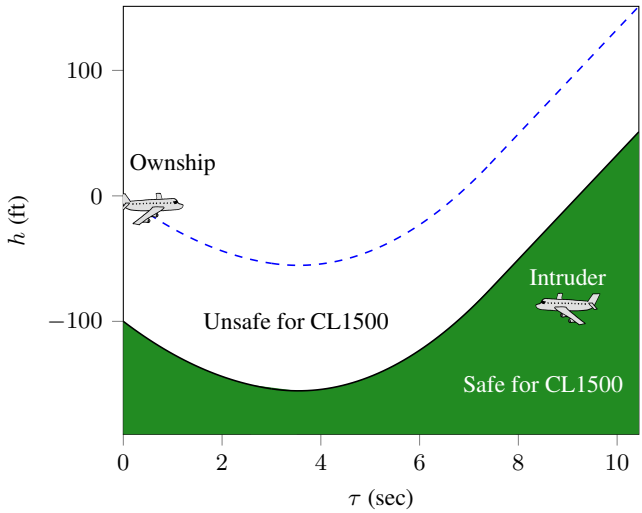

The \textit{safe region} is defined as the region in space where an intruder aircraft will be safe (i.e. will not enter the ownship puck), given the ownship aircraft is following a single advisory (shown in \cref{fig:SafeRegion1}). The safe region is described by the ownship travelling along a nominal trajectory. This nominal trajectory is described by the ownship following an advisory exactly, i.e., if the advisory issued allows a range of velocities $[1500, +\infty)$, the nominal trajectory will be defined by the ownship assuming a velocity of \US{1500}{\foot\per\minute}. From our earlier definition of $\tau=\frac{r-r_p}{r_v}$,  $\tau=0$ of the nominal trajectory is at $r=r_p$.

The nominal trajectory of the ownship is simply a parabolic trajectory due to constant vertical acceleration. The trajectory can be written as follows~\cite{jeannin2017formally}:
\begin{equation} \label{equation:nominal}
    h_n= 
    \begin{cases}
        \frac{a_{lo}}{2}\tau^2+v_{O}\tau,& \text{if } 0 \leq \tau < \frac{v_{lo}-v_{O}}{a_{lo}} \\
        v_{lo}\tau - \frac{(v_{lo}-v_{O})^2}{2a_{lo}}, & \text{if } \frac{v_{lo}-v_{O}}{a_{lo}} \leq \tau
    \end{cases}
\end{equation}
where $h_n$ is the altitude of the ownship in a coordinate frame centered in the starting position of the ownship \cref{fig:VerticalCAS_var} (Note: the subscript $n$ denotes the nominal trajectory). The piecewise \cref{equation:nominal} describes the dynamics when the ownship velocity is less than $v_{lo}$ and when the ownship velocity is greater than $v_{lo}$. Once the ownship climb rate reaches $v_{lo}$, the aircraft is compliant with the advisory and continues climbing with no vertical acceleration. 

For the example of CL1500, the safe region will be the region below the `puck' of the ownship flying along the CL1500 nominal trajectory. If an intruder is in this region below the ownship, it will be safe from collision for an ownship following the CL1500 advisory. Therefore, this region is defined as the \textit{safe region} for a particular advisory.

Safe regions have to be able to be represented in terms of network variables to define a search space in the state-space of the network. Representing safety bounds solely in terms of the five network variables poses some challenges which are discussed in the next section.

One limitation with using safe regions to verify safety properties is that safe regions assume that the ownship follows a single advisory throughout the encounter. In reality, multiple advisories can be issued during an encounter, giving the system an opportunity to change the advisory. Therefore, an advisory that was initially \textit{unsafe} can be made \textit{safe} with a change later in the encounter. Safeable regions, described below, build on the safe region concept and tackle this shortcoming.

\section{Worst-Case Scenario Approach} \label{worst-case}
An NMAC is defined as an intruder aircraft coming inside the ownship \textit{puck}, as depicted in \cref{fig:VerticalCAS_var}. The network is trying to prevent NMAC's, so the safe region is described by this puck around the ownship aircraft. 

In \cref{fig:SafeRegion1}, when the ownship is descending, the safe region bounds are described by the `back' of the ownship puck \cite{jeannin2017formally}, where the `back' of the ownship puck can be represented as $\tau_{back} = \tau - \frac{2r_p}{r_v}$, where $r_p$ is a known constant, but, $r_v$ is unknown because it is not an input to the network. Thus, a worst-case approximation must be made to define the safe region bounds described by the back of the ownship puck. 

At $\tau=0$ the horizontal separation between the intruder and ownship is $r_p$. After this point, horizontal separation between the ownship puck and the intruder will not be regained again until $t=\frac{2r_p}{r_v}$ seconds in the future when the intruder will cross the back of the ownship puck. In the worst-case, $r_v \to 0$, and horizontal separation may never be regained. As a result, the intruder must be at an altitude that is safeable for all time $t> \tau$. 

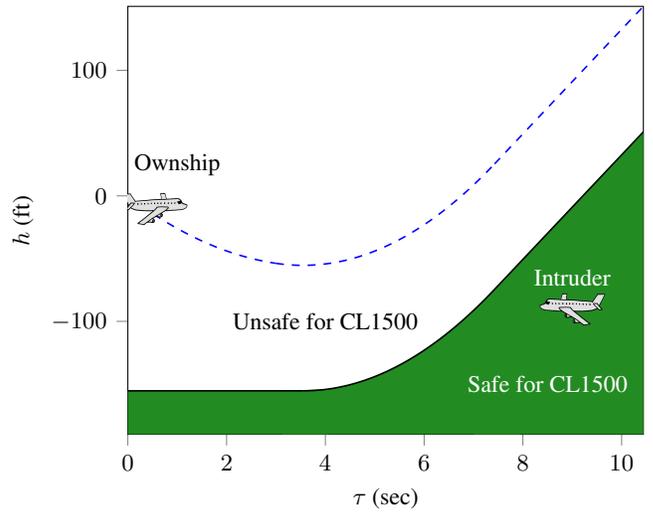
\begin{figure}[t]
\begin{center}
\input{SafeRegion2.tex}
\caption{Intruder in the worst-case safe region for the ownship advisory CL1500 and $v_O<0$}
\label{fig:wc_outer}
\end{center}
\end{figure}

The relative horizontal velocity of the two aircraft $r_v$ effectively dictates the width of the ownship puck in $\tau$-space. The worst-case safe region bound should include all other unsafe regions, which is achieved as $r_v\to 0$ or the ownship puck is infinitely wide. The worst-case safe region bounds can be seen in \cref{fig:wc_outer}. Using this approach, a worst-case safe region bound can be described where $\Omega_{\sf Unsafe} \subseteq \Omega_{\sf Unsafe (worst\,case)}$ i.e. all possible unsafe regions are subsets of the worst-case scenario unsafe region. 

\section{Safeable Regions}

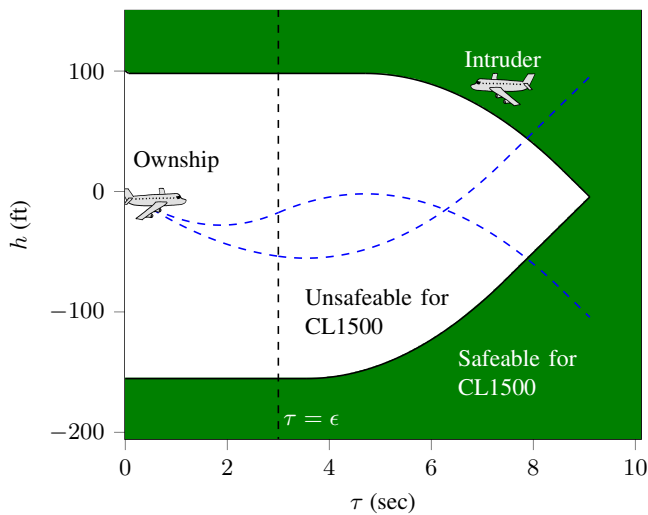
\begin{figure}[h]
\begin{center}
\input{Safeable_noUnsafe.tex}
\caption{Safeable region with strengthen and reversal alerts issued at $\tau = \epsilon$.}
\label{fig:safeable}
\end{center}
\end{figure}

Safeable regions are defined as regions which are currently safe or that can be made safe in the future. A safeable region is constructed by assuming two worst-case trajectories of an aircraft complying with an advisory for time $\epsilon$ (\system issues a new advisory every $\epsilon$ seconds). After $\epsilon$, these two trajectories represent the two extreme positions of the ownship that complies with the initial advisory. From this point, the strongest reversing and strengthening advisories that \system can issue are considered. If either of these advisories prevent a collision, then the intruder is in a safeable region. As a result, a collision with an intruder in the safeable region can always be avoided. For example, as seen in \Cref{fig:safeable}, if the intruder is located as shown, the system can safely issue a CL1500 advisory because a strong reversal at the next time step will ensure that the ownship descends before reaching the intruder. A more detailed explanation of safeable regions is provided in ~\cite{jeannin2017formally}.

\begin{figure}[h]
\begin{center}
\input{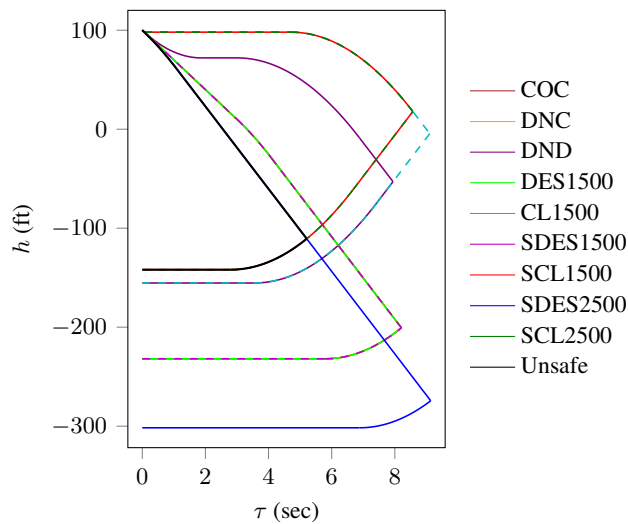}
\caption{All unsafeable regions and the region that is unsafeable for all advisories}
\label{fig:safeable_all}
\end{center}
\end{figure}

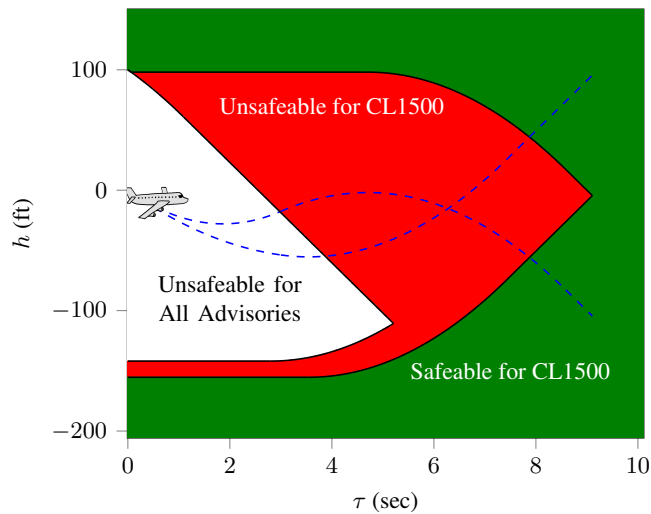
\begin{figure}[h!]
    \centering
    \input{Safeable_CheckRegion.tex}
    \caption{Safeable and unsafeable regions for CL1500 advisory} 
    \label{fig:Safeable_CheckRegion}
\end{figure}

If the system always gives safeable advisories whenever possible, then an intruder beginning in the safeable region will always be avoided. As a result, ensuring safety when using the neural network system requires checking for any instances when the neural network gives an unsafeable advisory when a safeable advisory exists. To generate these regions, the region that is unsafeable for all advisories must be computed, which can be done by generating the intersection of all possible advisories, as illustrated in \cref{fig:safeable_all}. The region to verify is shown in red in \cref{fig:Safeable_CheckRegion} because this region is unsafeable for CL1500 but would be safeable for another advisory such as SCL2500 or SDES2500. The next section describes how these safeable regions are adapted for use with the Reluplex neural network verification tool.

\section{Checking Safeable with Reluplex}
Reluplex extends the simplex method to verify neural network properties by representing neural networks, activation functions, and constraints as piecewise linear equations. Linear bounds are placed on the input variables to define the search region, and the output variables are constrained such that the advisory of interest must be associated with the largest valued output from the network. Reluplex systematically searches for an input to satisfy both input and output constraints.~\cite{katz2017reluplex}. The red unsafeable region in \cref{fig:Safeable_CheckRegion} is nonlinear and non-convex, so the region cannot be verified using Reluplex in the current form.

\begin{figure}[h!]
\begin{center}
\input{Safeable_Velocities2.tex}
\caption{Safeable regions for different initial climbrates for the ownship}
\label{fig:Safeable_Velocities}
\end{center}
\end{figure}
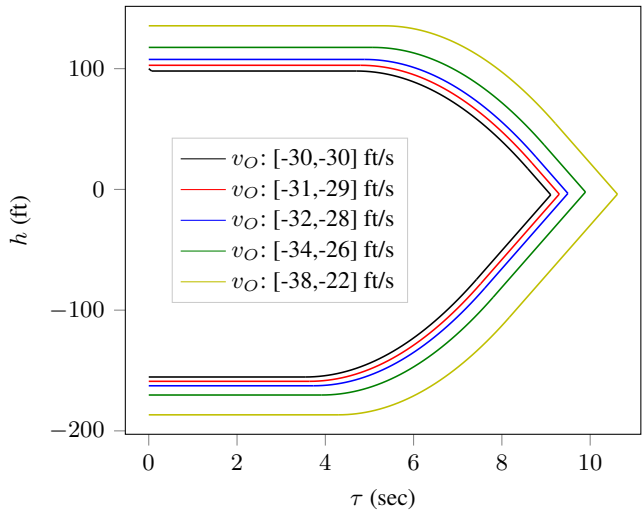

There are three adjustments made to the safeable regions to prepare the regions for Reluplex. First, the safeable regions are functions of the ownship's initial climbrate, which can vary from \US{-100}{\foot\per\second} to \US{100}{\foot\per\second}. In order to avoid verifying every possible region generated by all floating point values of ownship climb rate, the regions are generated assuming a small range of climb rates instead of a single climb rate. To generate the safeable boundaries, the upper and lower trajectories are generated assuming the worst case initial climb rate. As a result, the unsafeable boundaries grow outwards, as seen in \cref{fig:Safeable_Velocities}, which shows the safeable region boundaries for different ranges of climb rates.

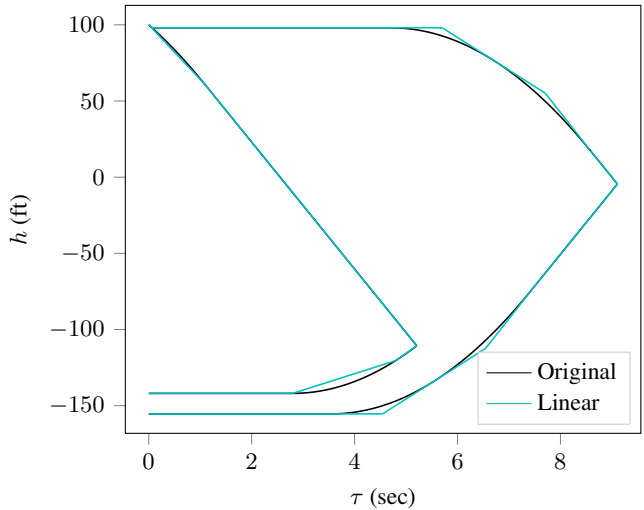
\begin{figure}[h!]
\begin{center}
\input{OverApprox.tex}
\caption{Over-approximation of the search region}
\label{fig:linear_safeable}
\end{center}
\end{figure}

Next, the safeable regions are linearized so that boundaries can be represented in Reluplex. The linearization over-approximates the unsafeable region by approximating quadratic bounds as a piecewise linear function. The approximation uses either an inner approximation connecting points on the curve, or an outer approximation using line segments tangent to the curve. The type of approximation used is chosen to over-approximate the unsafeable region. \Cref{fig:linear_safeable} shows the linearization of the safeable region that over-approximates the unsafeable region.

\begin{figure}[h]
\begin{center}
\input{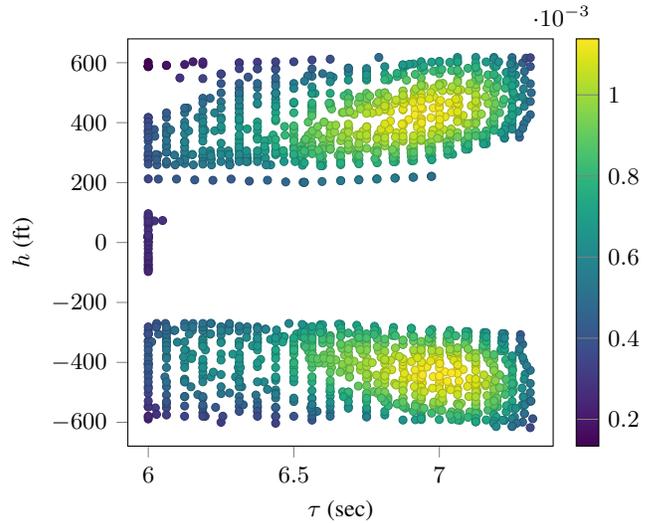}
\caption{Heat map of counterexamples for $a_\text{prev}$: Clear of Conflict}
\label{fig:linear_heatmap}
\end{center}
\end{figure}

Lastly, the region checked by Reluplex is split into small slices that are defined by a lower and upper bound on $\tau$ as well as a single linear lower bound on $h$ and a single linear upper bound on $h$. Because the neural network uses $\tau=6$ for inputs where $\tau=6$, the $\tau$ bounds are adjusted to ensure the network is evaluated at $\tau=6$ for inputs where $\tau<6$. Each small slice is checked as a separate query with Reluplex. A satisfiable set of inputs found by Reluplex represents a counterexample, or a set of network inputs that produce an unsafeable advisory when a safeable advisory exists. Because Reluplex is sound and complete, if Reluplex cannot find a counterexample for a query, then no counterexample exists.

\section{Results}
 To verify the unsafeable regions in all of the neural networks, each of the nine neural networks associated with one of the previous advisories is evaluated for all allowed advisories. Using a $\Delta v_O$ of 2 ft/s, there are 100 velocity ranges to verify. After slicing up each unsafeable region into small regions with linear bounds, a total of $42,032$ separate queries were generated and evaluated with Reluplex, which required 11 hours when using 9 independent threads. Each query was run with $\epsilon=1$ second (in all the figures $\epsilon=3$ seconds just for illustration purposes). As a result, $3,957$ counterexamples were discovered, about 9.14\% of all queries. A table of when these counterexamples occurred is shown in \cref{tab:counterexamples}, where N/A is used for advisories that are not allowed given the previous advisory. Most counterexamples occur for the COC advisory, but many other counterexamples exist for other advisories as well.\shivam{we should probably describe what each query is. 9 advisories, deltav etc.}\kyle{Added description}
\begin{figure} [t]
    \centering
    \input{Safeable_SAT_COC.tex}
    \caption{Unsafeable region for COC containing a counterexample}
    \label{fig:Safeable_SAT_COC}
\end{figure}
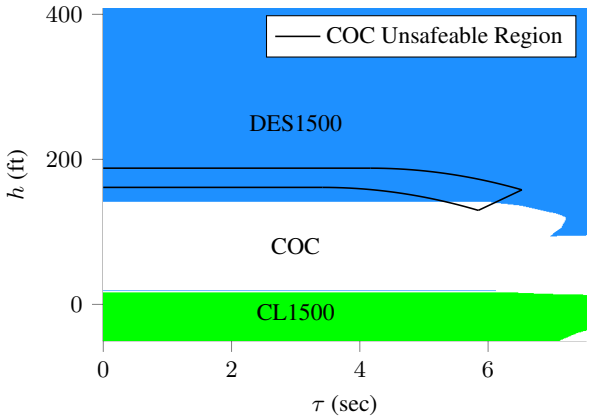

{\center
\input{ResultsTable.tex}
}

Visualizing the counterexamples in the form of a heat map allows for analysis of the network's performance. \cref{fig:linear_heatmap} plots all the counterexamples found by Reluplex for advisories issued after a clear of conflict advisory. No counterexamples are found in the white region in the middle of the plot because this region is unsafeable for all advisories and is omitted from the search region, as illustrated in \cref{fig:Safeable_CheckRegion}. The lighter points represent a higher probability density of counterexamples. The figure illustrates that counterexamples are most prevalent at around $\tau=\US{7}{\second}$. This information can be useful for tweaking networks to perform safely. Also, \cref{fig:linear_heatmap} shows rough vertical stripes, which is due to the preference of Reluplex to return SAT points that occur along the boundary of a region rather than somewhere in the middle of a region. \shivam{we should talk about why there are those vertical lines of SAT points} \kyle{Added a sentence to explain vertical striping.}

In addition to the $42,032$ separate queries run that are summarized in \cref{tab:counterexamples}, we ran $143,048$ queries on 25 independent threads to study the effect of the linearization approximation on the number of counterexamples generated. All advisories were checked for $a_\text{prev}$ = COC, linearized with line segment lengths of $\tau$= $0.125, 0.25, 0.5, 1.0,$ and $2.0$ seconds for both under and over-approximation. All linear segments were split into small regions of the same size so that the number of regions generated remained the same for all cases. Neither the method of linearization nor the level of discretization had any affect on the number of counterexamples found. For each level of discretization $1,476$ counterexamples were found for both the under-approximation and over-approximation method. This is most likely due to the fact that counterexamples are usually found around linear parts of the safeable bounds, so finer linearization had no impact.

Some of these counterexamples are informative, and visualizing the policy at these points reveals problems that need to be addressed. For example, \cref{fig:Safeable_SAT_COC} shows the unsafeable region for COC, which extends into a large area of COC. Given that the unsafeable region appears at low $\tau$ and $h$, a collision is imminent, and COC is not safe to give. This information can be used to refine the policy and network to discourage COC advisories in these situations. 

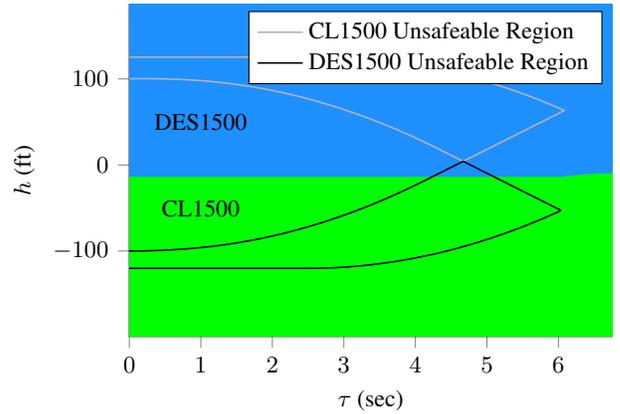
\begin{figure}
    \centering
    \input{Safeable_SAT_Boundary.tex}
    \caption{Unsafeable regions for CL1500 and DES1500 with DES1500 counterexample}
    \label{fig:Safeable_SAT_Boundary}
\end{figure}

Many counterexamples are found at the boundary between two alerting regions. As shown in \Cref{fig:Safeable_SAT_Boundary}, the regions being checked for DES1500 and CL1500 meet at a point. In order to avoid any counterexamples, the boundary between DES1500 and CL1500 must pass exactly through the point that divides the two unsafe regions. However, because the neural network is an approximation, the boundary is a little off, and a counterexample is discovered. In addition, no other advisory is safeable around the meeting point, so there is no other advisory the network could give to avoid a counterexample. Requiring the network to change advisories at an exact point in order to prove safety is too strict, so more work is needed to relax this requirement while still guaranteeing safety.

\section{Conclusions and Future Work}
\shivam{
    (1) Introduce safeable2 here maybe? Discuss how we are going to better handle the "invalid" counter examples. \\
}
After generating collision avoidance networks, linear safeable regions were defined for all possible advisories. The safeable regions define when an advisory can be made safe in the future, so that advisory is safe to give in the safeable region. If the system always gives safeable advisories when possible, then safety is guaranteed assuming the intruder begins in the safeable region. The safeable regions were checked with Reluplex, resulting in the discovery of thousands of counterexamples. The counterexamples can be used to refine the neural networks to improve safety.

A primary issue with proving safety using safeable regions is the hard safety requirement imposed on neural networks. The safeable property requires that the boundary between advisories given by the neural network must pass through an exact point in the state space. In reality, no neural network will be able to satisfy such a hard requirement in all situations.

To overcome this challenge, we have been exploring an extension to safeable, which we call safeable2. A safeable2 region is defined as a region that is safeable by at least two advisories. Verifying safety with safeable2 removes the hard requirement of the neural network having to switch advisories at a single point, but rather allows a small region to switch advisories. In addition, safeable2 omits a small region of uncertain behavior (the region that is safeable by only a single advisory) around the unsafeable region where a lot of counterexamples are found. It will be interesting to explore the implications of using safeable2 to verify safety and whether this method eliminates spurious counterexamples to safe operation. Furthermore, future work will model pilot delay to ensure safety can be guaranteed with realistic pilot compliance. \shivam{I'm not sure what else to include here - work on safeable2 is very premature, so I'm not sure what other claims we can make. }

\bibliographystyle{aaai}
\bibliography{references}

\end{document}

%% file: variables.tex
\tikzset{every picture/.style={line width=0.75pt}} 

\begin{tikzpicture}[x=0.75pt,y=0.75pt,yscale=-0.6,xscale=0.6]

\draw (135.63,119.04) node  {\includegraphics[width=48.91pt,height=28.82pt]{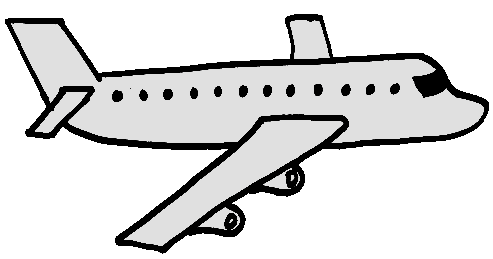}};
\draw  [fill={rgb, 255:red, 155; green, 155; blue, 155 }  ,fill opacity=0.65 ] (54.95,70.99) .. controls (54.95,65.1) and (91.37,60.32) .. (136.3,60.32) .. controls (181.23,60.32) and (217.65,65.1) .. (217.65,70.99) .. controls (217.65,76.89) and (181.23,81.67) .. (136.3,81.67) .. controls (91.37,81.67) and (54.95,76.89) .. (54.95,70.99) -- cycle ;
\draw    (54.95,70.99) -- (54.68,160.41) ;

\draw    (217.65,70.99) -- (216.7,161.74) ;

\draw  [draw opacity=0] (54.6,160.22) .. controls (67.11,171.13) and (100.23,178.48) .. (138.96,177.74) .. controls (172.75,177.1) and (202.03,170.47) .. (216.97,161.23) -- (138.41,148.96) -- cycle ; \draw   (54.6,160.22) .. controls (67.11,171.13) and (100.23,178.48) .. (138.96,177.74) .. controls (172.75,177.1) and (202.03,170.47) .. (216.97,161.23) ;
\draw (397.48,221.22) node [rotate=-358.9,xslant=0,xscale=-1] {\includegraphics[width=48.91pt,height=28.82pt]{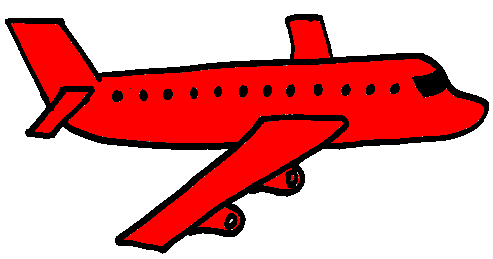}};
\draw [line width=1.5]  [dash pattern={on 5.63pt off 4.5pt}]  (134.34,115.03) -- (233.96,113.74) ;
\draw [shift={(236.96,113.7)}, rotate = 539.25] [fill={rgb, 255:red, 0; green, 0; blue, 0 }  ][line width=1.5]  [draw opacity=0] (11.61,-5.58) -- (0,0) -- (11.61,5.58) -- cycle    ;

\draw [line width=1.5]  [dash pattern={on 5.63pt off 4.5pt}]  (134.34,115.03) -- (134.3,7.2) ;
\draw [shift={(134.3,4.2)}, rotate = 449.98] [fill={rgb, 255:red, 0; green, 0; blue, 0 }  ][line width=1.5]  [draw opacity=0] (11.61,-5.58) -- (0,0) -- (11.61,5.58) -- cycle    ;

\draw [color={rgb, 255:red, 208; green, 2; blue, 27 }  ,draw opacity=1 ][fill={rgb, 255:red, 208; green, 2; blue, 27 }  ,fill opacity=1 ]   (134.34,115.03) -- (134.3,30.2) ;
\draw [shift={(134.3,28.2)}, rotate = 449.97] [fill={rgb, 255:red, 208; green, 2; blue, 27 }  ,fill opacity=1 ][line width=0.75]  [draw opacity=0] (8.93,-4.29) -- (0,0) -- (8.93,4.29) -- cycle    ;

\draw [color={rgb, 255:red, 208; green, 2; blue, 27 }  ,draw opacity=1 ][fill={rgb, 255:red, 208; green, 2; blue, 27 }  ,fill opacity=1 ]   (396.28,213.79) -- (396.3,285.2) ;
\draw [shift={(396.3,287.2)}, rotate = 269.98] [fill={rgb, 255:red, 208; green, 2; blue, 27 }  ,fill opacity=1 ][line width=0.75]  [draw opacity=0] (8.93,-4.29) -- (0,0) -- (8.93,4.29) -- cycle    ;

\draw [color={rgb, 255:red, 208; green, 2; blue, 27 }  ,draw opacity=1 ]   (292.31,111.03) -- (292.31,212.46) ;
\draw [shift={(292.31,212.46)}, rotate = 270] [color={rgb, 255:red, 208; green, 2; blue, 27 }  ,draw opacity=1 ][line width=0.75]    (0,5.59) -- (0,-5.59)(10.93,-3.29) .. controls (6.95,-1.4) and (3.31,-0.3) .. (0,0) .. controls (3.31,0.3) and (6.95,1.4) .. (10.93,3.29)   ;
\draw [shift={(292.31,111.03)}, rotate = 90] [color={rgb, 255:red, 208; green, 2; blue, 27 }  ,draw opacity=1 ][line width=0.75]    (0,5.59) -- (0,-5.59)(10.93,-3.29) .. controls (6.95,-1.4) and (3.31,-0.3) .. (0,0) .. controls (3.31,0.3) and (6.95,1.4) .. (10.93,3.29)   ;
\draw    (136.3,70.99) -- (56.95,70.99) ;
\draw [shift={(54.95,70.99)}, rotate = 360] [color={rgb, 255:red, 0; green, 0; blue, 0 }  ][line width=0.75]    (10.93,-3.29) .. controls (6.95,-1.4) and (3.31,-0.3) .. (0,0) .. controls (3.31,0.3) and (6.95,1.4) .. (10.93,3.29)   ;

\draw    (33.08,164.41) -- (33.08,68.32) ;
\draw [shift={(33.08,68.32)}, rotate = 450] [color={rgb, 255:red, 0; green, 0; blue, 0 }  ][line width=0.75]    (0,5.59) -- (0,-5.59)(10.93,-3.29) .. controls (6.95,-1.4) and (3.31,-0.3) .. (0,0) .. controls (3.31,0.3) and (6.95,1.4) .. (10.93,3.29)   ;
\draw [shift={(33.08,164.41)}, rotate = 270] [color={rgb, 255:red, 0; green, 0; blue, 0 }  ][line width=0.75]    (0,5.59) -- (0,-5.59)(10.93,-3.29) .. controls (6.95,-1.4) and (3.31,-0.3) .. (0,0) .. controls (3.31,0.3) and (6.95,1.4) .. (10.93,3.29)   ;

\draw (147.64,12.27) node  [align=left] {{\fontfamily{ptm}\selectfont y}};
\draw (251.4,112.36) node  [align=left] {{\fontfamily{ptm}\selectfont x}};
\draw (260.55,278.86) node  [align=left] {{\fontfamily{ptm}\selectfont Vertical intruder velocity }$\displaystyle ( v_{I})$};
\draw (280.31,35.61) node  [align=left] {{\fontfamily{ptm}\selectfont Vertical ownship velocity (}$\displaystyle v_{o}${\fontfamily{ptm}\selectfont )}};
\draw (400.42,150.06) node  [align=left] {{\fontfamily{ptm}\selectfont Vertical separation (}$\displaystyle h)$};
\draw (13.72,115.03) node  [align=left] {$\displaystyle 2h_{p}$};
\draw (89.38,48.64) node  [align=left] {$\displaystyle r_{p}$};

\end{tikzpicture}

%% file: Table_advisories.tex
\begin{table*}[t]
    \centering
    \label{tab:table1}
        \caption{\system advisories
        \label{tab:advisories}}
    \begin{tabular} {llcccc} 
    \toprule
      \textbf{Advisory} &\textbf{Description} &\textbf{Vertical Range} & \textbf{Strength}	&\textbf{Sign} &\textbf{Advisory} \\
       & & (Min, Max) $[\us{\foot\per\minute}]$ & $a_{lo}$ & $w$ & $v_{lo} \  [\us{\foot\per\minute}]$ \\ \midrule
      COC & Clear of Conflict & $(-\infty, +\infty)$ & $g/4$ & N/A & N/A\\ 
      DNC & Do Not Climb & $(-\infty, 0]$ & $g/4$ & $-1$ & $0$\\ 
      DND & Do Not Descend & $[0, +\infty)$ & $g/4$ & $+1$ & $0$\\
      DES1500 & Descend at least \US{1500}{\foot\per\minute} & $(-\infty, -1500]$ & $g/4$ & $-1$ & $-1500$\\
      CL1500 & Climb at least \US{1500}{\foot\per\minute} & $[+1500, +\infty)$ & $g/4$ & $+1$ & $+1500$ \\
      SDES1500 & Strengthen Descend to at least \US{1500}{\foot\per\minute} & $(-\infty, -1500]$ & $g/3$ & $-1$ & $-1500$ \\
      SCL1500 & Strengthen Climb to at least \US{1500}{\foot\per\minute} & $[+1500, +\infty)$ & $g/3$ & $+1$ & $+1500$ \\
      SDES2500 & Strengthen Descend to at least \US{2500}{\foot\per\minute} & $(-\infty, -2500]$ & $g/3$ & $-1$ & $-2500$ \\
      SCL2500 & Strengthen Climb to at least \US{2500}{\foot\per\minute} & $[+2500, +\infty)$ & $g/3$ & $+1$ & $+2500$ \\\bottomrule
    \end{tabular}
\end{table*}

%% file: PolicyPlots_Example.tex
\begin{tikzpicture}[]
\begin{groupplot}[group style={vertical sep=1.1cm, group size=1 by 2}]
\nextgroupplot [height = {6.5cm}, ylabel = {$h$ (ft)}, xmin = {0.0}, xmax = {40.0}, ymax = {1000.0}, ymin = {-1000.0}, width = {8cm}, enlargelimits = false, axis on top]\addplot [point meta min=1, point meta max=9] graphics [xmin=0.0, xmax=40.0, ymin=-1000.0, ymax=1000.0] {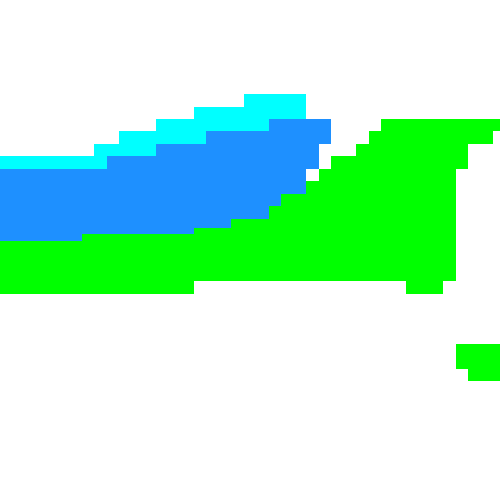};

\node[text=black] at (30,100)  {\footnotesize CL1500};
\node[text=black] at (15,250)  {\footnotesize DES1500};
\node[text=black] at (22,580)  {\footnotesize DNC};
\node[text=black] at (15,-600)  {\footnotesize COC};

\nextgroupplot [height = {6.5cm}, ylabel = {$h$ (ft)}, xmin = {0.0}, xmax = {40.0}, ymax = {1000.0}, xlabel = {$\tau$ (sec)}, ymin = {-1000.0}, width = {8cm}, enlargelimits = false, axis on top]\addplot [point meta min=1, point meta max=9] graphics [xmin=0.0, xmax=40.0, ymin=-1000.0, ymax=1000.0] {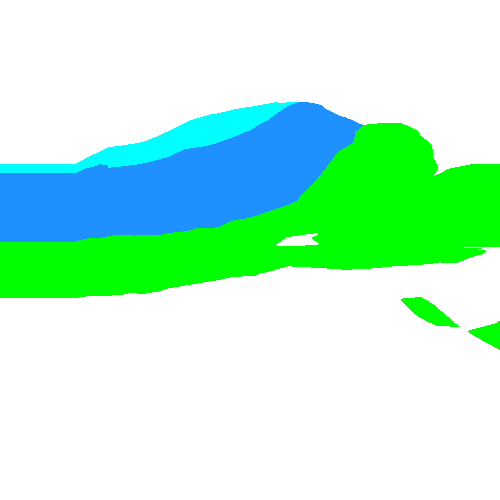};

\end{groupplot}
\end{tikzpicture}

%% file: SafeRegion1.tex
\begin{tikzpicture}

\definecolor{color0}{rgb}{0.133333333333333,0.545098039215686,0.133333333333333}

\begin{axis}[
tick align=outside,
tick pos=left,
x grid style={white!69.01960784313725!black},
xlabel={$\tau$ (sec)},
xmin=0, xmax=10.4399895962733,
y grid style={white!69.01960784313725!black},
ylabel={$h$ (ft)},
ymin=-190, ymax=151.045982142857,
legend style={xshift=-1cm}
]
\path [fill=color0] (axis cs:0,-190)
--(axis cs:0,-100)
--(axis cs:0.0303030303030303,-100.896473354232)
--(axis cs:0.0606060606060606,-101.792946708464)
--(axis cs:0.0909090909090909,-102.689420062696)
--(axis cs:0.121212121212121,-103.571100421576)
--(axis cs:0.151515151515152,-104.44233866609)
--(axis cs:0.181818181818182,-105.313576910604)
--(axis cs:0.212121212121212,-106.180464274133)
--(axis cs:0.242424242424242,-107.026467408929)
--(axis cs:0.272727272727273,-107.872470543725)
--(axis cs:0.303030303030303,-108.718473678521)
--(axis cs:0.333333333333333,-109.54533293698)
--(axis cs:0.363636363636364,-110.366100962058)
--(axis cs:0.393939393939394,-111.186868987137)
--(axis cs:0.424242424242424,-111.998935250243)
--(axis cs:0.454545454545455,-112.794468165604)
--(axis cs:0.484848484848485,-113.590001080964)
--(axis cs:0.515151515151515,-114.385533996325)
--(axis cs:0.545454545454545,-115.157572154362)
--(axis cs:0.575757575757576,-115.927869960004)
--(axis cs:0.606060606060606,-116.698167765647)
--(axis cs:0.636363636363636,-117.455412928332)
--(axis cs:0.666666666666667,-118.200475624257)
--(axis cs:0.696969696969697,-118.945538320182)
--(axis cs:0.727272727272727,-119.687990487515)
--(axis cs:0.757575757575758,-120.407818073722)
--(axis cs:0.787878787878788,-121.127645659929)
--(axis cs:0.818181818181818,-121.847473246136)
--(axis cs:0.848484848484849,-122.549897308399)
--(axis cs:0.878787878787879,-123.244489784888)
--(axis cs:0.909090909090909,-123.939082261377)
--(axis cs:0.939393939393939,-124.626713328289)
--(axis cs:0.96969696969697,-125.29607069506)
--(axis cs:1,-125.965428061831)
--(axis cs:1.03030303030303,-126.634785428602)
--(axis cs:1.06060606060606,-127.282388390444)
--(axis cs:1.09090909090909,-127.926510647498)
--(axis cs:1.12121212121212,-128.570632904551)
--(axis cs:1.15151515151515,-129.203442871041)
--(axis cs:1.18181818181818,-129.822330018376)
--(axis cs:1.21212121212121,-130.441217165712)
--(axis cs:1.24242424242424,-131.05923413685)
--(axis cs:1.27272727272727,-131.652886174468)
--(axis cs:1.3030303030303,-132.246538212085)
--(axis cs:1.33333333333333,-132.840190249703)
--(axis cs:1.36363636363636,-133.418179115771)
--(axis cs:1.39393939393939,-133.986596043671)
--(axis cs:1.42424242424242,-134.555012971571)
--(axis cs:1.45454545454545,-135.118208842287)
--(axis cs:1.48484848484848,-135.661390660469)
--(axis cs:1.51515151515152,-136.204572478651)
--(axis cs:1.54545454545455,-136.747754296833)
--(axis cs:1.57575757575758,-137.27092206248)
--(axis cs:1.60606060606061,-137.788868770944)
--(axis cs:1.63636363636364,-138.306815479408)
--(axis cs:1.66666666666667,-138.815190249703)
--(axis cs:1.6969696969697,-139.307901848449)
--(axis cs:1.72727272727273,-139.800613447195)
--(axis cs:1.75757575757576,-140.293325045941)
--(axis cs:1.78787878787879,-140.761671711166)
--(axis cs:1.81818181818182,-141.229148200195)
--(axis cs:1.84848484848485,-141.696624689223)
--(axis cs:1.87878787878788,-142.150178359096)
--(axis cs:1.90909090909091,-142.592419738407)
--(axis cs:1.93939393939394,-143.034661117717)
--(axis cs:1.96969696969697,-143.473421792239)
--(axis cs:2,-143.890428061831)
--(axis cs:2.03030303030303,-144.307434331424)
--(axis cs:2.06060606060606,-144.724440601016)
--(axis cs:2.09090909090909,-145.123173170468)
--(axis cs:2.12121212121212,-145.514944330343)
--(axis cs:2.15151515151515,-145.906715490217)
--(axis cs:2.18181818181818,-146.290655064317)
--(axis cs:2.21212121212121,-146.657191114474)
--(axis cs:2.24242424242424,-147.023727164631)
--(axis cs:2.27272727272727,-147.390263214788)
--(axis cs:2.3030303030303,-147.734174683818)
--(axis cs:2.33333333333333,-148.075475624257)
--(axis cs:2.36363636363636,-148.416776564696)
--(axis cs:2.39393939393939,-148.745895038374)
--(axis cs:2.42424242424242,-149.061960869095)
--(axis cs:2.45454545454545,-149.378026699816)
--(axis cs:2.48484848484848,-149.692352178143)
--(axis cs:2.51515151515152,-149.983182899146)
--(axis cs:2.54545454545455,-150.274013620149)
--(axis cs:2.57575757575758,-150.564844341152)
--(axis cs:2.60606060606061,-150.839141714409)
--(axis cs:2.63636363636364,-151.104737325695)
--(axis cs:2.66666666666667,-151.37033293698)
--(axis cs:2.6969696969697,-151.629837314885)
--(axis cs:2.72727272727273,-151.870197816452)
--(axis cs:2.75757575757576,-152.11055831802)
--(axis cs:2.78787878787879,-152.350918819587)
--(axis cs:2.81818181818182,-152.570395092422)
--(axis cs:2.84848484848485,-152.785520484272)
--(axis cs:2.87878787878788,-153.000645876122)
--(axis cs:2.90909090909091,-153.205329153605)
--(axis cs:2.93939393939394,-153.395219435737)
--(axis cs:2.96969696969697,-153.585109717868)
--(axis cs:3,-153.775)
--(axis cs:3,-190)
--(axis cs:3,-190)
--(axis cs:2.96969696969697,-190)
--(axis cs:2.93939393939394,-190)
--(axis cs:2.90909090909091,-190)
--(axis cs:2.87878787878788,-190)
--(axis cs:2.84848484848485,-190)
--(axis cs:2.81818181818182,-190)
--(axis cs:2.78787878787879,-190)
--(axis cs:2.75757575757576,-190)
--(axis cs:2.72727272727273,-190)
--(axis cs:2.6969696969697,-190)
--(axis cs:2.66666666666667,-190)
--(axis cs:2.63636363636364,-190)
--(axis cs:2.60606060606061,-190)
--(axis cs:2.57575757575758,-190)
--(axis cs:2.54545454545455,-190)
--(axis cs:2.51515151515152,-190)
--(axis cs:2.48484848484848,-190)
--(axis cs:2.45454545454545,-190)
--(axis cs:2.42424242424242,-190)
--(axis cs:2.39393939393939,-190)
--(axis cs:2.36363636363636,-190)
--(axis cs:2.33333333333333,-190)
--(axis cs:2.3030303030303,-190)
--(axis cs:2.27272727272727,-190)
--(axis cs:2.24242424242424,-190)
--(axis cs:2.21212121212121,-190)
--(axis cs:2.18181818181818,-190)
--(axis cs:2.15151515151515,-190)
--(axis cs:2.12121212121212,-190)
--(axis cs:2.09090909090909,-190)
--(axis cs:2.06060606060606,-190)
--(axis cs:2.03030303030303,-190)
--(axis cs:2,-190)
--(axis cs:1.96969696969697,-190)
--(axis cs:1.93939393939394,-190)
--(axis cs:1.90909090909091,-190)
--(axis cs:1.87878787878788,-190)
--(axis cs:1.84848484848485,-190)
--(axis cs:1.81818181818182,-190)
--(axis cs:1.78787878787879,-190)
--(axis cs:1.75757575757576,-190)
--(axis cs:1.72727272727273,-190)
--(axis cs:1.6969696969697,-190)
--(axis cs:1.66666666666667,-190)
--(axis cs:1.63636363636364,-190)
--(axis cs:1.60606060606061,-190)
--(axis cs:1.57575757575758,-190)
--(axis cs:1.54545454545455,-190)
--(axis cs:1.51515151515152,-190)
--(axis cs:1.48484848484848,-190)
--(axis cs:1.45454545454545,-190)
--(axis cs:1.42424242424242,-190)
--(axis cs:1.39393939393939,-190)
--(axis cs:1.36363636363636,-190)
--(axis cs:1.33333333333333,-190)
--(axis cs:1.3030303030303,-190)
--(axis cs:1.27272727272727,-190)
--(axis cs:1.24242424242424,-190)
--(axis cs:1.21212121212121,-190)
--(axis cs:1.18181818181818,-190)
--(axis cs:1.15151515151515,-190)
--(axis cs:1.12121212121212,-190)
--(axis cs:1.09090909090909,-190)
--(axis cs:1.06060606060606,-190)
--(axis cs:1.03030303030303,-190)
--(axis cs:1,-190)
--(axis cs:0.96969696969697,-190)
--(axis cs:0.939393939393939,-190)
--(axis cs:0.909090909090909,-190)
--(axis cs:0.878787878787879,-190)
--(axis cs:0.848484848484849,-190)
--(axis cs:0.818181818181818,-190)
--(axis cs:0.787878787878788,-190)
--(axis cs:0.757575757575758,-190)
--(axis cs:0.727272727272727,-190)
--(axis cs:0.696969696969697,-190)
--(axis cs:0.666666666666667,-190)
--(axis cs:0.636363636363636,-190)
--(axis cs:0.606060606060606,-190)
--(axis cs:0.575757575757576,-190)
--(axis cs:0.545454545454545,-190)
--(axis cs:0.515151515151515,-190)
--(axis cs:0.484848484848485,-190)
--(axis cs:0.454545454545455,-190)
--(axis cs:0.424242424242424,-190)
--(axis cs:0.393939393939394,-190)
--(axis cs:0.363636363636364,-190)
--(axis cs:0.333333333333333,-190)
--(axis cs:0.303030303030303,-190)
--(axis cs:0.272727272727273,-190)
--(axis cs:0.242424242424242,-190)
--(axis cs:0.212121212121212,-190)
--(axis cs:0.181818181818182,-190)
--(axis cs:0.151515151515152,-190)
--(axis cs:0.121212121212121,-190)
--(axis cs:0.0909090909090909,-190)
--(axis cs:0.0606060606060606,-190)
--(axis cs:0.0303030303030303,-190)
--(axis cs:0,-190)
--cycle;

\path [fill=color0] (axis cs:2.58620689655172,-190)
--(axis cs:2.58620689655172,-150.665130796671)
--(axis cs:3.51476960855121,-155.364301285626)
--(axis cs:3.51476960855121,-190)
--(axis cs:3.51476960855121,-190)
--(axis cs:2.58620689655172,-190)
--cycle;

\path [fill=color0] (axis cs:3,-190)
--(axis cs:3,-153.775)
--(axis cs:3.04471735993475,-154.011889347639)
--(axis cs:3.0894347198695,-154.248778695278)
--(axis cs:3.13415207980425,-154.451192868992)
--(axis cs:3.178869439739,-154.638667800672)
--(axis cs:3.22358679967376,-154.806606800461)
--(axis cs:3.26830415960851,-154.944667316182)
--(axis cs:3.31302151954326,-155.078131142046)
--(axis cs:3.35773887947801,-155.166777241808)
--(axis cs:3.40245623941276,-155.255423341571)
--(axis cs:3.44717359934751,-155.304997577551)
--(axis cs:3.49189095928226,-155.344229261354)
--(axis cs:3.53660831921701,-155.35932832341)
--(axis cs:3.58132567915177,-155.349145591255)
--(axis cs:3.62604303908652,-155.329769479386)
--(axis cs:3.67076039902127,-155.270172331271)
--(axis cs:3.71547775895602,-155.210575183156)
--(axis cs:3.76019511889077,-155.107309481404)
--(axis cs:3.80491247882552,-154.99829791733)
--(axis cs:3.84962983876027,-154.860557041653)
--(axis cs:3.89434719869502,-154.70213106162)
--(axis cs:3.93906455862978,-154.529915012018)
--(axis cs:3.98378191856453,-154.322074616027)
--(axis cs:4.02849927849928,-154.114234220035)
--(axis cs:4.07321663843403,-153.858128580549)
--(axis cs:4.11793399836878,-153.600873768599)
--(axis cs:4.16265135830353,-153.310292955188)
--(axis cs:4.20736871823828,-153.003623727279)
--(axis cs:4.25208607817303,-152.678567739943)
--(axis cs:4.29680343810779,-152.322484096075)
--(axis cs:4.34152079804254,-151.962952934815)
--(axis cs:4.38623815797729,-151.557454874988)
--(axis cs:4.43095551791204,-151.151956815161)
--(axis cs:4.47567287784679,-150.708536064017)
--(axis cs:4.52039023778154,-150.253623588231)
--(axis cs:4.56510759771629,-149.775727663162)
--(axis cs:4.60982495765104,-149.271400771417)
--(axis cs:4.6545423175858,-148.759029672423)
--(axis cs:4.69925967752055,-148.20528836472)
--(axis cs:4.7439770374553,-147.651547057016)
--(axis cs:4.78869439739005,-147.055286368138)
--(axis cs:4.8334117573248,-146.452130644476)
--(axis cs:4.87812911725955,-145.821394781674)
--(axis cs:4.9228464771943,-145.168824642052)
--(axis cs:4.96756383712905,-144.503613605325)
--(axis cs:5.01228119706381,-143.801629049745)
--(axis cs:5.05699855699856,-143.099644494164)
--(axis cs:5.10171591693331,-142.350543867553)
--(axis cs:5.14643327686806,-141.599144896014)
--(axis cs:5.19115063680281,-140.815569095478)
--(axis cs:5.23586799673756,-140.01475570798)
--(axis cs:5.28058535667231,-139.196704733519)
--(axis cs:5.32530271660706,-138.346476930062)
--(axis cs:5.37002007654182,-137.493950781677)
--(axis cs:5.41473743647657,-136.594308562261)
--(axis cs:5.45945479641132,-135.694666342845)
--(axis cs:5.50417215634607,-134.758250604576)
--(axis cs:5.54888951628082,-133.809193969201)
--(axis cs:5.59360687621557,-132.838303057007)
--(axis cs:5.63832423615032,-131.839832005673)
--(axis cs:5.68304159608507,-130.834465919555)
--(axis cs:5.72775895601983,-129.786580452262)
--(axis cs:5.77247631595458,-128.738694984969)
--(axis cs:5.81719367588933,-127.649439308967)
--(axis cs:5.86191103582408,-126.552139425715)
--(axis cs:5.90662839575883,-125.428408575788)
--(axis cs:5.95134575569358,-124.281694276577)
--(axis cs:5.99606311562833,-123.123488252725)
--(axis cs:6.04078047556308,-121.927359537556)
--(axis cs:6.08549783549784,-120.731230822386)
--(axis cs:6.13021519543259,-119.48913520865)
--(axis cs:6.17493255536734,-118.243592077522)
--(axis cs:6.21964991530209,-116.967021289861)
--(axis cs:6.26436727523684,-115.672063742774)
--(axis cs:6.30908463517159,-114.361017781189)
--(axis cs:6.35380199510634,-113.016645818143)
--(axis cs:6.39851935504109,-111.671124682632)
--(axis cs:6.44323671497585,-110.277338303627)
--(axis cs:6.4879540749106,-108.883551924622)
--(axis cs:6.53267143484535,-107.454141199228)
--(axis cs:6.5773887947801,-106.010940404264)
--(axis cs:6.62210615471485,-104.547054504946)
--(axis cs:6.6668235146496,-103.054439294023)
--(axis cs:6.71154087458435,-101.556078220779)
--(axis cs:6.7562582345191,-100.014048593897)
--(axis cs:6.80097559445386,-98.4720189670155)
--(axis cs:6.84569295438861,-96.8897683038881)
--(axis cs:6.89041031432336,-95.2983242610475)
--(axis cs:6.93512767425811,-93.6815984239953)
--(axis cs:6.97984503419286,-92.0407399651958)
--(axis cs:7.02456239412761,-90.3895389542187)
--(axis cs:7.06927975406236,-88.6992660794603)
--(axis cs:7.11399711399711,-87.0089932047018)
--(axis cs:7.15871447393187,-85.2739026038411)
--(axis cs:7.20343183386662,-83.5342153131237)
--(axis cs:7.24814919380137,-81.7646495383382)
--(axis cs:7.29286655373612,-79.9755478316619)
--(axis cs:7.33758391367087,-78.1715068829515)
--(axis cs:7.38230127360562,-76.3329907603163)
--(axis cs:7.42701863354037,-74.4944746376812)
--(axis cs:7.42701863354037,-190)
--(axis cs:7.42701863354037,-190)
--(axis cs:7.38230127360562,-190)
--(axis cs:7.33758391367087,-190)
--(axis cs:7.29286655373612,-190)
--(axis cs:7.24814919380137,-190)
--(axis cs:7.20343183386662,-190)
--(axis cs:7.15871447393187,-190)
--(axis cs:7.11399711399711,-190)
--(axis cs:7.06927975406236,-190)
--(axis cs:7.02456239412761,-190)
--(axis cs:6.97984503419286,-190)
--(axis cs:6.93512767425811,-190)
--(axis cs:6.89041031432336,-190)
--(axis cs:6.84569295438861,-190)
--(axis cs:6.80097559445386,-190)
--(axis cs:6.7562582345191,-190)
--(axis cs:6.71154087458435,-190)
--(axis cs:6.6668235146496,-190)
--(axis cs:6.62210615471485,-190)
--(axis cs:6.5773887947801,-190)
--(axis cs:6.53267143484535,-190)
--(axis cs:6.4879540749106,-190)
--(axis cs:6.44323671497585,-190)
--(axis cs:6.39851935504109,-190)
--(axis cs:6.35380199510634,-190)
--(axis cs:6.30908463517159,-190)
--(axis cs:6.26436727523684,-190)
--(axis cs:6.21964991530209,-190)
--(axis cs:6.17493255536734,-190)
--(axis cs:6.13021519543259,-190)
--(axis cs:6.08549783549784,-190)
--(axis cs:6.04078047556308,-190)
--(axis cs:5.99606311562833,-190)
--(axis cs:5.95134575569358,-190)
--(axis cs:5.90662839575883,-190)
--(axis cs:5.86191103582408,-190)
--(axis cs:5.81719367588933,-190)
--(axis cs:5.77247631595458,-190)
--(axis cs:5.72775895601983,-190)
--(axis cs:5.68304159608507,-190)
--(axis cs:5.63832423615032,-190)
--(axis cs:5.59360687621557,-190)
--(axis cs:5.54888951628082,-190)
--(axis cs:5.50417215634607,-190)
--(axis cs:5.45945479641132,-190)
--(axis cs:5.41473743647657,-190)
--(axis cs:5.37002007654182,-190)
--(axis cs:5.32530271660706,-190)
--(axis cs:5.28058535667231,-190)
--(axis cs:5.23586799673756,-190)
--(axis cs:5.19115063680281,-190)
--(axis cs:5.14643327686806,-190)
--(axis cs:5.10171591693331,-190)
--(axis cs:5.05699855699856,-190)
--(axis cs:5.01228119706381,-190)
--(axis cs:4.96756383712905,-190)
--(axis cs:4.9228464771943,-190)
--(axis cs:4.87812911725955,-190)
--(axis cs:4.8334117573248,-190)
--(axis cs:4.78869439739005,-190)
--(axis cs:4.7439770374553,-190)
--(axis cs:4.69925967752055,-190)
--(axis cs:4.6545423175858,-190)
--(axis cs:4.60982495765104,-190)
--(axis cs:4.56510759771629,-190)
--(axis cs:4.52039023778154,-190)
--(axis cs:4.47567287784679,-190)
--(axis cs:4.43095551791204,-190)
--(axis cs:4.38623815797729,-190)
--(axis cs:4.34152079804254,-190)
--(axis cs:4.29680343810779,-190)
--(axis cs:4.25208607817303,-190)
--(axis cs:4.20736871823828,-190)
--(axis cs:4.16265135830353,-190)
--(axis cs:4.11793399836878,-190)
--(axis cs:4.07321663843403,-190)
--(axis cs:4.02849927849928,-190)
--(axis cs:3.98378191856453,-190)
--(axis cs:3.93906455862978,-190)
--(axis cs:3.89434719869502,-190)
--(axis cs:3.84962983876027,-190)
--(axis cs:3.80491247882552,-190)
--(axis cs:3.76019511889077,-190)
--(axis cs:3.71547775895602,-190)
--(axis cs:3.67076039902127,-190)
--(axis cs:3.62604303908652,-190)
--(axis cs:3.58132567915177,-190)
--(axis cs:3.53660831921701,-190)
--(axis cs:3.49189095928226,-190)
--(axis cs:3.44717359934751,-190)
--(axis cs:3.40245623941276,-190)
--(axis cs:3.35773887947801,-190)
--(axis cs:3.31302151954326,-190)
--(axis cs:3.26830415960851,-190)
--(axis cs:3.22358679967376,-190)
--(axis cs:3.178869439739,-190)
--(axis cs:3.13415207980425,-190)
--(axis cs:3.0894347198695,-190)
--(axis cs:3.04471735993475,-190)
--(axis cs:3,-190)
--cycle;

\path [fill=color0] (axis cs:7.01520294669941,-190)
--(axis cs:7.01520294669941,-90.7433169977729)
--(axis cs:7.9464963857357,-52.849568296209)
--(axis cs:7.9464963857357,-190)
--(axis cs:7.9464963857357,-190)
--(axis cs:7.01520294669941,-190)
--cycle;

\path [fill=color0] (axis cs:7.42701863354037,-190)
--(axis cs:7.42701863354037,-74.4944746376812)
--(axis cs:7.45745268366899,-73.2263892156555)
--(axis cs:7.4878867337976,-71.9583037936299)
--(axis cs:7.51832078392622,-70.6902183716043)
--(axis cs:7.54875483405483,-69.4221329495786)
--(axis cs:7.57918888418345,-68.154047527553)
--(axis cs:7.60962293431206,-66.8859621055273)
--(axis cs:7.64005698444068,-65.6178766835017)
--(axis cs:7.6704910345693,-64.3497912614761)
--(axis cs:7.70092508469791,-63.0817058394504)
--(axis cs:7.73135913482653,-61.8136204174248)
--(axis cs:7.76179318495514,-60.5455349953991)
--(axis cs:7.79222723508376,-59.2774495733735)
--(axis cs:7.82266128521237,-58.0093641513479)
--(axis cs:7.85309533534099,-56.7412787293222)
--(axis cs:7.8835293854696,-55.4731933072966)
--(axis cs:7.91396343559822,-54.2051078852709)
--(axis cs:7.94439748572683,-52.9370224632453)
--(axis cs:7.97483153585545,-51.6689370412196)
--(axis cs:8.00526558598406,-50.400851619194)
--(axis cs:8.03569963611268,-49.1327661971684)
--(axis cs:8.06613368624129,-47.8646807751428)
--(axis cs:8.09656773636991,-46.5965953531171)
--(axis cs:8.12700178649852,-45.3285099310915)
--(axis cs:8.15743583662714,-44.0604245090658)
--(axis cs:8.18786988675576,-42.7923390870402)
--(axis cs:8.21830393688437,-41.5242536650145)
--(axis cs:8.24873798701299,-40.2561682429889)
--(axis cs:8.2791720371416,-38.9880828209632)
--(axis cs:8.30960608727022,-37.7199973989376)
--(axis cs:8.34004013739883,-36.451911976912)
--(axis cs:8.37047418752745,-35.1838265548864)
--(axis cs:8.40090823765606,-33.9157411328607)
--(axis cs:8.43134228778468,-32.6476557108351)
--(axis cs:8.46177633791329,-31.3795702888094)
--(axis cs:8.49221038804191,-30.1114848667838)
--(axis cs:8.52264443817053,-28.8433994447581)
--(axis cs:8.55307848829914,-27.5753140227325)
--(axis cs:8.58351253842775,-26.3072286007069)
--(axis cs:8.61394658855637,-25.0391431786813)
--(axis cs:8.64438063868499,-23.7710577566556)
--(axis cs:8.6748146888136,-22.50297233463)
--(axis cs:8.70524873894222,-21.2348869126043)
--(axis cs:8.73568278907083,-19.9668014905787)
--(axis cs:8.76611683919945,-18.698716068553)
--(axis cs:8.79655088932806,-17.4306306465274)
--(axis cs:8.82698493945668,-16.1625452245018)
--(axis cs:8.85741898958529,-14.8944598024761)
--(axis cs:8.88785303971391,-13.6263743804505)
--(axis cs:8.91828708984252,-12.3582889584248)
--(axis cs:8.94872113997114,-11.0902035363992)
--(axis cs:8.97915519009976,-9.82211811437356)
--(axis cs:9.00958924022837,-8.55403269234791)
--(axis cs:9.04002329035699,-7.28594727032226)
--(axis cs:9.0704573404856,-6.01786184829668)
--(axis cs:9.10089139061422,-4.74977642627103)
--(axis cs:9.13132544074283,-3.48169100424539)
--(axis cs:9.16175949087145,-2.21360558221974)
--(axis cs:9.19219354100006,-0.945520160194093)
--(axis cs:9.22262759112868,0.322565261831555)
--(axis cs:9.25306164125729,1.5906506838572)
--(axis cs:9.28349569138591,2.85873610588285)
--(axis cs:9.31392974151452,4.1268215279085)
--(axis cs:9.34436379164314,5.39490694993408)
--(axis cs:9.37479784177175,6.66299237195973)
--(axis cs:9.40523189190037,7.93107779398538)
--(axis cs:9.43566594202898,9.19916321601102)
--(axis cs:9.4660999921576,10.4672486380367)
--(axis cs:9.49653404228622,11.7353340600623)
--(axis cs:9.52696809241483,13.003419482088)
--(axis cs:9.55740214254345,14.2715049041135)
--(axis cs:9.58783619267206,15.5395903261392)
--(axis cs:9.61827024280068,16.8076757481648)
--(axis cs:9.64870429292929,18.0757611701905)
--(axis cs:9.67913834305791,19.3438465922161)
--(axis cs:9.70957239318652,20.6119320142418)
--(axis cs:9.74000644331514,21.8800174362674)
--(axis cs:9.77044049344375,23.1481028582931)
--(axis cs:9.80087454357237,24.4161882803187)
--(axis cs:9.83130859370099,25.6842737023444)
--(axis cs:9.8617426438296,26.9523591243699)
--(axis cs:9.89217669395821,28.2204445463956)
--(axis cs:9.92261074408683,29.4885299684212)
--(axis cs:9.95304479421545,30.7566153904469)
--(axis cs:9.98347884434406,32.0247008124725)
--(axis cs:10.0139128944727,33.2927862344982)
--(axis cs:10.0443469446013,34.5608716565238)
--(axis cs:10.0747809947299,35.8289570785494)
--(axis cs:10.1052150448585,37.0970425005751)
--(axis cs:10.1356490949871,38.3651279226007)
--(axis cs:10.1660831451158,39.6332133446264)
--(axis cs:10.1965171952444,40.901298766652)
--(axis cs:10.226951245373,42.1693841886776)
--(axis cs:10.2573852955016,43.4374696107033)
--(axis cs:10.2878193456302,44.705555032729)
--(axis cs:10.3182533957588,45.9736404547546)
--(axis cs:10.3486874458874,47.2417258767803)
--(axis cs:10.3791214960161,48.5098112988058)
--(axis cs:10.4095555461447,49.7778967208315)
--(axis cs:10.4399895962733,51.0459821428571)
--(axis cs:10.4399895962733,-190)
--(axis cs:10.4399895962733,-190)
--(axis cs:10.4095555461447,-190)
--(axis cs:10.3791214960161,-190)
--(axis cs:10.3486874458874,-190)
--(axis cs:10.3182533957588,-190)
--(axis cs:10.2878193456302,-190)
--(axis cs:10.2573852955016,-190)
--(axis cs:10.226951245373,-190)
--(axis cs:10.1965171952444,-190)
--(axis cs:10.1660831451158,-190)
--(axis cs:10.1356490949871,-190)
--(axis cs:10.1052150448585,-190)
--(axis cs:10.0747809947299,-190)
--(axis cs:10.0443469446013,-190)
--(axis cs:10.0139128944727,-190)
--(axis cs:9.98347884434406,-190)
--(axis cs:9.95304479421545,-190)
--(axis cs:9.92261074408683,-190)
--(axis cs:9.89217669395821,-190)
--(axis cs:9.8617426438296,-190)
--(axis cs:9.83130859370099,-190)
--(axis cs:9.80087454357237,-190)
--(axis cs:9.77044049344375,-190)
--(axis cs:9.74000644331514,-190)
--(axis cs:9.70957239318652,-190)
--(axis cs:9.67913834305791,-190)
--(axis cs:9.64870429292929,-190)
--(axis cs:9.61827024280068,-190)
--(axis cs:9.58783619267206,-190)
--(axis cs:9.55740214254345,-190)
--(axis cs:9.52696809241483,-190)
--(axis cs:9.49653404228622,-190)
--(axis cs:9.4660999921576,-190)
--(axis cs:9.43566594202898,-190)
--(axis cs:9.40523189190037,-190)
--(axis cs:9.37479784177175,-190)
--(axis cs:9.34436379164314,-190)
--(axis cs:9.31392974151452,-190)
--(axis cs:9.28349569138591,-190)
--(axis cs:9.25306164125729,-190)
--(axis cs:9.22262759112868,-190)
--(axis cs:9.19219354100006,-190)
--(axis cs:9.16175949087145,-190)
--(axis cs:9.13132544074283,-190)
--(axis cs:9.10089139061422,-190)
--(axis cs:9.0704573404856,-190)
--(axis cs:9.04002329035699,-190)
--(axis cs:9.00958924022837,-190)
--(axis cs:8.97915519009976,-190)
--(axis cs:8.94872113997114,-190)
--(axis cs:8.91828708984252,-190)
--(axis cs:8.88785303971391,-190)
--(axis cs:8.85741898958529,-190)
--(axis cs:8.82698493945668,-190)
--(axis cs:8.79655088932806,-190)
--(axis cs:8.76611683919945,-190)
--(axis cs:8.73568278907083,-190)
--(axis cs:8.70524873894222,-190)
--(axis cs:8.6748146888136,-190)
--(axis cs:8.64438063868499,-190)
--(axis cs:8.61394658855637,-190)
--(axis cs:8.58351253842775,-190)
--(axis cs:8.55307848829914,-190)
--(axis cs:8.52264443817053,-190)
--(axis cs:8.49221038804191,-190)
--(axis cs:8.46177633791329,-190)
--(axis cs:8.43134228778468,-190)
--(axis cs:8.40090823765606,-190)
--(axis cs:8.37047418752745,-190)
--(axis cs:8.34004013739883,-190)
--(axis cs:8.30960608727022,-190)
--(axis cs:8.2791720371416,-190)
--(axis cs:8.24873798701299,-190)
--(axis cs:8.21830393688437,-190)
--(axis cs:8.18786988675576,-190)
--(axis cs:8.15743583662714,-190)
--(axis cs:8.12700178649852,-190)
--(axis cs:8.09656773636991,-190)
--(axis cs:8.06613368624129,-190)
--(axis cs:8.03569963611268,-190)
--(axis cs:8.00526558598406,-190)
--(axis cs:7.97483153585545,-190)
--(axis cs:7.94439748572683,-190)
--(axis cs:7.91396343559822,-190)
--(axis cs:7.8835293854696,-190)
--(axis cs:7.85309533534099,-190)
--(axis cs:7.82266128521237,-190)
--(axis cs:7.79222723508376,-190)
--(axis cs:7.76179318495514,-190)
--(axis cs:7.73135913482653,-190)
--(axis cs:7.70092508469791,-190)
--(axis cs:7.6704910345693,-190)
--(axis cs:7.64005698444068,-190)
--(axis cs:7.60962293431206,-190)
--(axis cs:7.57918888418345,-190)
--(axis cs:7.54875483405483,-190)
--(axis cs:7.51832078392622,-190)
--(axis cs:7.4878867337976,-190)
--(axis cs:7.45745268366899,-190)
--(axis cs:7.42701863354037,-190)
--cycle;

\addplot [semithick, black, forget plot]
table [row sep=\\]{%
0	-100 \\
0.103448275862069	-103.060374554102 \\
0.206896551724138	-106.034601664685 \\
0.310344827586207	-108.922681331748 \\
0.413793103448276	-111.724613555291 \\
0.517241379310345	-114.440398335315 \\
0.620689655172414	-117.070035671819 \\
0.724137931034483	-119.613525564804 \\
0.827586206896552	-122.070868014269 \\
0.931034482758621	-124.442063020214 \\
1.03448275862069	-126.72711058264 \\
1.13793103448276	-128.926010701546 \\
1.24137931034483	-131.038763376932 \\
1.3448275862069	-133.065368608799 \\
1.44827586206897	-135.005826397146 \\
1.55172413793103	-136.860136741974 \\
1.6551724137931	-138.628299643282 \\
1.75862068965517	-140.31031510107 \\
1.86206896551724	-141.906183115339 \\
1.96551724137931	-143.415903686088 \\
2.06896551724138	-144.839476813318 \\
2.17241379310345	-146.176902497027 \\
2.27586206896552	-147.428180737218 \\
2.37931034482759	-148.593311533888 \\
2.48275862068966	-149.672294887039 \\
2.58620689655172	-150.665130796671 \\
2.68965517241379	-151.571819262782 \\
2.79310344827586	-152.392360285375 \\
2.89655172413793	-153.126753864447 \\
3	-153.775 \\
};
\addplot [semithick, black, forget plot]
table [row sep=\\]{%
3	-153.775 \\
3.10295392171024	-154.320396405029 \\
3.20590784342048	-154.752024736106 \\
3.30886176513072	-155.069884993231 \\
3.41181568684096	-155.273977176404 \\
3.51476960855121	-155.364301285626 \\
3.61772353026145	-155.340857320895 \\
3.72067745197169	-155.203645282213 \\
3.82363137368193	-154.952665169579 \\
3.92658529539217	-154.587916982992 \\
4.02953921710241	-154.109400722454 \\
4.13249313881265	-153.517116387964 \\
4.23544706052289	-152.811063979522 \\
4.33840098223314	-151.991243497128 \\
4.44135490394338	-151.057654940783 \\
4.54430882565362	-150.010298310485 \\
4.64726274736386	-148.849173606236 \\
4.7502166690741	-147.574280828034 \\
4.85317059078434	-146.185619975881 \\
4.95612451249458	-144.683191049776 \\
5.05907843420482	-143.066994049719 \\
5.16203235591507	-141.33702897571 \\
5.26498627762531	-139.493295827749 \\
5.36794019933555	-137.535794605836 \\
5.47089412104579	-135.464525309971 \\
5.57384804275603	-133.279487940155 \\
5.67680196446627	-130.980682496386 \\
5.77975588617651	-128.568108978666 \\
5.88270980788675	-126.041767386993 \\
5.98566372959699	-123.401657721369 \\
6.08861765130724	-120.647779981793 \\
6.19157157301748	-117.780134168265 \\
6.29452549472772	-114.798720280785 \\
6.39747941643796	-111.703538319353 \\
6.5004333381482	-108.49458828397 \\
6.60338725985844	-105.171870174634 \\
6.70634118156868	-101.735383991347 \\
6.80929510327893	-98.1851297341073 \\
6.91224902498917	-94.521107402916 \\
7.01520294669941	-90.7433169977729 \\
7.11815686840965	-86.8517585186777 \\
7.22111079011989	-82.8464319656308 \\
7.32406471183013	-78.7273373386319 \\
7.42701863354037	-74.4944746376812 \\
};
\addplot [semithick, black, forget plot]
table [row sep=\\]{%
7.42701863354037	-74.4944746376812 \\
7.53091418397944	-70.1654933693867 \\
7.6348097344185	-65.8365121010923 \\
7.73870528485757	-61.5075308327979 \\
7.84260083529664	-57.1785495645035 \\
7.9464963857357	-52.849568296209 \\
8.05039193617477	-48.5205870279146 \\
8.15428748661384	-44.1916057596202 \\
8.2581830370529	-39.8626244913258 \\
8.36207858749197	-35.5336432230313 \\
8.46597413793103	-31.2046619547369 \\
8.5698696883701	-26.8756806864425 \\
8.67376523880917	-22.5466994181481 \\
8.77766078924823	-18.2177181498537 \\
8.8815563396873	-13.8887368815592 \\
8.98545189012636	-9.55975561326481 \\
9.08934744056543	-5.23077434497037 \\
9.1932429910045	-0.901793076675943 \\
9.29713854144356	3.42718819161841 \\
9.40103409188263	7.75616945991285 \\
9.5049296423217	12.0851507282073 \\
9.60882519276076	16.4141319965017 \\
9.71272074319983	20.7431132647962 \\
9.81661629363889	25.0720945330906 \\
9.92051184407796	29.401075801385 \\
10.024407394517	33.7300570696795 \\
10.1283029449561	38.0590383379738 \\
10.2321984953952	42.3880196062682 \\
10.3360940458342	46.7170008745627 \\
10.4399895962733	51.0459821428571 \\
};
\addplot [semithick, blue, dashed, forget plot]
table [row sep=\\]{%
0	0 \\
0.103448275862069	-3.06037455410226 \\
0.206896551724138	-6.0346016646849 \\
0.310344827586207	-8.92268133174792 \\
0.413793103448276	-11.7246135552913 \\
0.517241379310345	-14.4403983353151 \\
0.620689655172414	-17.0700356718193 \\
0.724137931034483	-19.6135255648038 \\
0.827586206896552	-22.0708680142687 \\
0.931034482758621	-24.442063020214 \\
1.03448275862069	-26.7271105826397 \\
1.13793103448276	-28.9260107015458 \\
1.24137931034483	-31.0387633769322 \\
1.3448275862069	-33.065368608799 \\
1.44827586206897	-35.0058263971463 \\
1.55172413793103	-36.8601367419738 \\
1.6551724137931	-38.6282996432818 \\
1.75862068965517	-40.3103151010702 \\
1.86206896551724	-41.9061831153389 \\
1.96551724137931	-43.415903686088 \\
2.06896551724138	-44.8394768133175 \\
2.17241379310345	-46.1769024970274 \\
2.27586206896552	-47.4281807372176 \\
2.37931034482759	-48.5933115338882 \\
2.48275862068966	-49.6722948870392 \\
2.58620689655172	-50.6651307966706 \\
2.68965517241379	-51.5718192627824 \\
2.79310344827586	-52.3923602853746 \\
2.89655172413793	-53.1267538644471 \\
3	-53.775 \\
};
\addplot [semithick, blue, dashed, forget plot]
table [row sep=\\]{%
3	-53.775 \\
3.10295392171024	-54.320396405029 \\
3.20590784342048	-54.752024736106 \\
3.30886176513072	-55.0698849932312 \\
3.41181568684096	-55.2739771764045 \\
3.51476960855121	-55.3643012856258 \\
3.61772353026145	-55.3408573208953 \\
3.72067745197169	-55.2036452822129 \\
3.82363137368193	-54.9526651695785 \\
3.92658529539217	-54.5879169829923 \\
4.02953921710241	-54.1094007224542 \\
4.13249313881265	-53.5171163879642 \\
4.23544706052289	-52.8110639795222 \\
4.33840098223314	-51.9912434971284 \\
4.44135490394338	-51.0576549407827 \\
4.54430882565362	-50.0102983104851 \\
4.64726274736386	-48.8491736062356 \\
4.7502166690741	-47.5742808280342 \\
4.85317059078434	-46.1856199758809 \\
4.95612451249458	-44.6831910497757 \\
5.05907843420482	-43.0669940497185 \\
5.16203235591507	-41.3370289757095 \\
5.26498627762531	-39.4932958277487 \\
5.36794019933555	-37.5357946058358 \\
5.47089412104579	-35.4645253099711 \\
5.57384804275603	-33.2794879401545 \\
5.67680196446627	-30.9806824963861 \\
5.77975588617651	-28.5681089786656 \\
5.88270980788675	-26.0417673869934 \\
5.98566372959699	-23.4016577213692 \\
6.08861765130724	-20.6477799817931 \\
6.19157157301748	-17.7801341682651 \\
6.29452549472772	-14.7987202807852 \\
6.39747941643796	-11.7035383193534 \\
6.5004333381482	-8.49458828396972 \\
6.60338725985844	-5.17187017463417 \\
6.70634118156868	-1.73538399134667 \\
6.80929510327893	1.81487026589272 \\
6.91224902498917	5.478892597084 \\
7.01520294669941	9.25668300222716 \\
7.11815686840965	13.1482414813223 \\
7.22111079011989	17.1535680343692 \\
7.32406471183013	21.2726626613681 \\
7.42701863354037	25.5055253623189 \\
};
\addplot [semithick, blue, dashed, forget plot]
table [row sep=\\]{%
7.42701863354037	25.5055253623189 \\
7.53091418397944	29.8345066306133 \\
7.6348097344185	34.1634878989077 \\
7.73870528485757	38.4924691672021 \\
7.84260083529664	42.8214504354966 \\
7.9464963857357	47.150431703791 \\
8.05039193617477	51.4794129720854 \\
8.15428748661384	55.8083942403798 \\
8.2581830370529	60.1373755086743 \\
8.36207858749197	64.4663567769687 \\
8.46597413793103	68.7953380452631 \\
8.5698696883701	73.1243193135576 \\
8.67376523880917	77.4533005818519 \\
8.77766078924823	81.7822818501463 \\
8.8815563396873	86.1112631184408 \\
8.98545189012636	90.4402443867352 \\
9.08934744056543	94.7692256550297 \\
9.1932429910045	99.0982069233241 \\
9.29713854144356	103.427188191618 \\
9.40103409188263	107.756169459913 \\
9.5049296423217	112.085150728207 \\
9.60882519276076	116.414131996502 \\
9.71272074319983	120.743113264796 \\
9.81661629363889	125.072094533091 \\
9.92051184407796	129.401075801385 \\
10.024407394517	133.73005706968 \\
10.1283029449561	138.059038337974 \\
10.2321984953952	142.388019606268 \\
10.3360940458342	146.717000874563 \\
10.4399895962733	151.045982142857 \\
};
  \node (fig2) at (0.55,-10)
   {\includegraphics[scale=0.05]{jbplane.png}};
   
     \node (fig3) at (9,-90)
   {\includegraphics[scale=0.05]{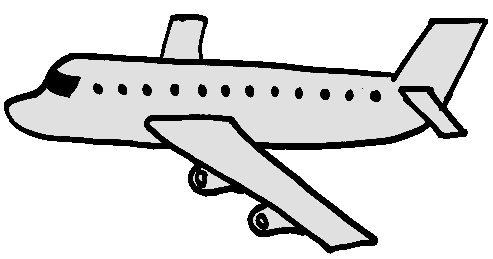}};
   
  \node[text=black] at (3.5,-100)  {\footnotesize Unsafe for CL1500};
   \node[text=white] at (8.5,-150) {\footnotesize Safe for CL1500};
  \node[text=black] at (1,25)  {\footnotesize Ownship};
  \node[text=white] at (9,-65)  {\footnotesize Intruder};

\end{axis}

\end{tikzpicture}

%% file: SafeRegion2.tex
\begin{tikzpicture}

\definecolor{color0}{rgb}{0.133333333333333,0.545098039215686,0.133333333333333}

\begin{axis}[
tick align=outside,
tick pos=left,
x grid style={white!69.01960784313725!black},
xlabel={$\tau$ (sec)},
xmin=0, xmax=10.4399895962733,
y grid style={white!69.01960784313725!black},
ylabel={$h$ (ft)},
ymin=-190, ymax=151.045982142857,
legend style={xshift=-1cm}
]
\path [fill=color0] (axis cs:0,-190)
--(axis cs:0,-155.369215838509)
--(axis cs:0.0358083945040467,-155.369215838509)
--(axis cs:0.0716167890080933,-155.369215838509)
--(axis cs:0.10742518351214,-155.369215838509)
--(axis cs:0.143233578016187,-155.369215838509)
--(axis cs:0.179041972520233,-155.369215838509)
--(axis cs:0.21485036702428,-155.369215838509)
--(axis cs:0.250658761528327,-155.369215838509)
--(axis cs:0.286467156032373,-155.369215838509)
--(axis cs:0.32227555053642,-155.369215838509)
--(axis cs:0.358083945040467,-155.369215838509)
--(axis cs:0.393892339544513,-155.369215838509)
--(axis cs:0.42970073404856,-155.369215838509)
--(axis cs:0.465509128552607,-155.369215838509)
--(axis cs:0.501317523056653,-155.369215838509)
--(axis cs:0.5371259175607,-155.369215838509)
--(axis cs:0.572934312064747,-155.369215838509)
--(axis cs:0.608742706568793,-155.369215838509)
--(axis cs:0.64455110107284,-155.369215838509)
--(axis cs:0.680359495576887,-155.369215838509)
--(axis cs:0.716167890080933,-155.369215838509)
--(axis cs:0.75197628458498,-155.369215838509)
--(axis cs:0.787784679089027,-155.369215838509)
--(axis cs:0.823593073593073,-155.369215838509)
--(axis cs:0.85940146809712,-155.369215838509)
--(axis cs:0.895209862601167,-155.369215838509)
--(axis cs:0.931018257105213,-155.369215838509)
--(axis cs:0.96682665160926,-155.369215838509)
--(axis cs:1.00263504611331,-155.369215838509)
--(axis cs:1.03844344061735,-155.369215838509)
--(axis cs:1.0742518351214,-155.369215838509)
--(axis cs:1.11006022962545,-155.369215838509)
--(axis cs:1.14586862412949,-155.369215838509)
--(axis cs:1.18167701863354,-155.369215838509)
--(axis cs:1.21748541313759,-155.369215838509)
--(axis cs:1.25329380764163,-155.369215838509)
--(axis cs:1.28910220214568,-155.369215838509)
--(axis cs:1.32491059664973,-155.369215838509)
--(axis cs:1.36071899115377,-155.369215838509)
--(axis cs:1.39652738565782,-155.369215838509)
--(axis cs:1.43233578016187,-155.369215838509)
--(axis cs:1.46814417466591,-155.369215838509)
--(axis cs:1.50395256916996,-155.369215838509)
--(axis cs:1.53976096367401,-155.369215838509)
--(axis cs:1.57556935817805,-155.369215838509)
--(axis cs:1.6113777526821,-155.369215838509)
--(axis cs:1.64718614718615,-155.369215838509)
--(axis cs:1.68299454169019,-155.369215838509)
--(axis cs:1.71880293619424,-155.369215838509)
--(axis cs:1.75461133069829,-155.369215838509)
--(axis cs:1.79041972520233,-155.369215838509)
--(axis cs:1.82622811970638,-155.369215838509)
--(axis cs:1.86203651421043,-155.369215838509)
--(axis cs:1.89784490871447,-155.369215838509)
--(axis cs:1.93365330321852,-155.369215838509)
--(axis cs:1.96946169772257,-155.369215838509)
--(axis cs:2.00527009222661,-155.369215838509)
--(axis cs:2.04107848673066,-155.369215838509)
--(axis cs:2.07688688123471,-155.369215838509)
--(axis cs:2.11269527573875,-155.369215838509)
--(axis cs:2.1485036702428,-155.369215838509)
--(axis cs:2.18431206474685,-155.369215838509)
--(axis cs:2.22012045925089,-155.369215838509)
--(axis cs:2.25592885375494,-155.369215838509)
--(axis cs:2.29173724825899,-155.369215838509)
--(axis cs:2.32754564276303,-155.369215838509)
--(axis cs:2.36335403726708,-155.369215838509)
--(axis cs:2.39916243177113,-155.369215838509)
--(axis cs:2.43497082627517,-155.369215838509)
--(axis cs:2.47077922077922,-155.369215838509)
--(axis cs:2.50658761528327,-155.369215838509)
--(axis cs:2.54239600978731,-155.369215838509)
--(axis cs:2.57820440429136,-155.369215838509)
--(axis cs:2.61401279879541,-155.369215838509)
--(axis cs:2.64982119329945,-155.369215838509)
--(axis cs:2.6856295878035,-155.369215838509)
--(axis cs:2.72143798230755,-155.369215838509)
--(axis cs:2.75724637681159,-155.369215838509)
--(axis cs:2.79305477131564,-155.369215838509)
--(axis cs:2.82886316581969,-155.369215838509)
--(axis cs:2.86467156032373,-155.369215838509)
--(axis cs:2.90047995482778,-155.369215838509)
--(axis cs:2.93628834933183,-155.369215838509)
--(axis cs:2.97209674383587,-155.369215838509)
--(axis cs:3.00790513833992,-155.369215838509)
--(axis cs:3.04371353284397,-155.369215838509)
--(axis cs:3.07952192734801,-155.369215838509)
--(axis cs:3.11533032185206,-155.369215838509)
--(axis cs:3.15113871635611,-155.369215838509)
--(axis cs:3.18694711086015,-155.369215838509)
--(axis cs:3.2227555053642,-155.369215838509)
--(axis cs:3.25856389986825,-155.369215838509)
--(axis cs:3.29437229437229,-155.369215838509)
--(axis cs:3.33018068887634,-155.369215838509)
--(axis cs:3.36598908338039,-155.369215838509)
--(axis cs:3.40179747788443,-155.369215838509)
--(axis cs:3.43760587238848,-155.369215838509)
--(axis cs:3.47341426689253,-155.369215838509)
--(axis cs:3.50922266139657,-155.369215838509)
--(axis cs:3.54503105590062,-155.369215838509)
--(axis cs:3.54503105590062,-190)
--(axis cs:3.54503105590062,-190)
--(axis cs:3.50922266139657,-190)
--(axis cs:3.47341426689253,-190)
--(axis cs:3.43760587238848,-190)
--(axis cs:3.40179747788443,-190)
--(axis cs:3.36598908338039,-190)
--(axis cs:3.33018068887634,-190)
--(axis cs:3.29437229437229,-190)
--(axis cs:3.25856389986825,-190)
--(axis cs:3.2227555053642,-190)
--(axis cs:3.18694711086015,-190)
--(axis cs:3.15113871635611,-190)
--(axis cs:3.11533032185206,-190)
--(axis cs:3.07952192734801,-190)
--(axis cs:3.04371353284397,-190)
--(axis cs:3.00790513833992,-190)
--(axis cs:2.97209674383587,-190)
--(axis cs:2.93628834933183,-190)
--(axis cs:2.90047995482778,-190)
--(axis cs:2.86467156032373,-190)
--(axis cs:2.82886316581969,-190)
--(axis cs:2.79305477131564,-190)
--(axis cs:2.75724637681159,-190)
--(axis cs:2.72143798230755,-190)
--(axis cs:2.6856295878035,-190)
--(axis cs:2.64982119329945,-190)
--(axis cs:2.61401279879541,-190)
--(axis cs:2.57820440429136,-190)
--(axis cs:2.54239600978731,-190)
--(axis cs:2.50658761528327,-190)
--(axis cs:2.47077922077922,-190)
--(axis cs:2.43497082627517,-190)
--(axis cs:2.39916243177113,-190)
--(axis cs:2.36335403726708,-190)
--(axis cs:2.32754564276303,-190)
--(axis cs:2.29173724825899,-190)
--(axis cs:2.25592885375494,-190)
--(axis cs:2.22012045925089,-190)
--(axis cs:2.18431206474685,-190)
--(axis cs:2.1485036702428,-190)
--(axis cs:2.11269527573875,-190)
--(axis cs:2.07688688123471,-190)
--(axis cs:2.04107848673066,-190)
--(axis cs:2.00527009222661,-190)
--(axis cs:1.96946169772257,-190)
--(axis cs:1.93365330321852,-190)
--(axis cs:1.89784490871447,-190)
--(axis cs:1.86203651421043,-190)
--(axis cs:1.82622811970638,-190)
--(axis cs:1.79041972520233,-190)
--(axis cs:1.75461133069829,-190)
--(axis cs:1.71880293619424,-190)
--(axis cs:1.68299454169019,-190)
--(axis cs:1.64718614718615,-190)
--(axis cs:1.6113777526821,-190)
--(axis cs:1.57556935817805,-190)
--(axis cs:1.53976096367401,-190)
--(axis cs:1.50395256916996,-190)
--(axis cs:1.46814417466591,-190)
--(axis cs:1.43233578016187,-190)
--(axis cs:1.39652738565782,-190)
--(axis cs:1.36071899115377,-190)
--(axis cs:1.32491059664973,-190)
--(axis cs:1.28910220214568,-190)
--(axis cs:1.25329380764163,-190)
--(axis cs:1.21748541313759,-190)
--(axis cs:1.18167701863354,-190)
--(axis cs:1.14586862412949,-190)
--(axis cs:1.11006022962545,-190)
--(axis cs:1.0742518351214,-190)
--(axis cs:1.03844344061735,-190)
--(axis cs:1.00263504611331,-190)
--(axis cs:0.96682665160926,-190)
--(axis cs:0.931018257105213,-190)
--(axis cs:0.895209862601167,-190)
--(axis cs:0.85940146809712,-190)
--(axis cs:0.823593073593073,-190)
--(axis cs:0.787784679089027,-190)
--(axis cs:0.75197628458498,-190)
--(axis cs:0.716167890080933,-190)
--(axis cs:0.680359495576887,-190)
--(axis cs:0.64455110107284,-190)
--(axis cs:0.608742706568793,-190)
--(axis cs:0.572934312064747,-190)
--(axis cs:0.5371259175607,-190)
--(axis cs:0.501317523056653,-190)
--(axis cs:0.465509128552607,-190)
--(axis cs:0.42970073404856,-190)
--(axis cs:0.393892339544513,-190)
--(axis cs:0.358083945040467,-190)
--(axis cs:0.32227555053642,-190)
--(axis cs:0.286467156032373,-190)
--(axis cs:0.250658761528327,-190)
--(axis cs:0.21485036702428,-190)
--(axis cs:0.179041972520233,-190)
--(axis cs:0.143233578016187,-190)
--(axis cs:0.10742518351214,-190)
--(axis cs:0.0716167890080933,-190)
--(axis cs:0.0358083945040467,-190)
--(axis cs:0,-190)
--cycle;

\path [fill=color0] (axis cs:3.12796857873584,-190)
--(axis cs:3.12796857873584,-155.369215838509)
--(axis cs:4.06962397179788,-153.892321368078)
--(axis cs:4.06962397179788,-190)
--(axis cs:4.06962397179788,-190)
--(axis cs:3.12796857873584,-190)
--cycle;

\path [fill=color0] (axis cs:3.54503105590062,-190)
--(axis cs:3.54503105590062,-155.369215838509)
--(axis cs:3.58424305163436,-155.347137012083)
--(axis cs:3.62345504736809,-155.325058185656)
--(axis cs:3.66266704310183,-155.288657958304)
--(axis cs:3.70187903883556,-155.222421479024)
--(axis cs:3.74109103456929,-155.156184999744)
--(axis cs:3.78030303030303,-155.061305718613)
--(axis cs:3.81951502603676,-154.95091158648)
--(axis cs:3.8587270217705,-154.840517454347)
--(axis cs:3.89793901750423,-154.687159119437)
--(axis cs:3.93715101323797,-154.532607334451)
--(axis cs:3.9763630089717,-154.364927598616)
--(axis cs:4.01557500470544,-154.166218160776)
--(axis cs:4.05478700043917,-153.967508722936)
--(axis cs:4.09399899617291,-153.741349933323)
--(axis cs:4.13321099190664,-153.498482842629)
--(axis cs:4.17242298764038,-153.255615751936)
--(axis cs:4.21163498337411,-152.970977908544)
--(axis cs:4.25084697910785,-152.683953164998)
--(axis cs:4.29005897484158,-152.38499392068)
--(axis cs:4.32927097057532,-152.05381152428)
--(axis cs:4.36848296630905,-151.72262912788)
--(axis cs:4.40769496204279,-151.365190829784)
--(axis cs:4.44690695777652,-150.989850780531)
--(axis cs:4.48611895351026,-150.614510731278)
--(axis cs:4.52533094924399,-150.198593379403)
--(axis cs:4.56454294497773,-149.779095677296)
--(axis cs:4.60375494071146,-149.348856924496)
--(axis cs:4.6429669364452,-148.885201569536)
--(axis cs:4.68217893217893,-148.421546214576)
--(axis cs:4.72139092791267,-147.932828407997)
--(axis cs:4.7606029236464,-147.425015400184)
--(axis cs:4.79981491938014,-146.917202392371)
--(axis cs:4.83902691511387,-146.370005532013)
--(axis cs:4.87823891084761,-145.818034871347)
--(axis cs:4.91745090658134,-145.256516610064)
--(axis cs:4.95666290231508,-144.660388296544)
--(axis cs:4.99587489804881,-144.064259983025)
--(axis cs:5.03508689378255,-143.444262667963)
--(axis cs:5.07429888951628,-142.80397670159)
--(axis cs:5.11351088525002,-142.163690735217)
--(axis cs:5.15272288098375,-141.485214366376)
--(axis cs:5.19193487671749,-140.80077074715)
--(axis cs:5.23114687245122,-140.107972977384)
--(axis cs:5.27035886818495,-139.379371705304)
--(axis cs:5.30957086391869,-138.650770433225)
--(axis cs:5.34878285965242,-137.89949360968)
--(axis cs:5.38799485538616,-137.126734684747)
--(axis cs:5.42720685111989,-136.353975759814)
--(axis cs:5.46641884685363,-135.544219882491)
--(axis cs:5.50563084258736,-134.727303304705)
--(axis cs:5.5448428383211,-133.903226026456)
--(axis cs:5.58405483405483,-133.042151795817)
--(axis cs:5.62326682978857,-132.181077565177)
--(axis cs:5.6624788255223,-131.29852123315)
--(axis cs:5.70169082125604,-130.393289349657)
--(axis cs:5.74090281698977,-129.488057466164)
--(axis cs:5.78011481272351,-128.547022080358)
--(axis cs:5.81932680845724,-127.597632544012)
--(axis cs:5.85853880419098,-126.64227575728)
--(axis cs:5.89775079992471,-125.648728568081)
--(axis cs:5.93696279565845,-124.655181378881)
--(axis cs:5.97617479139218,-123.641345538371)
--(axis cs:6.01538678712592,-122.603640696318)
--(axis cs:6.05459878285965,-121.565935854266)
--(axis cs:6.09381077859339,-120.493620959977)
--(axis cs:6.13302277432712,-119.411758465071)
--(axis cs:6.17223477006086,-118.325122169856)
--(axis cs:6.21144676579459,-117.199102022097)
--(axis cs:6.25065876152833,-116.073081874338)
--(axis cs:6.28987075726206,-114.927966525345)
--(axis cs:6.3290827529958,-113.757788724732)
--(axis cs:6.36829474872953,-112.587610924119)
--(axis cs:6.40750674446327,-111.384016521348)
--(axis cs:6.446718740197,-110.169681067882)
--(axis cs:6.48593073593074,-108.951765264184)
--(axis cs:6.52514273166447,-107.693272157865)
--(axis cs:6.56435472739821,-106.434779051546)
--(axis cs:6.60356672313194,-105.15838419407)
--(axis cs:6.64277871886568,-103.855733434898)
--(axis cs:6.68199071459941,-102.553082675725)
--(axis cs:6.72120271033315,-101.218208764471)
--(axis cs:6.76041470606688,-99.8714003524447)
--(axis cs:6.79962670180061,-98.5222050402647)
--(axis cs:6.83883869753435,-97.1312389753856)
--(axis cs:6.87805069326808,-95.7402729105064)
--(axis cs:6.91726268900182,-94.3325985445477)
--(axis cs:6.95647468473555,-92.8974748268153)
--(axis cs:6.99568668046929,-91.4623511090829)
--(axis cs:7.03489867620302,-89.9961976893455)
--(axis cs:7.07411067193676,-88.5169163187597)
--(axis cs:7.11332266767049,-87.0364414980969)
--(axis cs:7.15253466340423,-85.5130024746579)
--(axis cs:7.19174665913796,-83.9895634512188)
--(axis cs:7.2309586548717,-82.4506095767773)
--(axis cs:7.27017065060543,-80.883012900485)
--(axis cs:7.30938264633917,-79.3154162241926)
--(axis cs:7.3485946420729,-77.7179832959724)
--(axis cs:7.38780663780664,-76.1062289668268)
--(axis cs:7.42701863354037,-74.4944746376812)
--(axis cs:7.42701863354037,-190)
--(axis cs:7.42701863354037,-190)
--(axis cs:7.38780663780664,-190)
--(axis cs:7.3485946420729,-190)
--(axis cs:7.30938264633917,-190)
--(axis cs:7.27017065060543,-190)
--(axis cs:7.2309586548717,-190)
--(axis cs:7.19174665913796,-190)
--(axis cs:7.15253466340423,-190)
--(axis cs:7.11332266767049,-190)
--(axis cs:7.07411067193676,-190)
--(axis cs:7.03489867620302,-190)
--(axis cs:6.99568668046929,-190)
--(axis cs:6.95647468473555,-190)
--(axis cs:6.91726268900182,-190)
--(axis cs:6.87805069326808,-190)
--(axis cs:6.83883869753435,-190)
--(axis cs:6.79962670180061,-190)
--(axis cs:6.76041470606688,-190)
--(axis cs:6.72120271033315,-190)
--(axis cs:6.68199071459941,-190)
--(axis cs:6.64277871886568,-190)
--(axis cs:6.60356672313194,-190)
--(axis cs:6.56435472739821,-190)
--(axis cs:6.52514273166447,-190)
--(axis cs:6.48593073593074,-190)
--(axis cs:6.446718740197,-190)
--(axis cs:6.40750674446327,-190)
--(axis cs:6.36829474872953,-190)
--(axis cs:6.3290827529958,-190)
--(axis cs:6.28987075726206,-190)
--(axis cs:6.25065876152833,-190)
--(axis cs:6.21144676579459,-190)
--(axis cs:6.17223477006086,-190)
--(axis cs:6.13302277432712,-190)
--(axis cs:6.09381077859339,-190)
--(axis cs:6.05459878285965,-190)
--(axis cs:6.01538678712592,-190)
--(axis cs:5.97617479139218,-190)
--(axis cs:5.93696279565845,-190)
--(axis cs:5.89775079992471,-190)
--(axis cs:5.85853880419098,-190)
--(axis cs:5.81932680845724,-190)
--(axis cs:5.78011481272351,-190)
--(axis cs:5.74090281698977,-190)
--(axis cs:5.70169082125604,-190)
--(axis cs:5.6624788255223,-190)
--(axis cs:5.62326682978857,-190)
--(axis cs:5.58405483405483,-190)
--(axis cs:5.5448428383211,-190)
--(axis cs:5.50563084258736,-190)
--(axis cs:5.46641884685363,-190)
--(axis cs:5.42720685111989,-190)
--(axis cs:5.38799485538616,-190)
--(axis cs:5.34878285965242,-190)
--(axis cs:5.30957086391869,-190)
--(axis cs:5.27035886818495,-190)
--(axis cs:5.23114687245122,-190)
--(axis cs:5.19193487671749,-190)
--(axis cs:5.15272288098375,-190)
--(axis cs:5.11351088525002,-190)
--(axis cs:5.07429888951628,-190)
--(axis cs:5.03508689378255,-190)
--(axis cs:4.99587489804881,-190)
--(axis cs:4.95666290231508,-190)
--(axis cs:4.91745090658134,-190)
--(axis cs:4.87823891084761,-190)
--(axis cs:4.83902691511387,-190)
--(axis cs:4.79981491938014,-190)
--(axis cs:4.7606029236464,-190)
--(axis cs:4.72139092791267,-190)
--(axis cs:4.68217893217893,-190)
--(axis cs:4.6429669364452,-190)
--(axis cs:4.60375494071146,-190)
--(axis cs:4.56454294497773,-190)
--(axis cs:4.52533094924399,-190)
--(axis cs:4.48611895351026,-190)
--(axis cs:4.44690695777652,-190)
--(axis cs:4.40769496204279,-190)
--(axis cs:4.36848296630905,-190)
--(axis cs:4.32927097057532,-190)
--(axis cs:4.29005897484158,-190)
--(axis cs:4.25084697910785,-190)
--(axis cs:4.21163498337411,-190)
--(axis cs:4.17242298764038,-190)
--(axis cs:4.13321099190664,-190)
--(axis cs:4.09399899617291,-190)
--(axis cs:4.05478700043917,-190)
--(axis cs:4.01557500470544,-190)
--(axis cs:3.9763630089717,-190)
--(axis cs:3.93715101323797,-190)
--(axis cs:3.89793901750423,-190)
--(axis cs:3.8587270217705,-190)
--(axis cs:3.81951502603676,-190)
--(axis cs:3.78030303030303,-190)
--(axis cs:3.74109103456929,-190)
--(axis cs:3.70187903883556,-190)
--(axis cs:3.66266704310183,-190)
--(axis cs:3.62345504736809,-190)
--(axis cs:3.58424305163436,-190)
--(axis cs:3.54503105590062,-190)
--cycle;

\path [fill=color0] (axis cs:7.00734430082256,-190)
--(axis cs:7.00734430082256,-91.0356927065138)
--(axis cs:7.9464963857357,-52.849568296209)
--(axis cs:7.9464963857357,-190)
--(axis cs:7.9464963857357,-190)
--(axis cs:7.00734430082256,-190)
--cycle;

\path [fill=color0] (axis cs:7.42701863354037,-190)
--(axis cs:7.42701863354037,-74.4944746376812)
--(axis cs:7.45745268366899,-73.2263892156555)
--(axis cs:7.4878867337976,-71.9583037936299)
--(axis cs:7.51832078392622,-70.6902183716043)
--(axis cs:7.54875483405483,-69.4221329495786)
--(axis cs:7.57918888418345,-68.154047527553)
--(axis cs:7.60962293431206,-66.8859621055273)
--(axis cs:7.64005698444068,-65.6178766835017)
--(axis cs:7.6704910345693,-64.3497912614761)
--(axis cs:7.70092508469791,-63.0817058394504)
--(axis cs:7.73135913482653,-61.8136204174248)
--(axis cs:7.76179318495514,-60.5455349953991)
--(axis cs:7.79222723508376,-59.2774495733735)
--(axis cs:7.82266128521237,-58.0093641513479)
--(axis cs:7.85309533534099,-56.7412787293222)
--(axis cs:7.8835293854696,-55.4731933072966)
--(axis cs:7.91396343559822,-54.2051078852709)
--(axis cs:7.94439748572683,-52.9370224632453)
--(axis cs:7.97483153585545,-51.6689370412196)
--(axis cs:8.00526558598406,-50.400851619194)
--(axis cs:8.03569963611268,-49.1327661971684)
--(axis cs:8.06613368624129,-47.8646807751428)
--(axis cs:8.09656773636991,-46.5965953531171)
--(axis cs:8.12700178649852,-45.3285099310915)
--(axis cs:8.15743583662714,-44.0604245090658)
--(axis cs:8.18786988675576,-42.7923390870402)
--(axis cs:8.21830393688437,-41.5242536650145)
--(axis cs:8.24873798701299,-40.2561682429889)
--(axis cs:8.2791720371416,-38.9880828209632)
--(axis cs:8.30960608727022,-37.7199973989376)
--(axis cs:8.34004013739883,-36.451911976912)
--(axis cs:8.37047418752745,-35.1838265548864)
--(axis cs:8.40090823765606,-33.9157411328607)
--(axis cs:8.43134228778468,-32.6476557108351)
--(axis cs:8.46177633791329,-31.3795702888094)
--(axis cs:8.49221038804191,-30.1114848667838)
--(axis cs:8.52264443817053,-28.8433994447581)
--(axis cs:8.55307848829914,-27.5753140227325)
--(axis cs:8.58351253842775,-26.3072286007069)
--(axis cs:8.61394658855637,-25.0391431786813)
--(axis cs:8.64438063868499,-23.7710577566556)
--(axis cs:8.6748146888136,-22.50297233463)
--(axis cs:8.70524873894222,-21.2348869126043)
--(axis cs:8.73568278907083,-19.9668014905787)
--(axis cs:8.76611683919945,-18.698716068553)
--(axis cs:8.79655088932806,-17.4306306465274)
--(axis cs:8.82698493945668,-16.1625452245018)
--(axis cs:8.85741898958529,-14.8944598024761)
--(axis cs:8.88785303971391,-13.6263743804505)
--(axis cs:8.91828708984252,-12.3582889584248)
--(axis cs:8.94872113997114,-11.0902035363992)
--(axis cs:8.97915519009976,-9.82211811437356)
--(axis cs:9.00958924022837,-8.55403269234791)
--(axis cs:9.04002329035699,-7.28594727032226)
--(axis cs:9.0704573404856,-6.01786184829668)
--(axis cs:9.10089139061422,-4.74977642627103)
--(axis cs:9.13132544074283,-3.48169100424539)
--(axis cs:9.16175949087145,-2.21360558221974)
--(axis cs:9.19219354100006,-0.945520160194093)
--(axis cs:9.22262759112868,0.322565261831555)
--(axis cs:9.25306164125729,1.5906506838572)
--(axis cs:9.28349569138591,2.85873610588285)
--(axis cs:9.31392974151452,4.1268215279085)
--(axis cs:9.34436379164314,5.39490694993408)
--(axis cs:9.37479784177175,6.66299237195973)
--(axis cs:9.40523189190037,7.93107779398538)
--(axis cs:9.43566594202898,9.19916321601102)
--(axis cs:9.4660999921576,10.4672486380367)
--(axis cs:9.49653404228622,11.7353340600623)
--(axis cs:9.52696809241483,13.003419482088)
--(axis cs:9.55740214254345,14.2715049041135)
--(axis cs:9.58783619267206,15.5395903261392)
--(axis cs:9.61827024280068,16.8076757481648)
--(axis cs:9.64870429292929,18.0757611701905)
--(axis cs:9.67913834305791,19.3438465922161)
--(axis cs:9.70957239318652,20.6119320142418)
--(axis cs:9.74000644331514,21.8800174362674)
--(axis cs:9.77044049344375,23.1481028582931)
--(axis cs:9.80087454357237,24.4161882803187)
--(axis cs:9.83130859370099,25.6842737023444)
--(axis cs:9.8617426438296,26.9523591243699)
--(axis cs:9.89217669395821,28.2204445463956)
--(axis cs:9.92261074408683,29.4885299684212)
--(axis cs:9.95304479421545,30.7566153904469)
--(axis cs:9.98347884434406,32.0247008124725)
--(axis cs:10.0139128944727,33.2927862344982)
--(axis cs:10.0443469446013,34.5608716565238)
--(axis cs:10.0747809947299,35.8289570785494)
--(axis cs:10.1052150448585,37.0970425005751)
--(axis cs:10.1356490949871,38.3651279226007)
--(axis cs:10.1660831451158,39.6332133446264)
--(axis cs:10.1965171952444,40.901298766652)
--(axis cs:10.226951245373,42.1693841886776)
--(axis cs:10.2573852955016,43.4374696107033)
--(axis cs:10.2878193456302,44.705555032729)
--(axis cs:10.3182533957588,45.9736404547546)
--(axis cs:10.3486874458874,47.2417258767803)
--(axis cs:10.3791214960161,48.5098112988058)
--(axis cs:10.4095555461447,49.7778967208315)
--(axis cs:10.4399895962733,51.0459821428571)
--(axis cs:10.4399895962733,-190)
--(axis cs:10.4399895962733,-190)
--(axis cs:10.4095555461447,-190)
--(axis cs:10.3791214960161,-190)
--(axis cs:10.3486874458874,-190)
--(axis cs:10.3182533957588,-190)
--(axis cs:10.2878193456302,-190)
--(axis cs:10.2573852955016,-190)
--(axis cs:10.226951245373,-190)
--(axis cs:10.1965171952444,-190)
--(axis cs:10.1660831451158,-190)
--(axis cs:10.1356490949871,-190)
--(axis cs:10.1052150448585,-190)
--(axis cs:10.0747809947299,-190)
--(axis cs:10.0443469446013,-190)
--(axis cs:10.0139128944727,-190)
--(axis cs:9.98347884434406,-190)
--(axis cs:9.95304479421545,-190)
--(axis cs:9.92261074408683,-190)
--(axis cs:9.89217669395821,-190)
--(axis cs:9.8617426438296,-190)
--(axis cs:9.83130859370099,-190)
--(axis cs:9.80087454357237,-190)
--(axis cs:9.77044049344375,-190)
--(axis cs:9.74000644331514,-190)
--(axis cs:9.70957239318652,-190)
--(axis cs:9.67913834305791,-190)
--(axis cs:9.64870429292929,-190)
--(axis cs:9.61827024280068,-190)
--(axis cs:9.58783619267206,-190)
--(axis cs:9.55740214254345,-190)
--(axis cs:9.52696809241483,-190)
--(axis cs:9.49653404228622,-190)
--(axis cs:9.4660999921576,-190)
--(axis cs:9.43566594202898,-190)
--(axis cs:9.40523189190037,-190)
--(axis cs:9.37479784177175,-190)
--(axis cs:9.34436379164314,-190)
--(axis cs:9.31392974151452,-190)
--(axis cs:9.28349569138591,-190)
--(axis cs:9.25306164125729,-190)
--(axis cs:9.22262759112868,-190)
--(axis cs:9.19219354100006,-190)
--(axis cs:9.16175949087145,-190)
--(axis cs:9.13132544074283,-190)
--(axis cs:9.10089139061422,-190)
--(axis cs:9.0704573404856,-190)
--(axis cs:9.04002329035699,-190)
--(axis cs:9.00958924022837,-190)
--(axis cs:8.97915519009976,-190)
--(axis cs:8.94872113997114,-190)
--(axis cs:8.91828708984252,-190)
--(axis cs:8.88785303971391,-190)
--(axis cs:8.85741898958529,-190)
--(axis cs:8.82698493945668,-190)
--(axis cs:8.79655088932806,-190)
--(axis cs:8.76611683919945,-190)
--(axis cs:8.73568278907083,-190)
--(axis cs:8.70524873894222,-190)
--(axis cs:8.6748146888136,-190)
--(axis cs:8.64438063868499,-190)
--(axis cs:8.61394658855637,-190)
--(axis cs:8.58351253842775,-190)
--(axis cs:8.55307848829914,-190)
--(axis cs:8.52264443817053,-190)
--(axis cs:8.49221038804191,-190)
--(axis cs:8.46177633791329,-190)
--(axis cs:8.43134228778468,-190)
--(axis cs:8.40090823765606,-190)
--(axis cs:8.37047418752745,-190)
--(axis cs:8.34004013739883,-190)
--(axis cs:8.30960608727022,-190)
--(axis cs:8.2791720371416,-190)
--(axis cs:8.24873798701299,-190)
--(axis cs:8.21830393688437,-190)
--(axis cs:8.18786988675576,-190)
--(axis cs:8.15743583662714,-190)
--(axis cs:8.12700178649852,-190)
--(axis cs:8.09656773636991,-190)
--(axis cs:8.06613368624129,-190)
--(axis cs:8.03569963611268,-190)
--(axis cs:8.00526558598406,-190)
--(axis cs:7.97483153585545,-190)
--(axis cs:7.94439748572683,-190)
--(axis cs:7.91396343559822,-190)
--(axis cs:7.8835293854696,-190)
--(axis cs:7.85309533534099,-190)
--(axis cs:7.82266128521237,-190)
--(axis cs:7.79222723508376,-190)
--(axis cs:7.76179318495514,-190)
--(axis cs:7.73135913482653,-190)
--(axis cs:7.70092508469791,-190)
--(axis cs:7.6704910345693,-190)
--(axis cs:7.64005698444068,-190)
--(axis cs:7.60962293431206,-190)
--(axis cs:7.57918888418345,-190)
--(axis cs:7.54875483405483,-190)
--(axis cs:7.51832078392622,-190)
--(axis cs:7.4878867337976,-190)
--(axis cs:7.45745268366899,-190)
--(axis cs:7.42701863354037,-190)
--cycle;

\addplot [semithick, black, forget plot]
table [row sep=\\]{%
0	-155.369215838509 \\
0.104265619291195	-155.369215838509 \\
0.208531238582389	-155.369215838509 \\
0.312796857873584	-155.369215838509 \\
0.417062477164779	-155.369215838509 \\
0.521328096455974	-155.369215838509 \\
0.625593715747168	-155.369215838509 \\
0.729859335038363	-155.369215838509 \\
0.834124954329558	-155.369215838509 \\
0.938390573620752	-155.369215838509 \\
1.04265619291195	-155.369215838509 \\
1.14692181220314	-155.369215838509 \\
1.25118743149434	-155.369215838509 \\
1.35545305078553	-155.369215838509 \\
1.45971867007673	-155.369215838509 \\
1.56398428936792	-155.369215838509 \\
1.66824990865912	-155.369215838509 \\
1.77251552795031	-155.369215838509 \\
1.8767811472415	-155.369215838509 \\
1.9810467665327	-155.369215838509 \\
2.08531238582389	-155.369215838509 \\
2.18957800511509	-155.369215838509 \\
2.29384362440628	-155.369215838509 \\
2.39810924369748	-155.369215838509 \\
2.50237486298867	-155.369215838509 \\
2.60664048227987	-155.369215838509 \\
2.71090610157106	-155.369215838509 \\
2.81517172086226	-155.369215838509 \\
2.91943734015345	-155.369215838509 \\
3.02370295944465	-155.369215838509 \\
3.12796857873584	-155.369215838509 \\
3.23223419802704	-155.369215838509 \\
3.33649981731823	-155.369215838509 \\
3.44076543660943	-155.369215838509 \\
3.54503105590062	-155.369215838509 \\
};
\addplot [semithick, black, forget plot]
table [row sep=\\]{%
3.54503105590062	-155.369215838509 \\
3.64994963908007	-155.310140059692 \\
3.75486822225953	-155.13291272324 \\
3.85978680543898	-154.837533829154 \\
3.96470538861843	-154.424003377433 \\
4.06962397179788	-153.892321368078 \\
4.17454255497734	-153.242487801088 \\
4.27946113815679	-152.474502676464 \\
4.38437972133624	-151.588365994205 \\
4.4892983045157	-150.584077754311 \\
4.59421688769515	-149.461637956783 \\
4.6991354708746	-148.221046601621 \\
4.80405405405405	-146.862303688824 \\
4.90897263723351	-145.385409218392 \\
5.01389122041296	-143.790363190326 \\
5.11880980359241	-142.077165604626 \\
5.22372838677186	-140.245816461291 \\
5.32864696995132	-138.296315760321 \\
5.43356555313077	-136.228663501717 \\
5.53848413631022	-134.042859685479 \\
5.64340271948967	-131.738904311606 \\
5.74832130266913	-129.316797380098 \\
5.85323988584858	-126.776538890956 \\
5.95815846902803	-124.118128844179 \\
6.06307705220749	-121.341567239768 \\
6.16799563538694	-118.446854077722 \\
6.27291421856639	-115.433989358042 \\
6.37783280174584	-112.302973080727 \\
6.4827513849253	-109.053805245778 \\
6.58766996810475	-105.686485853194 \\
6.6925885512842	-102.201014902976 \\
6.79750713446366	-98.597392395123 \\
6.90242571764311	-94.8756183296357 \\
7.00734430082256	-91.0356927065138 \\
7.11226288400201	-87.0776155257574 \\
7.21718146718147	-83.0013867873665 \\
7.32210005036092	-78.8070064913411 \\
7.42701863354037	-74.4944746376812 \\
};
\addplot [semithick, black, forget plot]
table [row sep=\\]{%
7.42701863354037	-74.4944746376812 \\
7.53091418397944	-70.1654933693867 \\
7.6348097344185	-65.8365121010923 \\
7.73870528485757	-61.5075308327979 \\
7.84260083529664	-57.1785495645035 \\
7.9464963857357	-52.849568296209 \\
8.05039193617477	-48.5205870279146 \\
8.15428748661384	-44.1916057596202 \\
8.2581830370529	-39.8626244913258 \\
8.36207858749197	-35.5336432230313 \\
8.46597413793103	-31.2046619547369 \\
8.5698696883701	-26.8756806864425 \\
8.67376523880917	-22.5466994181481 \\
8.77766078924823	-18.2177181498537 \\
8.8815563396873	-13.8887368815592 \\
8.98545189012636	-9.55975561326481 \\
9.08934744056543	-5.23077434497037 \\
9.1932429910045	-0.901793076675943 \\
9.29713854144356	3.42718819161841 \\
9.40103409188263	7.75616945991285 \\
9.5049296423217	12.0851507282073 \\
9.60882519276076	16.4141319965017 \\
9.71272074319983	20.7431132647962 \\
9.81661629363889	25.0720945330906 \\
9.92051184407796	29.401075801385 \\
10.024407394517	33.7300570696795 \\
10.1283029449561	38.0590383379738 \\
10.2321984953952	42.3880196062682 \\
10.3360940458342	46.7170008745627 \\
10.4399895962733	51.0459821428571 \\
};
\addplot [semithick, blue, dashed, forget plot]
table [row sep=\\]{%
0	0 \\
0.103448275862069	-3.06037455410226 \\
0.206896551724138	-6.0346016646849 \\
0.310344827586207	-8.92268133174792 \\
0.413793103448276	-11.7246135552913 \\
0.517241379310345	-14.4403983353151 \\
0.620689655172414	-17.0700356718193 \\
0.724137931034483	-19.6135255648038 \\
0.827586206896552	-22.0708680142687 \\
0.931034482758621	-24.442063020214 \\
1.03448275862069	-26.7271105826397 \\
1.13793103448276	-28.9260107015458 \\
1.24137931034483	-31.0387633769322 \\
1.3448275862069	-33.065368608799 \\
1.44827586206897	-35.0058263971463 \\
1.55172413793103	-36.8601367419738 \\
1.6551724137931	-38.6282996432818 \\
1.75862068965517	-40.3103151010702 \\
1.86206896551724	-41.9061831153389 \\
1.96551724137931	-43.415903686088 \\
2.06896551724138	-44.8394768133175 \\
2.17241379310345	-46.1769024970274 \\
2.27586206896552	-47.4281807372176 \\
2.37931034482759	-48.5933115338882 \\
2.48275862068966	-49.6722948870392 \\
2.58620689655172	-50.6651307966706 \\
2.68965517241379	-51.5718192627824 \\
2.79310344827586	-52.3923602853746 \\
2.89655172413793	-53.1267538644471 \\
3	-53.775 \\
};
\addplot [semithick, blue, dashed, forget plot]
table [row sep=\\]{%
3	-53.775 \\
3.10295392171024	-54.320396405029 \\
3.20590784342048	-54.752024736106 \\
3.30886176513072	-55.0698849932312 \\
3.41181568684096	-55.2739771764045 \\
3.51476960855121	-55.3643012856258 \\
3.61772353026145	-55.3408573208953 \\
3.72067745197169	-55.2036452822129 \\
3.82363137368193	-54.9526651695785 \\
3.92658529539217	-54.5879169829923 \\
4.02953921710241	-54.1094007224542 \\
4.13249313881265	-53.5171163879642 \\
4.23544706052289	-52.8110639795222 \\
4.33840098223314	-51.9912434971284 \\
4.44135490394338	-51.0576549407827 \\
4.54430882565362	-50.0102983104851 \\
4.64726274736386	-48.8491736062356 \\
4.7502166690741	-47.5742808280342 \\
4.85317059078434	-46.1856199758809 \\
4.95612451249458	-44.6831910497757 \\
5.05907843420482	-43.0669940497185 \\
5.16203235591507	-41.3370289757095 \\
5.26498627762531	-39.4932958277487 \\
5.36794019933555	-37.5357946058358 \\
5.47089412104579	-35.4645253099711 \\
5.57384804275603	-33.2794879401545 \\
5.67680196446627	-30.9806824963861 \\
5.77975588617651	-28.5681089786656 \\
5.88270980788675	-26.0417673869934 \\
5.98566372959699	-23.4016577213692 \\
6.08861765130724	-20.6477799817931 \\
6.19157157301748	-17.7801341682651 \\
6.29452549472772	-14.7987202807852 \\
6.39747941643796	-11.7035383193534 \\
6.5004333381482	-8.49458828396972 \\
6.60338725985844	-5.17187017463417 \\
6.70634118156868	-1.73538399134667 \\
6.80929510327893	1.81487026589272 \\
6.91224902498917	5.478892597084 \\
7.01520294669941	9.25668300222716 \\
7.11815686840965	13.1482414813223 \\
7.22111079011989	17.1535680343692 \\
7.32406471183013	21.2726626613681 \\
7.42701863354037	25.5055253623189 \\
};
\addplot [semithick, blue, dashed, forget plot]
table [row sep=\\]{%
7.42701863354037	25.5055253623189 \\
7.53091418397944	29.8345066306133 \\
7.6348097344185	34.1634878989077 \\
7.73870528485757	38.4924691672021 \\
7.84260083529664	42.8214504354966 \\
7.9464963857357	47.150431703791 \\
8.05039193617477	51.4794129720854 \\
8.15428748661384	55.8083942403798 \\
8.2581830370529	60.1373755086743 \\
8.36207858749197	64.4663567769687 \\
8.46597413793103	68.7953380452631 \\
8.5698696883701	73.1243193135576 \\
8.67376523880917	77.4533005818519 \\
8.77766078924823	81.7822818501463 \\
8.8815563396873	86.1112631184408 \\
8.98545189012636	90.4402443867352 \\
9.08934744056543	94.7692256550297 \\
9.1932429910045	99.0982069233241 \\
9.29713854144356	103.427188191618 \\
9.40103409188263	107.756169459913 \\
9.5049296423217	112.085150728207 \\
9.60882519276076	116.414131996502 \\
9.71272074319983	120.743113264796 \\
9.81661629363889	125.072094533091 \\
9.92051184407796	129.401075801385 \\
10.024407394517	133.73005706968 \\
10.1283029449561	138.059038337974 \\
10.2321984953952	142.388019606268 \\
10.3360940458342	146.717000874563 \\
10.4399895962733	151.045982142857 \\
};

    \node (fig2) at (0.55,-10)
   {\includegraphics[scale=0.05]{jbplane.png}};
    \node (fig3) at (9,-90)
   {\includegraphics[scale=0.05]{jbplane_intruder.png}};
   
    \node[text=black] at (4,-100)  {\footnotesize Unsafe for CL1500};
    \node[text=white] at (8.5,-150) {\footnotesize Safe for CL1500};
    \node[text=white] at (9,-65)  {\footnotesize Intruder};
    \node[text=black] at (1,25)  {\footnotesize Ownship};
\end{axis}

\end{tikzpicture}

%% file: Safeable_noUnsafe.tex
\begin{tikzpicture}

\begin{axis}[
axis line style = thick,
tick align=outside,
tick pos=left,
x grid style={lightgray!92.02614379084967!black},
xlabel={$\tau$ (sec)},
xmin=0, xmax=10.1070299689441,
y grid style={lightgray!92.02614379084967!black},
ylabel={$h$ (ft)},
ymin=-205.369215838509, ymax=150
]
\addplot graphics [includegraphics cmd=\pgfimage,xmin=0, xmax=10.1070299689441, ymin=-205.369215838509, ymax=150] {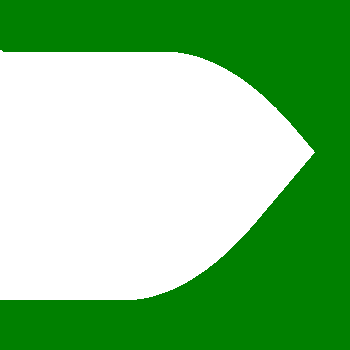};
\addplot [semithick, black, forget plot]
table [row sep=\\]{%
0	-155.369215838509 \\
0.104265619291195	-155.369215838509 \\
0.208531238582389	-155.369215838509 \\
0.312796857873584	-155.369215838509 \\
0.417062477164779	-155.369215838509 \\
0.521328096455974	-155.369215838509 \\
0.625593715747168	-155.369215838509 \\
0.729859335038363	-155.369215838509 \\
0.834124954329558	-155.369215838509 \\
0.938390573620752	-155.369215838509 \\
1.04265619291195	-155.369215838509 \\
1.14692181220314	-155.369215838509 \\
1.25118743149434	-155.369215838509 \\
1.35545305078553	-155.369215838509 \\
1.45971867007673	-155.369215838509 \\
1.56398428936792	-155.369215838509 \\
1.66824990865912	-155.369215838509 \\
1.77251552795031	-155.369215838509 \\
1.8767811472415	-155.369215838509 \\
1.9810467665327	-155.369215838509 \\
2.08531238582389	-155.369215838509 \\
2.18957800511509	-155.369215838509 \\
2.29384362440628	-155.369215838509 \\
2.39810924369748	-155.369215838509 \\
2.50237486298867	-155.369215838509 \\
2.60664048227987	-155.369215838509 \\
2.71090610157106	-155.369215838509 \\
2.81517172086226	-155.369215838509 \\
2.91943734015345	-155.369215838509 \\
3.02370295944465	-155.369215838509 \\
3.12796857873584	-155.369215838509 \\
3.23223419802704	-155.369215838509 \\
3.33649981731823	-155.369215838509 \\
3.44076543660943	-155.369215838509 \\
3.54503105590062	-155.369215838509 \\
};
\addplot [semithick, black, forget plot]
table [row sep=\\]{%
3.54503105590062	-155.369215838509 \\
3.64994963908007	-155.310140059692 \\
3.75486822225953	-155.13291272324 \\
3.85978680543898	-154.837533829154 \\
3.96470538861843	-154.424003377433 \\
4.06962397179788	-153.892321368078 \\
4.17454255497734	-153.242487801088 \\
4.27946113815679	-152.474502676464 \\
4.38437972133624	-151.588365994205 \\
4.4892983045157	-150.584077754311 \\
4.59421688769515	-149.461637956783 \\
4.6991354708746	-148.221046601621 \\
4.80405405405405	-146.862303688824 \\
4.90897263723351	-145.385409218392 \\
5.01389122041296	-143.790363190326 \\
5.11880980359241	-142.077165604626 \\
5.22372838677186	-140.245816461291 \\
5.32864696995132	-138.296315760321 \\
5.43356555313077	-136.228663501717 \\
5.53848413631022	-134.042859685479 \\
5.64340271948967	-131.738904311606 \\
5.74832130266913	-129.316797380098 \\
5.85323988584858	-126.776538890956 \\
5.95815846902803	-124.118128844179 \\
6.06307705220749	-121.341567239768 \\
6.16799563538694	-118.446854077722 \\
6.27291421856639	-115.433989358042 \\
6.37783280174584	-112.302973080727 \\
6.4827513849253	-109.053805245778 \\
6.58766996810475	-105.686485853194 \\
6.6925885512842	-102.201014902976 \\
6.79750713446366	-98.597392395123 \\
6.90242571764311	-94.8756183296357 \\
7.00734430082256	-91.0356927065138 \\
7.11226288400201	-87.0776155257574 \\
7.21718146718147	-83.0013867873665 \\
7.32210005036092	-78.8070064913411 \\
7.42701863354037	-74.4944746376812 \\
};
\addplot [semithick, black, forget plot]
table [row sep=\\]{%
7.42701863354037	-74.4944746376812 \\
7.53901938923395	-69.8277764837819 \\
7.65102014492754	-65.1610783298827 \\
7.76302090062112	-60.4943801759834 \\
7.8750216563147	-55.8276820220842 \\
7.98702241200828	-51.1609838681849 \\
8.09902316770186	-46.4942857142857 \\
8.21102392339545	-41.8275875603864 \\
8.32302467908903	-37.1608894064872 \\
8.43502543478261	-32.494191252588 \\
8.54702619047619	-27.8274930986887 \\
8.65902694616977	-23.1607949447895 \\
8.77102770186336	-18.4940967908902 \\
8.88302845755694	-13.8273986369909 \\
8.99502921325052	-9.16070048309174 \\
9.1070299689441	-4.49400232919248 \\
};
\addplot [semithick, black, forget plot]
table [row sep=\\]{%
0	100 \\
0.0330794847658176	99.0164241881401 \\
0.0661589695316352	98.0504658385093 \\
};
\addplot [semithick, black, forget plot]
table [row sep=\\]{%
0.0661589695316352	98.0504658385093 \\
0.169243635633141	98.0504658385093 \\
0.272328301734646	98.0504658385093 \\
0.375412967836151	98.0504658385093 \\
0.478497633937657	98.0504658385093 \\
0.581582300039162	98.0504658385093 \\
0.684666966140668	98.0504658385093 \\
0.787751632242173	98.0504658385093 \\
0.890836298343679	98.0504658385093 \\
0.993920964445184	98.0504658385093 \\
1.09700563054669	98.0504658385093 \\
1.20009029664819	98.0504658385093 \\
1.3031749627497	98.0504658385093 \\
1.40625962885121	98.0504658385093 \\
1.50934429495271	98.0504658385093 \\
1.61242896105422	98.0504658385093 \\
1.71551362715572	98.0504658385093 \\
1.81859829325723	98.0504658385093 \\
1.92168295935873	98.0504658385093 \\
2.02476762546024	98.0504658385093 \\
2.12785229156174	98.0504658385093 \\
2.23093695766325	98.0504658385093 \\
2.33402162376475	98.0504658385093 \\
2.43710628986626	98.0504658385093 \\
2.54019095596777	98.0504658385093 \\
2.64327562206927	98.0504658385093 \\
2.74636028817078	98.0504658385093 \\
2.84944495427228	98.0504658385093 \\
2.95252962037379	98.0504658385093 \\
3.05561428647529	98.0504658385093 \\
3.1586989525768	98.0504658385093 \\
3.2617836186783	98.0504658385093 \\
3.36486828477981	98.0504658385093 \\
3.46795295088131	98.0504658385093 \\
3.57103761698282	98.0504658385093 \\
3.67412228308432	98.0504658385093 \\
3.77720694918583	98.0504658385093 \\
3.88029161528734	98.0504658385093 \\
3.98337628138884	98.0504658385093 \\
4.08646094749035	98.0504658385093 \\
4.18954561359185	98.0504658385093 \\
4.29263027969336	98.0504658385093 \\
4.39571494579486	98.0504658385093 \\
4.49879961189637	98.0504658385093 \\
4.60188427799787	98.0504658385093 \\
4.70496894409938	98.0504658385093 \\
};
\addplot [semithick, black, forget plot]
table [row sep=\\]{%
4.70496894409938	98.0504658385093 \\
4.80988752727883	97.9913900596921 \\
4.91480611045828	97.8141627232403 \\
5.01972469363774	97.518783829154 \\
5.12464327681719	97.1052533774332 \\
5.22956185999664	96.5735713680778 \\
5.3344804431761	95.923737801088 \\
5.43939902635555	95.1557526764636 \\
5.544317609535	94.2696159942047 \\
5.64923619271445	93.2653277543113 \\
5.75415477589391	92.1428879567834 \\
5.85907335907336	90.9022966016209 \\
5.96399194225281	89.543553688824 \\
6.06891052543226	88.0666592183925 \\
6.17382910861172	86.4716131903265 \\
6.27874769179117	84.758415604626 \\
6.38366627497062	82.9270664612909 \\
6.48858485815007	80.9775657603214 \\
6.59350344132953	78.9099135017173 \\
6.69842202450898	76.7241096854787 \\
6.80334060768843	74.4201543116056 \\
6.90825919086789	71.9980473800979 \\
7.01317777404734	69.4577888909558 \\
7.11809635722679	66.7993788441791 \\
7.22301494040624	64.0228172397679 \\
7.3279335235857	61.1281040777222 \\
7.43285210676515	58.115239358042 \\
7.5377706899446	54.9842230807272 \\
7.64268927312405	51.735055245778 \\
7.74760785630351	48.3677358531942 \\
7.85252643948296	44.8822649029759 \\
7.95744502266241	41.2786423951231 \\
8.06236360584187	37.5568683296357 \\
8.16728218902132	33.7169427065139 \\
8.27220077220077	29.7588655257575 \\
8.37711935538022	25.6826367873666 \\
8.48203793855968	21.4882564913412 \\
8.58695652173913	17.1757246376812 \\
};
\addplot [semithick, black, forget plot]
table [row sep=\\]{%
8.58695652173913	17.1757246376812 \\
8.71697488354037	11.7582928959628 \\
8.84699324534161	6.34086115424431 \\
8.97701160714286	0.923429412525852 \\
9.1070299689441	-4.4940023291926 \\
};
\addplot [semithick, blue, dashed, forget plot]
table [row sep=\\]{%
0	0 \\
0.103448275862069	-3.06037455410226 \\
0.206896551724138	-6.0346016646849 \\
0.310344827586207	-8.92268133174792 \\
0.413793103448276	-11.7246135552913 \\
0.517241379310345	-14.4403983353151 \\
0.620689655172414	-17.0700356718193 \\
0.724137931034483	-19.6135255648038 \\
0.827586206896552	-22.0708680142687 \\
0.931034482758621	-24.442063020214 \\
1.03448275862069	-26.7271105826397 \\
1.13793103448276	-28.9260107015458 \\
1.24137931034483	-31.0387633769322 \\
1.3448275862069	-33.065368608799 \\
1.44827586206897	-35.0058263971463 \\
1.55172413793103	-36.8601367419738 \\
1.6551724137931	-38.6282996432818 \\
1.75862068965517	-40.3103151010702 \\
1.86206896551724	-41.9061831153389 \\
1.96551724137931	-43.415903686088 \\
2.06896551724138	-44.8394768133175 \\
2.17241379310345	-46.1769024970274 \\
2.27586206896552	-47.4281807372176 \\
2.37931034482759	-48.5933115338882 \\
2.48275862068966	-49.6722948870392 \\
2.58620689655172	-50.6651307966706 \\
2.68965517241379	-51.5718192627824 \\
2.79310344827586	-52.3923602853746 \\
2.89655172413793	-53.1267538644471 \\
3	-53.775 \\
};
\addplot [semithick, blue, dashed, forget plot]
table [row sep=\\]{%
3	-53.775 \\
3.10295392171024	-54.320396405029 \\
3.20590784342048	-54.752024736106 \\
3.30886176513072	-55.0698849932312 \\
3.41181568684096	-55.2739771764045 \\
3.51476960855121	-55.3643012856258 \\
3.61772353026145	-55.3408573208953 \\
3.72067745197169	-55.2036452822129 \\
3.82363137368193	-54.9526651695785 \\
3.92658529539217	-54.5879169829923 \\
4.02953921710241	-54.1094007224542 \\
4.13249313881265	-53.5171163879642 \\
4.23544706052289	-52.8110639795222 \\
4.33840098223314	-51.9912434971284 \\
4.44135490394338	-51.0576549407827 \\
4.54430882565362	-50.0102983104851 \\
4.64726274736386	-48.8491736062356 \\
4.7502166690741	-47.5742808280342 \\
4.85317059078434	-46.1856199758809 \\
4.95612451249458	-44.6831910497757 \\
5.05907843420482	-43.0669940497185 \\
5.16203235591507	-41.3370289757095 \\
5.26498627762531	-39.4932958277487 \\
5.36794019933555	-37.5357946058358 \\
5.47089412104579	-35.4645253099711 \\
5.57384804275603	-33.2794879401545 \\
5.67680196446627	-30.9806824963861 \\
5.77975588617651	-28.5681089786656 \\
5.88270980788675	-26.0417673869934 \\
5.98566372959699	-23.4016577213692 \\
6.08861765130724	-20.6477799817931 \\
6.19157157301748	-17.7801341682651 \\
6.29452549472772	-14.7987202807852 \\
6.39747941643796	-11.7035383193534 \\
6.5004333381482	-8.49458828396972 \\
6.60338725985844	-5.17187017463417 \\
6.70634118156868	-1.73538399134667 \\
6.80929510327893	1.81487026589272 \\
6.91224902498917	5.478892597084 \\
7.01520294669941	9.25668300222716 \\
7.11815686840965	13.1482414813223 \\
7.22111079011989	17.1535680343692 \\
7.32406471183013	21.2726626613681 \\
7.42701863354037	25.5055253623189 \\
};
\addplot [semithick, blue, dashed, forget plot]
table [row sep=\\]{%
7.42701863354037	25.5055253623189 \\
7.53901938923395	30.1722235162181 \\
7.65102014492754	34.8389216701174 \\
7.76302090062112	39.5056198240166 \\
7.8750216563147	44.1723179779158 \\
7.98702241200828	48.8390161318151 \\
8.09902316770186	53.5057142857143 \\
8.21102392339545	58.1724124396136 \\
8.32302467908903	62.8391105935128 \\
8.43502543478261	67.5058087474121 \\
8.54702619047619	72.1725069013114 \\
8.65902694616977	76.8392050552105 \\
8.77102770186336	81.5059032091098 \\
8.88302845755694	86.1726013630091 \\
8.99502921325052	90.8392995169083 \\
9.1070299689441	95.5059976708075 \\
};
\addplot [semithick, blue, dashed, forget plot]
table [row sep=\\]{%
0	0 \\
0.103448275862069	-3.01730083234245 \\
0.206896551724138	-5.86230677764566 \\
0.310344827586207	-8.53501783590963 \\
0.413793103448276	-11.0354340071344 \\
0.517241379310345	-13.3635552913199 \\
0.620689655172414	-15.5193816884661 \\
0.724137931034483	-17.5029131985731 \\
0.827586206896552	-19.3141498216409 \\
0.931034482758621	-20.9530915576694 \\
1.03448275862069	-22.4197384066587 \\
1.13793103448276	-23.7140903686088 \\
1.24137931034483	-24.8361474435196 \\
1.3448275862069	-25.7859096313912 \\
1.44827586206897	-26.5633769322235 \\
1.55172413793103	-27.1685493460166 \\
1.6551724137931	-27.6014268727705 \\
1.75862068965517	-27.8620095124851 \\
1.86206896551724	-27.9502972651605 \\
1.96551724137931	-27.8662901307967 \\
2.06896551724138	-27.6099881093936 \\
2.17241379310345	-27.1813912009513 \\
2.27586206896552	-26.5804994054697 \\
2.37931034482759	-25.8073127229489 \\
2.48275862068966	-24.8618311533888 \\
2.58620689655172	-23.7440546967895 \\
2.68965517241379	-22.453983353151 \\
2.79310344827586	-20.9916171224732 \\
2.89655172413793	-19.3569560047562 \\
3	-17.55 \\
};
\addplot [semithick, blue, dashed, forget plot]
table [row sep=\\]{%
3	-17.55 \\
3.10346215780998	-15.7140895558736 \\
3.20692431561997	-13.99307319934 \\
3.31038647342995	-12.386950930399 \\
3.41384863123994	-10.8957227490507 \\
3.51731078904992	-9.51938865529512 \\
3.6207729468599	-8.25794864913222 \\
3.72423510466989	-7.11140273056201 \\
3.82769726247987	-6.0797508995845 \\
3.93115942028985	-5.16299315619967 \\
4.03462157809984	-4.36112950040754 \\
4.13808373590982	-3.6741599322081 \\
4.24154589371981	-3.10208445160136 \\
4.34500805152979	-2.6449030585873 \\
4.44847020933977	-2.30261575316594 \\
4.55193236714976	-2.07522253533726 \\
4.65539452495974	-1.96272340510128 \\
4.75885668276973	-1.96511836245799 \\
4.86231884057971	-2.0824074074074 \\
4.96578099838969	-2.31459053994949 \\
5.06924315619968	-2.66166776008428 \\
5.17270531400966	-3.12363906781176 \\
5.27616747181965	-3.70050446313194 \\
5.37962962962963	-4.3922639460448 \\
5.48309178743961	-5.19891751655035 \\
5.5865539452496	-6.12046517464859 \\
5.69001610305958	-7.15690692033954 \\
5.79347826086956	-8.30824275362317 \\
5.89694041867955	-9.5744726744995 \\
6.00040257648953	-10.9555966829685 \\
6.10386473429952	-12.4516147790302 \\
6.2073268921095	-14.0625269626846 \\
6.31078904991948	-15.7883332339317 \\
6.41425120772947	-17.6290335927715 \\
6.51771336553945	-19.584628039204 \\
6.62117552334944	-21.6551165732291 \\
6.72463768115942	-23.840499194847 \\
6.8280998389694	-26.1407759040575 \\
6.93156199677939	-28.5559467008608 \\
7.03502415458937	-31.0860115852567 \\
7.13848631239935	-33.7309705572453 \\
7.24194847020934	-36.4908236168267 \\
7.34541062801932	-39.3655707640007 \\
7.44887278582931	-42.3552119987674 \\
7.55233494363929	-45.4597473211268 \\
7.65579710144927	-48.6791767310789 \\
7.75925925925926	-52.0135002286236 \\
7.86272141706924	-55.4627178137611 \\
7.96618357487923	-59.0268294864913 \\
8.06964573268921	-62.7058352468142 \\
8.17310789049919	-66.4997350947296 \\
8.27657004830918	-70.408529030238 \\
8.38003220611916	-74.4322170533388 \\
8.48349436392914	-78.5707991640324 \\
8.58695652173913	-82.8242753623188 \\
};
\addplot [semithick, blue, dashed, forget plot]
table [row sep=\\]{%
8.58695652173913	-82.8242753623188 \\
8.71697488354037	-88.2417071040372 \\
8.84699324534161	-93.6591388457557 \\
8.97701160714286	-99.0765705874741 \\
9.1070299689441	-104.494002329193 \\
};
\addplot [semithick, black, dashed, forget plot]
table [row sep=\\]{
3.0   -500.0 \\
3.0   500.0 \\
};

  \node (fig2) at (0.55,-10)
   {\includegraphics[scale=0.05]{jbplane.png}};
   
  \node (fig3) at (7.4,85)
   {\includegraphics[scale=0.05]{jbplane_intruder.png}};
   
    %
  \node[text=black, text width=2cm] at (5,-100)  {\footnotesize Unsafeable for CL1500};
  \node[text=white, text width=2cm] at (8,-150) {\footnotesize Safeable for CL1500};
  \node[text=black] at (1,25)  {\footnotesize Ownship};
  \node[text=white] at (7.4,110)  {\footnotesize Intruder};
  \node[text=white] at (3.65,-190)  {\footnotesize $\tau=\epsilon$};
   
\end{axis}

\end{tikzpicture}

%% file: Safeable_CheckRegion.tex
\begin{tikzpicture}

\begin{axis}[
axis line style = thick,
tick align=outside,
tick pos=left,
x grid style={white!69.01960784313725!black},
xlabel={$\tau$ (sec)},
xmin=0, xmax=10.1070299689441,
y grid style={white!69.01960784313725!black},
ylabel={$h$ (ft)},
ymin=-205.369215838509, ymax=150
]
\addplot graphics [includegraphics cmd=\pgfimage,xmin=0, xmax=10.1070299689441, ymin=-205.369215838509, ymax=150] {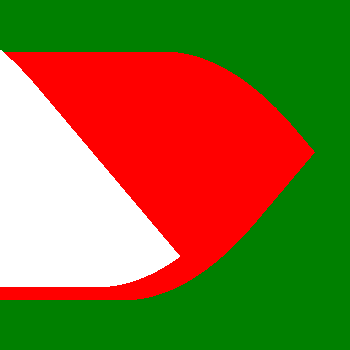};
\addplot [semithick, black, forget plot]
table [row sep=\\]{%
0	-155.369215838509 \\
0.104265619291195	-155.369215838509 \\
0.208531238582389	-155.369215838509 \\
0.312796857873584	-155.369215838509 \\
0.417062477164779	-155.369215838509 \\
0.521328096455974	-155.369215838509 \\
0.625593715747168	-155.369215838509 \\
0.729859335038363	-155.369215838509 \\
0.834124954329558	-155.369215838509 \\
0.938390573620752	-155.369215838509 \\
1.04265619291195	-155.369215838509 \\
1.14692181220314	-155.369215838509 \\
1.25118743149434	-155.369215838509 \\
1.35545305078553	-155.369215838509 \\
1.45971867007673	-155.369215838509 \\
1.56398428936792	-155.369215838509 \\
1.66824990865912	-155.369215838509 \\
1.77251552795031	-155.369215838509 \\
1.8767811472415	-155.369215838509 \\
1.9810467665327	-155.369215838509 \\
2.08531238582389	-155.369215838509 \\
2.18957800511509	-155.369215838509 \\
2.29384362440628	-155.369215838509 \\
2.39810924369748	-155.369215838509 \\
2.50237486298867	-155.369215838509 \\
2.60664048227987	-155.369215838509 \\
2.71090610157106	-155.369215838509 \\
2.81517172086226	-155.369215838509 \\
2.91943734015345	-155.369215838509 \\
3.02370295944465	-155.369215838509 \\
3.12796857873584	-155.369215838509 \\
3.23223419802704	-155.369215838509 \\
3.33649981731823	-155.369215838509 \\
3.44076543660943	-155.369215838509 \\
3.54503105590062	-155.369215838509 \\
};
\addplot [semithick, black, forget plot]
table [row sep=\\]{%
3.54503105590062	-155.369215838509 \\
3.64994963908007	-155.310140059692 \\
3.75486822225953	-155.13291272324 \\
3.85978680543898	-154.837533829154 \\
3.96470538861843	-154.424003377433 \\
4.06962397179788	-153.892321368078 \\
4.17454255497734	-153.242487801088 \\
4.27946113815679	-152.474502676464 \\
4.38437972133624	-151.588365994205 \\
4.4892983045157	-150.584077754311 \\
4.59421688769515	-149.461637956783 \\
4.6991354708746	-148.221046601621 \\
4.80405405405405	-146.862303688824 \\
4.90897263723351	-145.385409218392 \\
5.01389122041296	-143.790363190326 \\
5.11880980359241	-142.077165604626 \\
5.22372838677186	-140.245816461291 \\
5.32864696995132	-138.296315760321 \\
5.43356555313077	-136.228663501717 \\
5.53848413631022	-134.042859685479 \\
5.64340271948967	-131.738904311606 \\
5.74832130266913	-129.316797380098 \\
5.85323988584858	-126.776538890956 \\
5.95815846902803	-124.118128844179 \\
6.06307705220749	-121.341567239768 \\
6.16799563538694	-118.446854077722 \\
6.27291421856639	-115.433989358042 \\
6.37783280174584	-112.302973080727 \\
6.4827513849253	-109.053805245778 \\
6.58766996810475	-105.686485853194 \\
6.6925885512842	-102.201014902976 \\
6.79750713446366	-98.597392395123 \\
6.90242571764311	-94.8756183296357 \\
7.00734430082256	-91.0356927065138 \\
7.11226288400201	-87.0776155257574 \\
7.21718146718147	-83.0013867873665 \\
7.32210005036092	-78.8070064913411 \\
7.42701863354037	-74.4944746376812 \\
};
\addplot [semithick, black, forget plot]
table [row sep=\\]{%
7.42701863354037	-74.4944746376812 \\
7.53901938923395	-69.8277764837819 \\
7.65102014492754	-65.1610783298827 \\
7.76302090062112	-60.4943801759834 \\
7.8750216563147	-55.8276820220842 \\
7.98702241200828	-51.1609838681849 \\
8.09902316770186	-46.4942857142857 \\
8.21102392339545	-41.8275875603864 \\
8.32302467908903	-37.1608894064872 \\
8.43502543478261	-32.494191252588 \\
8.54702619047619	-27.8274930986887 \\
8.65902694616977	-23.1607949447895 \\
8.77102770186336	-18.4940967908902 \\
8.88302845755694	-13.8273986369909 \\
8.99502921325052	-9.16070048309174 \\
9.1070299689441	-4.49400232919248 \\
};
\addplot [semithick, black, forget plot]
table [row sep=\\]{%
0	100 \\
0.0330794847658176	99.0164241881401 \\
0.0661589695316352	98.0504658385093 \\
};
\addplot [semithick, black, forget plot]
table [row sep=\\]{%
0.0661589695316352	98.0504658385093 \\
0.169243635633141	98.0504658385093 \\
0.272328301734646	98.0504658385093 \\
0.375412967836151	98.0504658385093 \\
0.478497633937657	98.0504658385093 \\
0.581582300039162	98.0504658385093 \\
0.684666966140668	98.0504658385093 \\
0.787751632242173	98.0504658385093 \\
0.890836298343679	98.0504658385093 \\
0.993920964445184	98.0504658385093 \\
1.09700563054669	98.0504658385093 \\
1.20009029664819	98.0504658385093 \\
1.3031749627497	98.0504658385093 \\
1.40625962885121	98.0504658385093 \\
1.50934429495271	98.0504658385093 \\
1.61242896105422	98.0504658385093 \\
1.71551362715572	98.0504658385093 \\
1.81859829325723	98.0504658385093 \\
1.92168295935873	98.0504658385093 \\
2.02476762546024	98.0504658385093 \\
2.12785229156174	98.0504658385093 \\
2.23093695766325	98.0504658385093 \\
2.33402162376475	98.0504658385093 \\
2.43710628986626	98.0504658385093 \\
2.54019095596777	98.0504658385093 \\
2.64327562206927	98.0504658385093 \\
2.74636028817078	98.0504658385093 \\
2.84944495427228	98.0504658385093 \\
2.95252962037379	98.0504658385093 \\
3.05561428647529	98.0504658385093 \\
3.1586989525768	98.0504658385093 \\
3.2617836186783	98.0504658385093 \\
3.36486828477981	98.0504658385093 \\
3.46795295088131	98.0504658385093 \\
3.57103761698282	98.0504658385093 \\
3.67412228308432	98.0504658385093 \\
3.77720694918583	98.0504658385093 \\
3.88029161528734	98.0504658385093 \\
3.98337628138884	98.0504658385093 \\
4.08646094749035	98.0504658385093 \\
4.18954561359185	98.0504658385093 \\
4.29263027969336	98.0504658385093 \\
4.39571494579486	98.0504658385093 \\
4.49879961189637	98.0504658385093 \\
4.60188427799787	98.0504658385093 \\
4.70496894409938	98.0504658385093 \\
};
\addplot [semithick, black, forget plot]
table [row sep=\\]{%
4.70496894409938	98.0504658385093 \\
4.80988752727883	97.9913900596921 \\
4.91480611045828	97.8141627232403 \\
5.01972469363774	97.518783829154 \\
5.12464327681719	97.1052533774332 \\
5.22956185999664	96.5735713680778 \\
5.3344804431761	95.923737801088 \\
5.43939902635555	95.1557526764636 \\
5.544317609535	94.2696159942047 \\
5.64923619271445	93.2653277543113 \\
5.75415477589391	92.1428879567834 \\
5.85907335907336	90.9022966016209 \\
5.96399194225281	89.543553688824 \\
6.06891052543226	88.0666592183925 \\
6.17382910861172	86.4716131903265 \\
6.27874769179117	84.758415604626 \\
6.38366627497062	82.9270664612909 \\
6.48858485815007	80.9775657603214 \\
6.59350344132953	78.9099135017173 \\
6.69842202450898	76.7241096854787 \\
6.80334060768843	74.4201543116056 \\
6.90825919086789	71.9980473800979 \\
7.01317777404734	69.4577888909558 \\
7.11809635722679	66.7993788441791 \\
7.22301494040624	64.0228172397679 \\
7.3279335235857	61.1281040777222 \\
7.43285210676515	58.115239358042 \\
7.5377706899446	54.9842230807272 \\
7.64268927312405	51.735055245778 \\
7.74760785630351	48.3677358531942 \\
7.85252643948296	44.8822649029759 \\
7.95744502266241	41.2786423951231 \\
8.06236360584187	37.5568683296357 \\
8.16728218902132	33.7169427065139 \\
8.27220077220077	29.7588655257575 \\
8.37711935538022	25.6826367873666 \\
8.48203793855968	21.4882564913412 \\
8.58695652173913	17.1757246376812 \\
};
\addplot [semithick, black, forget plot]
table [row sep=\\]{%
8.58695652173913	17.1757246376812 \\
8.71697488354037	11.7582928959628 \\
8.84699324534161	6.34086115424431 \\
8.97701160714286	0.923429412525852 \\
9.1070299689441	-4.4940023291926 \\
};
\addplot [semithick, blue, dashed, forget plot]
table [row sep=\\]{%
0	0 \\
0.103448275862069	-3.06037455410226 \\
0.206896551724138	-6.0346016646849 \\
0.310344827586207	-8.92268133174792 \\
0.413793103448276	-11.7246135552913 \\
0.517241379310345	-14.4403983353151 \\
0.620689655172414	-17.0700356718193 \\
0.724137931034483	-19.6135255648038 \\
0.827586206896552	-22.0708680142687 \\
0.931034482758621	-24.442063020214 \\
1.03448275862069	-26.7271105826397 \\
1.13793103448276	-28.9260107015458 \\
1.24137931034483	-31.0387633769322 \\
1.3448275862069	-33.065368608799 \\
1.44827586206897	-35.0058263971463 \\
1.55172413793103	-36.8601367419738 \\
1.6551724137931	-38.6282996432818 \\
1.75862068965517	-40.3103151010702 \\
1.86206896551724	-41.9061831153389 \\
1.96551724137931	-43.415903686088 \\
2.06896551724138	-44.8394768133175 \\
2.17241379310345	-46.1769024970274 \\
2.27586206896552	-47.4281807372176 \\
2.37931034482759	-48.5933115338882 \\
2.48275862068966	-49.6722948870392 \\
2.58620689655172	-50.6651307966706 \\
2.68965517241379	-51.5718192627824 \\
2.79310344827586	-52.3923602853746 \\
2.89655172413793	-53.1267538644471 \\
3	-53.775 \\
};
\addplot [semithick, blue, dashed, forget plot]
table [row sep=\\]{%
3	-53.775 \\
3.10295392171024	-54.320396405029 \\
3.20590784342048	-54.752024736106 \\
3.30886176513072	-55.0698849932312 \\
3.41181568684096	-55.2739771764045 \\
3.51476960855121	-55.3643012856258 \\
3.61772353026145	-55.3408573208953 \\
3.72067745197169	-55.2036452822129 \\
3.82363137368193	-54.9526651695785 \\
3.92658529539217	-54.5879169829923 \\
4.02953921710241	-54.1094007224542 \\
4.13249313881265	-53.5171163879642 \\
4.23544706052289	-52.8110639795222 \\
4.33840098223314	-51.9912434971284 \\
4.44135490394338	-51.0576549407827 \\
4.54430882565362	-50.0102983104851 \\
4.64726274736386	-48.8491736062356 \\
4.7502166690741	-47.5742808280342 \\
4.85317059078434	-46.1856199758809 \\
4.95612451249458	-44.6831910497757 \\
5.05907843420482	-43.0669940497185 \\
5.16203235591507	-41.3370289757095 \\
5.26498627762531	-39.4932958277487 \\
5.36794019933555	-37.5357946058358 \\
5.47089412104579	-35.4645253099711 \\
5.57384804275603	-33.2794879401545 \\
5.67680196446627	-30.9806824963861 \\
5.77975588617651	-28.5681089786656 \\
5.88270980788675	-26.0417673869934 \\
5.98566372959699	-23.4016577213692 \\
6.08861765130724	-20.6477799817931 \\
6.19157157301748	-17.7801341682651 \\
6.29452549472772	-14.7987202807852 \\
6.39747941643796	-11.7035383193534 \\
6.5004333381482	-8.49458828396972 \\
6.60338725985844	-5.17187017463417 \\
6.70634118156868	-1.73538399134667 \\
6.80929510327893	1.81487026589272 \\
6.91224902498917	5.478892597084 \\
7.01520294669941	9.25668300222716 \\
7.11815686840965	13.1482414813223 \\
7.22111079011989	17.1535680343692 \\
7.32406471183013	21.2726626613681 \\
7.42701863354037	25.5055253623189 \\
};
\addplot [semithick, blue, dashed, forget plot]
table [row sep=\\]{%
7.42701863354037	25.5055253623189 \\
7.53901938923395	30.1722235162181 \\
7.65102014492754	34.8389216701174 \\
7.76302090062112	39.5056198240166 \\
7.8750216563147	44.1723179779158 \\
7.98702241200828	48.8390161318151 \\
8.09902316770186	53.5057142857143 \\
8.21102392339545	58.1724124396136 \\
8.32302467908903	62.8391105935128 \\
8.43502543478261	67.5058087474121 \\
8.54702619047619	72.1725069013114 \\
8.65902694616977	76.8392050552105 \\
8.77102770186336	81.5059032091098 \\
8.88302845755694	86.1726013630091 \\
8.99502921325052	90.8392995169083 \\
9.1070299689441	95.5059976708075 \\
};
\addplot [semithick, blue, dashed, forget plot]
table [row sep=\\]{%
0	0 \\
0.103448275862069	-3.01730083234245 \\
0.206896551724138	-5.86230677764566 \\
0.310344827586207	-8.53501783590963 \\
0.413793103448276	-11.0354340071344 \\
0.517241379310345	-13.3635552913199 \\
0.620689655172414	-15.5193816884661 \\
0.724137931034483	-17.5029131985731 \\
0.827586206896552	-19.3141498216409 \\
0.931034482758621	-20.9530915576694 \\
1.03448275862069	-22.4197384066587 \\
1.13793103448276	-23.7140903686088 \\
1.24137931034483	-24.8361474435196 \\
1.3448275862069	-25.7859096313912 \\
1.44827586206897	-26.5633769322235 \\
1.55172413793103	-27.1685493460166 \\
1.6551724137931	-27.6014268727705 \\
1.75862068965517	-27.8620095124851 \\
1.86206896551724	-27.9502972651605 \\
1.96551724137931	-27.8662901307967 \\
2.06896551724138	-27.6099881093936 \\
2.17241379310345	-27.1813912009513 \\
2.27586206896552	-26.5804994054697 \\
2.37931034482759	-25.8073127229489 \\
2.48275862068966	-24.8618311533888 \\
2.58620689655172	-23.7440546967895 \\
2.68965517241379	-22.453983353151 \\
2.79310344827586	-20.9916171224732 \\
2.89655172413793	-19.3569560047562 \\
3	-17.55 \\
};
\addplot [semithick, blue, dashed, forget plot]
table [row sep=\\]{%
3	-17.55 \\
3.10346215780998	-15.7140895558736 \\
3.20692431561997	-13.99307319934 \\
3.31038647342995	-12.386950930399 \\
3.41384863123994	-10.8957227490507 \\
3.51731078904992	-9.51938865529512 \\
3.6207729468599	-8.25794864913222 \\
3.72423510466989	-7.11140273056201 \\
3.82769726247987	-6.0797508995845 \\
3.93115942028985	-5.16299315619967 \\
4.03462157809984	-4.36112950040754 \\
4.13808373590982	-3.6741599322081 \\
4.24154589371981	-3.10208445160136 \\
4.34500805152979	-2.6449030585873 \\
4.44847020933977	-2.30261575316594 \\
4.55193236714976	-2.07522253533726 \\
4.65539452495974	-1.96272340510128 \\
4.75885668276973	-1.96511836245799 \\
4.86231884057971	-2.0824074074074 \\
4.96578099838969	-2.31459053994949 \\
5.06924315619968	-2.66166776008428 \\
5.17270531400966	-3.12363906781176 \\
5.27616747181965	-3.70050446313194 \\
5.37962962962963	-4.3922639460448 \\
5.48309178743961	-5.19891751655035 \\
5.5865539452496	-6.12046517464859 \\
5.69001610305958	-7.15690692033954 \\
5.79347826086956	-8.30824275362317 \\
5.89694041867955	-9.5744726744995 \\
6.00040257648953	-10.9555966829685 \\
6.10386473429952	-12.4516147790302 \\
6.2073268921095	-14.0625269626846 \\
6.31078904991948	-15.7883332339317 \\
6.41425120772947	-17.6290335927715 \\
6.51771336553945	-19.584628039204 \\
6.62117552334944	-21.6551165732291 \\
6.72463768115942	-23.840499194847 \\
6.8280998389694	-26.1407759040575 \\
6.93156199677939	-28.5559467008608 \\
7.03502415458937	-31.0860115852567 \\
7.13848631239935	-33.7309705572453 \\
7.24194847020934	-36.4908236168267 \\
7.34541062801932	-39.3655707640007 \\
7.44887278582931	-42.3552119987674 \\
7.55233494363929	-45.4597473211268 \\
7.65579710144927	-48.6791767310789 \\
7.75925925925926	-52.0135002286236 \\
7.86272141706924	-55.4627178137611 \\
7.96618357487923	-59.0268294864913 \\
8.06964573268921	-62.7058352468142 \\
8.17310789049919	-66.4997350947296 \\
8.27657004830918	-70.408529030238 \\
8.38003220611916	-74.4322170533388 \\
8.48349436392914	-78.5707991640324 \\
8.58695652173913	-82.8242753623188 \\
};
\addplot [semithick, blue, dashed, forget plot]
table [row sep=\\]{%
8.58695652173913	-82.8242753623188 \\
8.71697488354037	-88.2417071040372 \\
8.84699324534161	-93.6591388457557 \\
8.97701160714286	-99.0765705874741 \\
9.1070299689441	-104.494002329193 \\
};
\addplot [semithick, black, forget plot]
table [row sep=\\]{%
0	-141.925465838509 \\
0.107501194457716	-141.925465838509 \\
0.215002388915432	-141.925465838509 \\
0.322503583373149	-141.925465838509 \\
0.430004777830865	-141.925465838509 \\
0.537505972288581	-141.925465838509 \\
0.645007166746297	-141.925465838509 \\
0.752508361204013	-141.925465838509 \\
0.86000955566173	-141.925465838509 \\
0.967510750119446	-141.925465838509 \\
1.07501194457716	-141.925465838509 \\
1.18251313903488	-141.925465838509 \\
1.29001433349259	-141.925465838509 \\
1.39751552795031	-141.925465838509 \\
1.50501672240803	-141.925465838509 \\
1.61251791686574	-141.925465838509 \\
1.72001911132346	-141.925465838509 \\
1.82752030578118	-141.925465838509 \\
1.93502150023889	-141.925465838509 \\
2.04252269469661	-141.925465838509 \\
2.15002388915432	-141.925465838509 \\
2.25752508361204	-141.925465838509 \\
2.36502627806976	-141.925465838509 \\
2.47252747252747	-141.925465838509 \\
2.58002866698519	-141.925465838509 \\
2.6875298614429	-141.925465838509 \\
2.79503105590062	-141.925465838509 \\
};
\addplot [semithick, black, forget plot]
table [row sep=\\]{%
2.79503105590062	-141.925465838509 \\
2.89995501572451	-141.866384004757 \\
3.0048789755484	-141.689138503501 \\
3.10980293537229	-141.39372933474 \\
3.21472689519618	-140.980156498475 \\
3.31965085502008	-140.448419994705 \\
3.42457481484397	-139.798519823431 \\
3.52949877466786	-139.030455984653 \\
3.63442273449175	-138.14422847837 \\
3.73934669431564	-137.139837304583 \\
3.84427065413953	-136.017282463292 \\
3.94919461396342	-134.776563954496 \\
4.05411857378731	-133.417681778196 \\
4.1590425336112	-131.940635934392 \\
4.26396649343509	-130.345426423083 \\
4.36889045325898	-128.63205324427 \\
4.47381441308287	-126.800516397953 \\
4.57873837290677	-124.850815884131 \\
4.68366233273066	-122.782951702805 \\
4.78858629255455	-120.596923853975 \\
4.89351025237844	-118.29273233764 \\
4.99843421220233	-115.870377153801 \\
5.10335817202622	-113.329858302457 \\
5.20828213185011	-110.67117578361 \\
};
\addplot [semithick, black, forget plot]
table [row sep=\\]{%
0	100 \\
0.120772946859903	96.2985328323493 \\
0.241545893719807	92.4405081409912 \\
0.36231884057971	88.4259259259259 \\
0.483091787439613	84.2547861871533 \\
0.603864734299517	79.9270889246735 \\
0.72463768115942	75.4428341384863 \\
0.845410628019323	70.8020218285919 \\
0.966183574879227	66.0046519949902 \\
1.08695652173913	61.0507246376812 \\
};
\addplot [semithick, black, forget plot]
table [row sep=\\]{%
1.08695652173913	61.0507246376812 \\
1.19323671497585	56.622383252818 \\
1.29951690821256	52.1940418679549 \\
1.40579710144928	47.7657004830918 \\
1.51207729468599	43.3373590982287 \\
1.6183574879227	38.9090177133656 \\
1.72463768115942	34.4806763285024 \\
1.83091787439613	30.0523349436393 \\
1.93719806763285	25.6239935587762 \\
2.04347826086956	21.1956521739131 \\
2.14975845410628	16.7673107890499 \\
2.25603864734299	12.3389694041868 \\
2.36231884057971	7.91062801932368 \\
2.46859903381642	3.48228663446056 \\
2.57487922705314	-0.946054750402567 \\
2.68115942028985	-5.3743961352657 \\
2.78743961352657	-9.80273752012882 \\
2.89371980676328	-14.2310789049919 \\
3	-18.6594202898551 \\
};
\addplot [semithick, black, forget plot]
table [row sep=\\]{%
3	-18.6594202898551 \\
3.10515629199286	-23.0409324562243 \\
3.21031258398572	-27.4224446225936 \\
3.31546887597859	-31.8039567889629 \\
3.42062516797145	-36.1854689553321 \\
3.52578145996431	-40.5669811217014 \\
3.63093775195717	-44.9484932880707 \\
3.73609404395004	-49.3300054544399 \\
3.8412503359429	-53.7115176208092 \\
3.94640662793576	-58.0930297871785 \\
4.05156291992862	-62.4745419535477 \\
4.15671921192149	-66.856054119917 \\
4.26187550391435	-71.2375662862863 \\
4.36703179590721	-75.6190784526555 \\
4.47218808790007	-80.0005906190248 \\
4.57734437989294	-84.3821027853941 \\
4.6825006718858	-88.7636149517633 \\
4.78765696387866	-93.1451271181326 \\
4.89281325587152	-97.5266392845019 \\
4.99796954786439	-101.908151450871 \\
5.10312583985725	-106.28966361724 \\
5.20828213185011	-110.67117578361 \\
};

  \node (fig2) at (0.55,-10)
   {\includegraphics[scale=0.05]{jbplane.png}};

\node[text=black,text width=2cm,align=center] at (2,-90)  {\footnotesize Unsafeable for All Advisories};
\node[text=white] at (7.5,-150)  {\footnotesize Safeable for CL1500};
\node[text=white] at (4,70)  {\footnotesize Unsafeable for CL1500};

\end{axis}

\end{tikzpicture}

%% file: Safeable_Velocities2.tex
\begin{tikzpicture}

\definecolor{color0}{rgb}{0.75,0.75,0}

\begin{axis}[
legend cell align={left},
legend entries={{$v_O$: [-30,-30] ft/s},{$v_O$: [-31,-29] ft/s},{$v_O$: [-32,-28] ft/s},{$v_O$: [-34,-26] ft/s},{$v_O$: [-38,-22] ft/s}},
legend style={at={(0.09,0.5)}, anchor=west, draw=white!80.0!black},
tick align=outside,
tick pos=left,
x grid style={lightgray!92.02614379084967!black},
xlabel={$\tau$ (sec)},
xmin=-0.53087277173913, xmax=11.1483282065217,
y grid style={lightgray!92.02614379084967!black},
ylabel={$h$ (ft)},
ymin=-202.821993400621, ymax=151.623585015528
]
\addlegendimage{no markers, black}
\addlegendimage{no markers, red}
\addlegendimage{no markers, blue}
\addlegendimage{no markers, green!50.0!black}
\addlegendimage{no markers, color0}
\addplot [semithick, black]
table [row sep=\\]{%
0	-155.369215838509 \\
0.104265619291195	-155.369215838509 \\
0.208531238582389	-155.369215838509 \\
0.312796857873584	-155.369215838509 \\
0.417062477164779	-155.369215838509 \\
0.521328096455974	-155.369215838509 \\
0.625593715747168	-155.369215838509 \\
0.729859335038363	-155.369215838509 \\
0.834124954329558	-155.369215838509 \\
0.938390573620752	-155.369215838509 \\
1.04265619291195	-155.369215838509 \\
1.14692181220314	-155.369215838509 \\
1.25118743149434	-155.369215838509 \\
1.35545305078553	-155.369215838509 \\
1.45971867007673	-155.369215838509 \\
1.56398428936792	-155.369215838509 \\
1.66824990865912	-155.369215838509 \\
1.77251552795031	-155.369215838509 \\
1.8767811472415	-155.369215838509 \\
1.9810467665327	-155.369215838509 \\
2.08531238582389	-155.369215838509 \\
2.18957800511509	-155.369215838509 \\
2.29384362440628	-155.369215838509 \\
2.39810924369748	-155.369215838509 \\
2.50237486298867	-155.369215838509 \\
2.60664048227987	-155.369215838509 \\
2.71090610157106	-155.369215838509 \\
2.81517172086226	-155.369215838509 \\
2.91943734015345	-155.369215838509 \\
3.02370295944465	-155.369215838509 \\
3.12796857873584	-155.369215838509 \\
3.23223419802704	-155.369215838509 \\
3.33649981731823	-155.369215838509 \\
3.44076543660943	-155.369215838509 \\
3.54503105590062	-155.369215838509 \\
};
\addplot [semithick, black, forget plot]
table [row sep=\\]{%
3.54503105590062	-155.369215838509 \\
3.64994963908007	-155.310140059692 \\
3.75486822225953	-155.13291272324 \\
3.85978680543898	-154.837533829154 \\
3.96470538861843	-154.424003377433 \\
4.06962397179788	-153.892321368078 \\
4.17454255497734	-153.242487801088 \\
4.27946113815679	-152.474502676464 \\
4.38437972133624	-151.588365994205 \\
4.4892983045157	-150.584077754311 \\
4.59421688769515	-149.461637956783 \\
4.6991354708746	-148.221046601621 \\
4.80405405405405	-146.862303688824 \\
4.90897263723351	-145.385409218392 \\
5.01389122041296	-143.790363190326 \\
5.11880980359241	-142.077165604626 \\
5.22372838677186	-140.245816461291 \\
5.32864696995132	-138.296315760321 \\
5.43356555313077	-136.228663501717 \\
5.53848413631022	-134.042859685479 \\
5.64340271948967	-131.738904311606 \\
5.74832130266913	-129.316797380098 \\
5.85323988584858	-126.776538890956 \\
5.95815846902803	-124.118128844179 \\
6.06307705220749	-121.341567239768 \\
6.16799563538694	-118.446854077722 \\
6.27291421856639	-115.433989358042 \\
6.37783280174584	-112.302973080727 \\
6.4827513849253	-109.053805245778 \\
6.58766996810475	-105.686485853194 \\
6.6925885512842	-102.201014902976 \\
6.79750713446366	-98.597392395123 \\
6.90242571764311	-94.8756183296357 \\
7.00734430082256	-91.0356927065138 \\
7.11226288400201	-87.0776155257574 \\
7.21718146718147	-83.0013867873665 \\
7.32210005036092	-78.8070064913411 \\
7.42701863354037	-74.4944746376812 \\
};
\addplot [semithick, black, forget plot]
table [row sep=\\]{%
7.42701863354037	-74.4944746376812 \\
7.53901938923395	-69.8277764837819 \\
7.65102014492754	-65.1610783298827 \\
7.76302090062112	-60.4943801759834 \\
7.8750216563147	-55.8276820220842 \\
7.98702241200828	-51.1609838681849 \\
8.09902316770186	-46.4942857142857 \\
8.21102392339545	-41.8275875603864 \\
8.32302467908903	-37.1608894064872 \\
8.43502543478261	-32.494191252588 \\
8.54702619047619	-27.8274930986887 \\
8.65902694616977	-23.1607949447895 \\
8.77102770186336	-18.4940967908902 \\
8.88302845755694	-13.8273986369909 \\
8.99502921325052	-9.16070048309174 \\
9.1070299689441	-4.49400232919248 \\
};
\addplot [semithick, black, forget plot]
table [row sep=\\]{%
0	100 \\
0.0330794847658176	99.0164241881401 \\
0.0661589695316352	98.0504658385093 \\
};
\addplot [semithick, black, forget plot]
table [row sep=\\]{%
0.0661589695316352	98.0504658385093 \\
0.169243635633141	98.0504658385093 \\
0.272328301734646	98.0504658385093 \\
0.375412967836151	98.0504658385093 \\
0.478497633937657	98.0504658385093 \\
0.581582300039162	98.0504658385093 \\
0.684666966140668	98.0504658385093 \\
0.787751632242173	98.0504658385093 \\
0.890836298343679	98.0504658385093 \\
0.993920964445184	98.0504658385093 \\
1.09700563054669	98.0504658385093 \\
1.20009029664819	98.0504658385093 \\
1.3031749627497	98.0504658385093 \\
1.40625962885121	98.0504658385093 \\
1.50934429495271	98.0504658385093 \\
1.61242896105422	98.0504658385093 \\
1.71551362715572	98.0504658385093 \\
1.81859829325723	98.0504658385093 \\
1.92168295935873	98.0504658385093 \\
2.02476762546024	98.0504658385093 \\
2.12785229156174	98.0504658385093 \\
2.23093695766325	98.0504658385093 \\
2.33402162376475	98.0504658385093 \\
2.43710628986626	98.0504658385093 \\
2.54019095596777	98.0504658385093 \\
2.64327562206927	98.0504658385093 \\
2.74636028817078	98.0504658385093 \\
2.84944495427228	98.0504658385093 \\
2.95252962037379	98.0504658385093 \\
3.05561428647529	98.0504658385093 \\
3.1586989525768	98.0504658385093 \\
3.2617836186783	98.0504658385093 \\
3.36486828477981	98.0504658385093 \\
3.46795295088131	98.0504658385093 \\
3.57103761698282	98.0504658385093 \\
3.67412228308432	98.0504658385093 \\
3.77720694918583	98.0504658385093 \\
3.88029161528734	98.0504658385093 \\
3.98337628138884	98.0504658385093 \\
4.08646094749035	98.0504658385093 \\
4.18954561359185	98.0504658385093 \\
4.29263027969336	98.0504658385093 \\
4.39571494579486	98.0504658385093 \\
4.49879961189637	98.0504658385093 \\
4.60188427799787	98.0504658385093 \\
4.70496894409938	98.0504658385093 \\
};
\addplot [semithick, black, forget plot]
table [row sep=\\]{%
4.70496894409938	98.0504658385093 \\
4.80988752727883	97.9913900596921 \\
4.91480611045828	97.8141627232403 \\
5.01972469363774	97.518783829154 \\
5.12464327681719	97.1052533774332 \\
5.22956185999664	96.5735713680778 \\
5.3344804431761	95.923737801088 \\
5.43939902635555	95.1557526764636 \\
5.544317609535	94.2696159942047 \\
5.64923619271445	93.2653277543113 \\
5.75415477589391	92.1428879567834 \\
5.85907335907336	90.9022966016209 \\
5.96399194225281	89.543553688824 \\
6.06891052543226	88.0666592183925 \\
6.17382910861172	86.4716131903265 \\
6.27874769179117	84.758415604626 \\
6.38366627497062	82.9270664612909 \\
6.48858485815007	80.9775657603214 \\
6.59350344132953	78.9099135017173 \\
6.69842202450898	76.7241096854787 \\
6.80334060768843	74.4201543116056 \\
6.90825919086789	71.9980473800979 \\
7.01317777404734	69.4577888909558 \\
7.11809635722679	66.7993788441791 \\
7.22301494040624	64.0228172397679 \\
7.3279335235857	61.1281040777222 \\
7.43285210676515	58.115239358042 \\
7.5377706899446	54.9842230807272 \\
7.64268927312405	51.735055245778 \\
7.74760785630351	48.3677358531942 \\
7.85252643948296	44.8822649029759 \\
7.95744502266241	41.2786423951231 \\
8.06236360584187	37.5568683296357 \\
8.16728218902132	33.7169427065139 \\
8.27220077220077	29.7588655257575 \\
8.37711935538022	25.6826367873666 \\
8.48203793855968	21.4882564913412 \\
8.58695652173913	17.1757246376812 \\
};
\addplot [semithick, black, forget plot]
table [row sep=\\]{%
8.58695652173913	17.1757246376812 \\
8.71697488354037	11.7582928959628 \\
8.84699324534161	6.34086115424431 \\
8.97701160714286	0.923429412525852 \\
9.1070299689441	-4.4940023291926 \\
};
\addplot [semithick, red]
table [row sep=\\]{%
0	-158.960830745342 \\
0.103948535936114	-158.960830745342 \\
0.207897071872227	-158.960830745342 \\
0.311845607808341	-158.960830745342 \\
0.415794143744454	-158.960830745342 \\
0.519742679680568	-158.960830745342 \\
0.623691215616681	-158.960830745342 \\
0.727639751552795	-158.960830745342 \\
0.831588287488909	-158.960830745342 \\
0.935536823425022	-158.960830745342 \\
1.03948535936114	-158.960830745342 \\
1.14343389529725	-158.960830745342 \\
1.24738243123336	-158.960830745342 \\
1.35133096716948	-158.960830745342 \\
1.45527950310559	-158.960830745342 \\
1.5592280390417	-158.960830745342 \\
1.66317657497782	-158.960830745342 \\
1.76712511091393	-158.960830745342 \\
1.87107364685004	-158.960830745342 \\
1.97502218278616	-158.960830745342 \\
2.07897071872227	-158.960830745342 \\
2.18291925465839	-158.960830745342 \\
2.2868677905945	-158.960830745342 \\
2.39081632653061	-158.960830745342 \\
2.49476486246673	-158.960830745342 \\
2.59871339840284	-158.960830745342 \\
2.70266193433895	-158.960830745342 \\
2.80661047027507	-158.960830745342 \\
2.91055900621118	-158.960830745342 \\
3.01450754214729	-158.960830745342 \\
3.11845607808341	-158.960830745342 \\
3.22240461401952	-158.960830745342 \\
3.32635314995563	-158.960830745342 \\
3.43030168589175	-158.960830745342 \\
3.53425022182786	-158.960830745342 \\
3.63819875776397	-158.960830745342 \\
};
\addplot [semithick, red, forget plot]
table [row sep=\\]{%
3.63819875776397	-158.960830745342 \\
3.74311734094343	-158.901754966524 \\
3.84803592412288	-158.724527630073 \\
3.95295450730233	-158.429148735986 \\
4.05787309048179	-158.015618284265 \\
4.16279167366124	-157.48393627491 \\
4.26771025684069	-156.83410270792 \\
4.37262884002014	-156.066117583296 \\
4.4775474231996	-155.179980901037 \\
4.58246600637905	-154.175692661144 \\
4.6873845895585	-153.053252863616 \\
4.79230317273796	-151.812661508453 \\
4.89722175591741	-150.453918595656 \\
5.00214033909686	-148.977024125225 \\
5.10705892227631	-147.381978097159 \\
5.21197750545577	-145.668780511458 \\
5.31689608863522	-143.837431368123 \\
5.42181467181467	-141.887930667154 \\
5.52673325499412	-139.82027840855 \\
5.63165183817358	-137.634474592311 \\
5.73657042135303	-135.330519218438 \\
5.84148900453248	-132.90841228693 \\
5.94640758771193	-130.368153797788 \\
6.05132617089139	-127.709743751011 \\
6.15624475407084	-124.9331821466 \\
6.26116333725029	-122.038468984554 \\
6.36608192042975	-119.025604264874 \\
6.4710005036092	-115.894587987559 \\
6.57591908678865	-112.64542015261 \\
6.6808376699681	-109.278100760026 \\
6.78575625314756	-105.792629809808 \\
6.89067483632701	-102.189007301955 \\
6.99559341950646	-98.4672332364679 \\
7.10051200268592	-94.6273076133461 \\
7.20543058586537	-90.6692304325897 \\
7.31034916904482	-86.5930016941988 \\
7.41526775222427	-82.3986213981734 \\
7.52018633540373	-78.0860895445135 \\
};
\addplot [semithick, red, forget plot]
table [row sep=\\]{%
7.52018633540373	-78.0860895445135 \\
7.63144441964286	-73.4503360345497 \\
7.74270250388199	-68.814582524586 \\
7.85396058812112	-64.1788290146222 \\
7.96521867236025	-59.5430755046584 \\
8.07647675659938	-54.9073219946946 \\
8.18773484083851	-50.2715684847308 \\
8.29899292507764	-45.6358149747671 \\
8.41025100931677	-41.0000614648034 \\
8.5215090935559	-36.3643079548396 \\
8.63276717779503	-31.7285544448758 \\
8.74402526203416	-27.092800934912 \\
8.85528334627329	-22.4570474249483 \\
8.96654143051242	-17.8212939149845 \\
9.07779951475155	-13.1855404050208 \\
9.18905759899068	-8.549786895057 \\
9.30031568322981	-3.91403338509322 \\
};
\addplot [semithick, red, forget plot]
table [row sep=\\]{%
0	102.80201863354 \\
0.104307318390494	102.80201863354 \\
0.208614636780988	102.80201863354 \\
0.312921955171483	102.80201863354 \\
0.417229273561977	102.80201863354 \\
0.521536591952471	102.80201863354 \\
0.625843910342965	102.80201863354 \\
0.730151228733459	102.80201863354 \\
0.834458547123954	102.80201863354 \\
0.938765865514448	102.80201863354 \\
1.04307318390494	102.80201863354 \\
1.14738050229544	102.80201863354 \\
1.25168782068593	102.80201863354 \\
1.35599513907642	102.80201863354 \\
1.46030245746692	102.80201863354 \\
1.56460977585741	102.80201863354 \\
1.66891709424791	102.80201863354 \\
1.7732244126384	102.80201863354 \\
1.8775317310289	102.80201863354 \\
1.98183904941939	102.80201863354 \\
2.08614636780988	102.80201863354 \\
2.19045368620038	102.80201863354 \\
2.29476100459087	102.80201863354 \\
2.39906832298137	102.80201863354 \\
2.50337564137186	102.80201863354 \\
2.60768295976235	102.80201863354 \\
2.71199027815285	102.80201863354 \\
2.81629759654334	102.80201863354 \\
2.92060491493384	102.80201863354 \\
3.02491223332433	102.80201863354 \\
3.12921955171483	102.80201863354 \\
3.23352687010532	102.80201863354 \\
3.33783418849581	102.80201863354 \\
3.44214150688631	102.80201863354 \\
3.5464488252768	102.80201863354 \\
3.6507561436673	102.80201863354 \\
3.75506346205779	102.80201863354 \\
3.85937078044829	102.80201863354 \\
3.96367809883878	102.80201863354 \\
4.06798541722927	102.80201863354 \\
4.17229273561977	102.80201863354 \\
4.27660005401026	102.80201863354 \\
4.38090737240076	102.80201863354 \\
4.48521469079125	102.80201863354 \\
4.58952200918174	102.80201863354 \\
4.69382932757224	102.80201863354 \\
4.79813664596273	102.80201863354 \\
};
\addplot [semithick, red, forget plot]
table [row sep=\\]{%
4.79813664596273	102.80201863354 \\
4.90305522914219	102.742942854723 \\
5.00797381232164	102.565715518271 \\
5.11289239550109	102.270336624185 \\
5.21781097868054	101.856806172464 \\
5.32272956186	101.325124163109 \\
5.42764814503945	100.675290596119 \\
5.5325667282189	99.9073054714947 \\
5.63748531139835	99.0211687892358 \\
5.74240389457781	98.0168805493424 \\
5.84732247775726	96.8944407518144 \\
5.95224106093671	95.653849396652 \\
6.05715964411617	94.295106483855 \\
6.16207822729562	92.8182120134236 \\
6.26699681047507	91.2231659853576 \\
6.37191539365452	89.509968399657 \\
6.47683397683398	87.678619256322 \\
6.58175256001343	85.7291185553524 \\
6.68667114319288	83.6614662967483 \\
6.79158972637233	81.4756624805098 \\
6.89650830955179	79.1717071066366 \\
7.00142689273124	76.749600175129 \\
7.10634547591069	74.2093416859868 \\
7.21126405909015	71.5509316392101 \\
7.3161826422696	68.7743700347989 \\
7.42110122544905	65.8796568727532 \\
7.5260198086285	62.866792153073 \\
7.63093839180796	59.7357758757583 \\
7.73585697498741	56.486608040809 \\
7.84077555816686	53.1192886482252 \\
7.94569414134631	49.633817698007 \\
8.05061272452577	46.0301951901541 \\
8.15553130770522	42.3084211246668 \\
8.26044989088467	38.4684955015449 \\
8.36536847406412	34.5104183207885 \\
8.47028705724358	30.4341895823976 \\
8.57520564042303	26.2398092863722 \\
8.68012422360248	21.9272774327123 \\
};
\addplot [semithick, red, forget plot]
table [row sep=\\]{%
8.68012422360248	21.9272774327123 \\
8.80416251552795	16.7590152691512 \\
8.92820080745341	11.5907531055901 \\
9.05223909937888	6.42249094202905 \\
9.17627739130435	1.25422877846799 \\
9.30031568322981	-3.91403338509308 \\
};
\addplot [semithick, blue]
table [row sep=\\]{%
0	-162.645613354037 \\
0.103649068322981	-162.645613354037 \\
0.207298136645963	-162.645613354037 \\
0.310947204968944	-162.645613354037 \\
0.414596273291925	-162.645613354037 \\
0.518245341614907	-162.645613354037 \\
0.621894409937888	-162.645613354037 \\
0.725543478260869	-162.645613354037 \\
0.829192546583851	-162.645613354037 \\
0.932841614906832	-162.645613354037 \\
1.03649068322981	-162.645613354037 \\
1.14013975155279	-162.645613354037 \\
1.24378881987578	-162.645613354037 \\
1.34743788819876	-162.645613354037 \\
1.45108695652174	-162.645613354037 \\
1.55473602484472	-162.645613354037 \\
1.6583850931677	-162.645613354037 \\
1.76203416149068	-162.645613354037 \\
1.86568322981366	-162.645613354037 \\
1.96933229813665	-162.645613354037 \\
2.07298136645963	-162.645613354037 \\
2.17663043478261	-162.645613354037 \\
2.28027950310559	-162.645613354037 \\
2.38392857142857	-162.645613354037 \\
2.48757763975155	-162.645613354037 \\
2.59122670807453	-162.645613354037 \\
2.69487577639751	-162.645613354037 \\
2.7985248447205	-162.645613354037 \\
2.90217391304348	-162.645613354037 \\
3.00582298136646	-162.645613354037 \\
3.10947204968944	-162.645613354037 \\
3.21312111801242	-162.645613354037 \\
3.3167701863354	-162.645613354037 \\
3.42041925465838	-162.645613354037 \\
3.52406832298137	-162.645613354037 \\
3.62771739130435	-162.645613354037 \\
3.73136645962733	-162.645613354037 \\
};
\addplot [semithick, blue, forget plot]
table [row sep=\\]{%
3.73136645962733	-162.645613354037 \\
3.83628504280678	-162.58653757522 \\
3.94120362598623	-162.409310238768 \\
4.04612220916569	-162.113931344682 \\
4.15104079234514	-161.700400892961 \\
4.25595937552459	-161.168718883606 \\
4.36087795870405	-160.518885316616 \\
4.4657965418835	-159.750900191992 \\
4.57071512506295	-158.864763509733 \\
4.6756337082424	-157.860475269839 \\
4.78055229142186	-156.738035472311 \\
4.88547087460131	-155.497444117149 \\
4.99038945778076	-154.138701204352 \\
5.09530804096021	-152.66180673392 \\
5.20022662413967	-151.066760705854 \\
5.30514520731912	-149.353563120154 \\
5.41006379049857	-147.522213976819 \\
5.51498237367803	-145.572713275849 \\
5.61990095685748	-143.505061017245 \\
5.72481954003693	-141.319257201007 \\
5.82973812321638	-139.015301827133 \\
5.93465670639584	-136.593194895626 \\
6.03957528957529	-134.052936406484 \\
6.14449387275474	-131.394526359707 \\
6.24941245593419	-128.617964755296 \\
6.35433103911365	-125.72325159325 \\
6.4592496222931	-122.71038687357 \\
6.56416820547255	-119.579370596255 \\
6.66908678865201	-116.330202761306 \\
6.77400537183146	-112.962883368722 \\
6.87892395501091	-109.477412418504 \\
6.98384253819036	-105.873789910651 \\
7.08876112136982	-102.152015845164 \\
7.19367970454927	-98.3120902220417 \\
7.29859828772872	-94.3540130412853 \\
7.40351687090818	-90.2777843028944 \\
7.50843545408763	-86.083404006869 \\
7.61335403726708	-81.7708721532091 \\
};
\addplot [semithick, blue, forget plot]
table [row sep=\\]{%
7.61335403726708	-81.7708721532091 \\
7.72408835403727	-77.1569422877847 \\
7.83482267080745	-72.5430124223603 \\
7.94555698757764	-67.9290825569358 \\
8.05629130434783	-63.3151526915114 \\
8.16702562111801	-58.7012228260869 \\
8.2777599378882	-54.0872929606625 \\
8.38849425465838	-49.4733630952381 \\
8.49922857142857	-44.8594332298136 \\
8.60996288819876	-40.2455033643892 \\
8.72069720496894	-35.6315734989648 \\
8.83143152173913	-31.0176436335404 \\
8.94216583850932	-26.4037137681159 \\
9.0529001552795	-21.7897839026915 \\
9.16363447204969	-17.175854037267 \\
9.27436878881988	-12.5619241718425 \\
9.38510310559006	-7.94799430641814 \\
9.49583742236025	-3.33406444099373 \\
};
\addplot [semithick, blue, forget plot]
table [row sep=\\]{%
0	107.646739130435 \\
0.104070305272895	107.646739130435 \\
0.208140610545791	107.646739130435 \\
0.312210915818686	107.646739130435 \\
0.416281221091582	107.646739130435 \\
0.520351526364477	107.646739130435 \\
0.624421831637373	107.646739130435 \\
0.728492136910268	107.646739130435 \\
0.832562442183164	107.646739130435 \\
0.936632747456059	107.646739130435 \\
1.04070305272895	107.646739130435 \\
1.14477335800185	107.646739130435 \\
1.24884366327475	107.646739130435 \\
1.35291396854764	107.646739130435 \\
1.45698427382054	107.646739130435 \\
1.56105457909343	107.646739130435 \\
1.66512488436633	107.646739130435 \\
1.76919518963922	107.646739130435 \\
1.87326549491212	107.646739130435 \\
1.97733580018501	107.646739130435 \\
2.08140610545791	107.646739130435 \\
2.1854764107308	107.646739130435 \\
2.2895467160037	107.646739130435 \\
2.3936170212766	107.646739130435 \\
2.49768732654949	107.646739130435 \\
2.60175763182239	107.646739130435 \\
2.70582793709528	107.646739130435 \\
2.80989824236818	107.646739130435 \\
2.91396854764107	107.646739130435 \\
3.01803885291397	107.646739130435 \\
3.12210915818686	107.646739130435 \\
3.22617946345976	107.646739130435 \\
3.33024976873266	107.646739130435 \\
3.43432007400555	107.646739130435 \\
3.53839037927845	107.646739130435 \\
3.64246068455134	107.646739130435 \\
3.74653098982424	107.646739130435 \\
3.85060129509713	107.646739130435 \\
3.95467160037003	107.646739130435 \\
4.05874190564292	107.646739130435 \\
4.16281221091582	107.646739130435 \\
4.26688251618871	107.646739130435 \\
4.37095282146161	107.646739130435 \\
4.4750231267345	107.646739130435 \\
4.5790934320074	107.646739130435 \\
4.6831637372803	107.646739130435 \\
4.78723404255319	107.646739130435 \\
4.89130434782609	107.646739130435 \\
};
\addplot [semithick, blue, forget plot]
table [row sep=\\]{%
4.89130434782609	107.646739130435 \\
4.99622293100554	107.587663351618 \\
5.10114151418499	107.410436015166 \\
5.20606009736445	107.115057121079 \\
5.3109786805439	106.701526669359 \\
5.41589726372335	106.169844660003 \\
5.5208158469028	105.520011093013 \\
5.62573443008226	104.752025968389 \\
5.73065301326171	103.86588928613 \\
5.83557159644116	102.861601046237 \\
5.94049017962061	101.739161248709 \\
6.04540876280007	100.498569893546 \\
6.15032734597952	99.1398269807494 \\
6.25524592915897	97.662932510318 \\
6.36016451233843	96.067886482252 \\
6.46508309551788	94.3546888965514 \\
6.57000167869733	92.5233397532164 \\
6.67492026187678	90.5738390522468 \\
6.77983884505624	88.5061867936427 \\
6.88475742823569	86.3203829774041 \\
6.98967601141514	84.016427603531 \\
7.09459459459459	81.5943206720234 \\
7.19951317777405	79.0540621828812 \\
7.3044317609535	76.3956521361046 \\
7.40935034413295	73.6190905316934 \\
7.51426892731241	70.7243773696477 \\
7.61918751049186	67.7115126499674 \\
7.72410609367131	64.5804963726527 \\
7.82902467685076	61.3313285377034 \\
7.93394326003022	57.9640091451197 \\
8.03886184320967	54.4785381949014 \\
8.14378042638912	50.8749156870485 \\
8.24869900956857	47.1531416215612 \\
8.35361759274803	43.3132159984393 \\
8.45853617592748	39.3551388176829 \\
8.56345475910693	35.278910079292 \\
8.66837334228638	31.0845297832666 \\
8.77329192546584	26.7719979296067 \\
};
\addplot [semithick, blue, forget plot]
table [row sep=\\]{%
8.77329192546584	26.7719979296067 \\
8.89371617494824	21.7543208678399 \\
9.01414042443064	16.7366438060732 \\
9.13456467391304	11.7189667443064 \\
9.25498892339545	6.70128968253965 \\
9.37541317287785	1.68361262077295 \\
9.49583742236025	-3.33406444099382 \\
};
\addplot [semithick, green!50.0!black]
table [row sep=\\]{%
0	-170.294681677019 \\
0.103097417456685	-170.294681677019 \\
0.20619483491337	-170.294681677019 \\
0.309292252370056	-170.294681677019 \\
0.412389669826741	-170.294681677019 \\
0.515487087283426	-170.294681677019 \\
0.618584504740111	-170.294681677019 \\
0.721681922196796	-170.294681677019 \\
0.824779339653481	-170.294681677019 \\
0.927876757110167	-170.294681677019 \\
1.03097417456685	-170.294681677019 \\
1.13407159202354	-170.294681677019 \\
1.23716900948022	-170.294681677019 \\
1.34026642693691	-170.294681677019 \\
1.44336384439359	-170.294681677019 \\
1.54646126185028	-170.294681677019 \\
1.64955867930696	-170.294681677019 \\
1.75265609676365	-170.294681677019 \\
1.85575351422033	-170.294681677019 \\
1.95885093167702	-170.294681677019 \\
2.0619483491337	-170.294681677019 \\
2.16504576659039	-170.294681677019 \\
2.26814318404707	-170.294681677019 \\
2.37124060150376	-170.294681677019 \\
2.47433801896044	-170.294681677019 \\
2.57743543641713	-170.294681677019 \\
2.68053285387381	-170.294681677019 \\
2.7836302713305	-170.294681677019 \\
2.88672768878718	-170.294681677019 \\
2.98982510624387	-170.294681677019 \\
3.09292252370056	-170.294681677019 \\
3.19601994115724	-170.294681677019 \\
3.29911735861393	-170.294681677019 \\
3.40221477607061	-170.294681677019 \\
3.5053121935273	-170.294681677019 \\
3.60840961098398	-170.294681677019 \\
3.71150702844067	-170.294681677019 \\
3.81460444589735	-170.294681677019 \\
3.91770186335404	-170.294681677019 \\
};
\addplot [semithick, green!50.0!black, forget plot]
table [row sep=\\]{%
3.91770186335404	-170.294681677019 \\
4.02262044653349	-170.235605898201 \\
4.12753902971294	-170.05837856175 \\
4.23245761289239	-169.762999667663 \\
4.33737619607185	-169.349469215942 \\
4.4422947792513	-168.817787206587 \\
4.54721336243075	-168.167953639597 \\
4.65213194561021	-167.399968514973 \\
4.75705052878966	-166.513831832714 \\
4.86196911196911	-165.509543592821 \\
4.96688769514856	-164.387103795293 \\
5.07180627832802	-163.14651244013 \\
5.17672486150747	-161.787769527333 \\
5.28164344468692	-160.310875056902 \\
5.38656202786638	-158.715829028836 \\
5.49148061104583	-157.002631443135 \\
5.59639919422528	-155.1712822998 \\
5.70131777740473	-153.221781598831 \\
5.80623636058419	-151.154129340227 \\
5.91115494376364	-148.968325523988 \\
6.01607352694309	-146.664370150115 \\
6.12099211012254	-144.242263218607 \\
6.225910693302	-141.702004729465 \\
6.33082927648145	-139.043594682688 \\
6.4357478596609	-136.267033078277 \\
6.54066644284036	-133.372319916232 \\
6.64558502601981	-130.359455196551 \\
6.75050360919926	-127.228438919237 \\
6.85542219237871	-123.979271084287 \\
6.96034077555817	-120.611951691703 \\
7.06525935873762	-117.126480741485 \\
7.17017794191707	-113.522858233632 \\
7.27509652509652	-109.801084168145 \\
7.38001510827598	-105.961158545023 \\
7.48493369145543	-102.003081364267 \\
7.58985227463488	-97.9268526258759 \\
7.69477085781434	-93.7324723298504 \\
7.79968944099379	-89.4199404761905 \\
};
\addplot [semithick, green!50.0!black, forget plot]
table [row sep=\\]{%
7.79968944099379	-89.4199404761905 \\
7.90989467963387	-84.8280555328539 \\
8.02009991827394	-80.2361705895173 \\
8.13030515691402	-75.6442856461807 \\
8.2405103955541	-71.0524007028441 \\
8.35071563419418	-66.4605157595075 \\
8.46092087283426	-61.8686308161709 \\
8.57112611147434	-57.2767458728343 \\
8.68133135011442	-52.6848609294977 \\
8.79153658875449	-48.0929759861611 \\
8.90174182739457	-43.5010910428245 \\
9.01194706603465	-38.9092060994879 \\
9.12215230467473	-34.3173211561513 \\
9.23235754331481	-29.7254362128147 \\
9.34256278195489	-25.1335512694781 \\
9.45276802059496	-20.5416663261415 \\
9.56297325923504	-15.9497813828049 \\
9.67317849787512	-11.3578964394683 \\
9.7833837365152	-6.76601149613175 \\
9.89358897515528	-2.17412655279513 \\
};
\addplot [semithick, green!50.0!black, forget plot]
table [row sep=\\]{%
0	117.615683229814 \\
0.103625301052098	117.615683229814 \\
0.207250602104196	117.615683229814 \\
0.310875903156294	117.615683229814 \\
0.414501204208391	117.615683229814 \\
0.518126505260489	117.615683229814 \\
0.621751806312587	117.615683229814 \\
0.725377107364685	117.615683229814 \\
0.829002408416783	117.615683229814 \\
0.932627709468881	117.615683229814 \\
1.03625301052098	117.615683229814 \\
1.13987831157308	117.615683229814 \\
1.24350361262517	117.615683229814 \\
1.34712891367727	117.615683229814 \\
1.45075421472937	117.615683229814 \\
1.55437951578147	117.615683229814 \\
1.65800481683357	117.615683229814 \\
1.76163011788566	117.615683229814 \\
1.86525541893776	117.615683229814 \\
1.96888071998986	117.615683229814 \\
2.07250602104196	117.615683229814 \\
2.17613132209406	117.615683229814 \\
2.27975662314615	117.615683229814 \\
2.38338192419825	117.615683229814 \\
2.48700722525035	117.615683229814 \\
2.59063252630245	117.615683229814 \\
2.69425782735454	117.615683229814 \\
2.79788312840664	117.615683229814 \\
2.90150842945874	117.615683229814 \\
3.00513373051084	117.615683229814 \\
3.10875903156294	117.615683229814 \\
3.21238433261503	117.615683229814 \\
3.31600963366713	117.615683229814 \\
3.41963493471923	117.615683229814 \\
3.52326023577133	117.615683229814 \\
3.62688553682343	117.615683229814 \\
3.73051083787552	117.615683229814 \\
3.83413613892762	117.615683229814 \\
3.93776143997972	117.615683229814 \\
4.04138674103182	117.615683229814 \\
4.14501204208391	117.615683229814 \\
4.24863734313601	117.615683229814 \\
4.35226264418811	117.615683229814 \\
4.45588794524021	117.615683229814 \\
4.55951324629231	117.615683229814 \\
4.6631385473444	117.615683229814 \\
4.7667638483965	117.615683229814 \\
4.8703891494486	117.615683229814 \\
4.9740144505007	117.615683229814 \\
5.0776397515528	117.615683229814 \\
};
\addplot [semithick, green!50.0!black, forget plot]
table [row sep=\\]{%
5.0776397515528	117.615683229814 \\
5.18255833473225	117.556607450996 \\
5.2874769179117	117.379380114545 \\
5.39239550109115	117.084001220458 \\
5.49731408427061	116.670470768738 \\
5.60223266745006	116.138788759382 \\
5.70715125062951	115.488955192392 \\
5.81206983380896	114.720970067768 \\
5.91698841698842	113.834833385509 \\
6.02190700016787	112.830545145616 \\
6.12682558334732	111.708105348088 \\
6.23174416652678	110.467513992925 \\
6.33666274970623	109.108771080128 \\
6.44158133288568	107.631876609697 \\
6.54649991606513	106.036830581631 \\
6.65141849924459	104.32363299593 \\
6.75633708242404	102.492283852595 \\
6.86125566560349	100.542783151626 \\
6.96617424878294	98.4751308930216 \\
7.0710928319624	96.289327076783 \\
7.17601141514185	93.9853717029099 \\
7.2809299983213	91.5632647714022 \\
7.38584858150076	89.0230062822601 \\
7.49076716468021	86.3645962354834 \\
7.59568574785966	83.5880346310722 \\
7.70060433103911	80.6933214690265 \\
7.80552291421857	77.6804567493463 \\
7.91044149739802	74.5494404720315 \\
8.01536008057747	71.3002726370822 \\
8.12027866375692	67.9329532444985 \\
8.22519724693638	64.4474822942802 \\
8.33011583011583	60.8438597864274 \\
8.43503441329528	57.12208572094 \\
8.53995299647474	53.2821600978182 \\
8.64487157965419	49.3240829170618 \\
8.74979016283364	45.2478541786709 \\
8.85470874601309	41.0534738826455 \\
8.95962732919255	36.7409420289855 \\
};
\addplot [semithick, green!50.0!black, forget plot]
table [row sep=\\]{%
8.95962732919255	36.7409420289855 \\
9.07637253493789	31.876558456263 \\
9.19311774068323	27.0121748835404 \\
9.30986294642857	22.1477913108179 \\
9.42660815217391	17.2834077380953 \\
9.54335335791925	12.4190241653728 \\
9.66009856366459	7.55464059265022 \\
9.77684376940994	2.69025701992762 \\
9.89358897515528	-2.1741265527949 \\
};
\addplot [semithick, color0]
table [row sep=\\]{%
0	-186.710830745342 \\
0.104643235873353	-186.710830745342 \\
0.209286471746705	-186.710830745342 \\
0.313929707620058	-186.710830745342 \\
0.41857294349341	-186.710830745342 \\
0.523216179366762	-186.710830745342 \\
0.627859415240115	-186.710830745342 \\
0.732502651113468	-186.710830745342 \\
0.83714588698682	-186.710830745342 \\
0.941789122860172	-186.710830745342 \\
1.04643235873352	-186.710830745342 \\
1.15107559460688	-186.710830745342 \\
1.25571883048023	-186.710830745342 \\
1.36036206635358	-186.710830745342 \\
1.46500530222694	-186.710830745342 \\
1.56964853810029	-186.710830745342 \\
1.67429177397364	-186.710830745342 \\
1.77893500984699	-186.710830745342 \\
1.88357824572034	-186.710830745342 \\
1.9882214815937	-186.710830745342 \\
2.09286471746705	-186.710830745342 \\
2.1975079533404	-186.710830745342 \\
2.30215118921376	-186.710830745342 \\
2.40679442508711	-186.710830745342 \\
2.51143766096046	-186.710830745342 \\
2.61608089683381	-186.710830745342 \\
2.72072413270716	-186.710830745342 \\
2.82536736858052	-186.710830745342 \\
2.93001060445387	-186.710830745342 \\
3.03465384032722	-186.710830745342 \\
3.13929707620058	-186.710830745342 \\
3.24394031207393	-186.710830745342 \\
3.34858354794728	-186.710830745342 \\
3.45322678382063	-186.710830745342 \\
3.55787001969398	-186.710830745342 \\
3.66251325556734	-186.710830745342 \\
3.76715649144069	-186.710830745342 \\
3.87179972731404	-186.710830745342 \\
3.9764429631874	-186.710830745342 \\
4.08108619906075	-186.710830745342 \\
4.1857294349341	-186.710830745342 \\
4.29037267080745	-186.710830745342 \\
};
\addplot [semithick, color0, forget plot]
table [row sep=\\]{%
4.29037267080745	-186.710830745342 \\
4.39529125398691	-186.651754966524 \\
4.50020983716636	-186.474527630073 \\
4.60512842034581	-186.179148735986 \\
4.71004700352526	-185.765618284265 \\
4.81496558670472	-185.23393627491 \\
4.91988416988417	-184.58410270792 \\
5.02480275306362	-183.816117583296 \\
5.12972133624308	-182.929980901037 \\
5.23463991942253	-181.925692661144 \\
5.33955850260198	-180.803252863616 \\
5.44447708578143	-179.562661508453 \\
5.54939566896089	-178.203918595656 \\
5.65431425214034	-176.727024125225 \\
5.75923283531979	-175.131978097159 \\
5.86415141849924	-173.418780511458 \\
5.9690700016787	-171.587431368123 \\
6.07398858485815	-169.637930667154 \\
6.1789071680376	-167.57027840855 \\
6.28382575121706	-165.384474592311 \\
6.38874433439651	-163.080519218438 \\
6.49366291757596	-160.65841228693 \\
6.59858150075541	-158.118153797788 \\
6.70350008393487	-155.459743751011 \\
6.80841866711432	-152.6831821466 \\
6.91333725029377	-149.788468984554 \\
7.01825583347322	-146.775604264874 \\
7.12317441665268	-143.644587987559 \\
7.22809299983213	-140.39542015261 \\
7.33301158301158	-137.028100760026 \\
7.43793016619104	-133.542629809808 \\
7.54284874937049	-129.939007301955 \\
7.64776733254994	-126.217233236468 \\
7.75268591572939	-122.377307613346 \\
7.85760449890885	-118.41923043259 \\
7.9625230820883	-114.343001694199 \\
8.06744166526775	-110.148621398173 \\
8.17236024844721	-105.836089544513 \\
};
\addplot [semithick, color0, forget plot]
table [row sep=\\]{%
8.17236024844721	-105.836089544513 \\
8.27866873480961	-101.406569279413 \\
8.38497722117202	-96.9770490143127 \\
8.49128570753443	-92.5475287492123 \\
8.59759419389684	-88.1180084841119 \\
8.70390268025925	-83.6884882190116 \\
8.81021116662166	-79.2589679539112 \\
8.91651965298407	-74.8294476888108 \\
9.02282813934648	-70.3999274237105 \\
9.12913662570888	-65.9704071586101 \\
9.23544511207129	-61.5408868935098 \\
9.3417535984337	-57.1113666284094 \\
9.44806208479611	-52.681846363309 \\
9.55437057115852	-48.2523260982086 \\
9.66067905752093	-43.8228058331083 \\
9.76698754388334	-39.3932855680079 \\
9.87329603024575	-34.9637653029075 \\
9.97960451660816	-30.5342450378072 \\
10.0859130029706	-26.1047247727068 \\
10.192221489333	-21.6752045076064 \\
10.2985299756954	-17.2456842425061 \\
10.4048384620578	-12.8161639774057 \\
10.5111469484202	-8.38664371230531 \\
10.6174554347826	-3.95712344720496 \\
};
\addplot [semithick, color0, forget plot]
table [row sep=\\]{%
0	135.512422360248 \\
0.102484472049689	135.512422360248 \\
0.204968944099379	135.512422360248 \\
0.307453416149068	135.512422360248 \\
0.409937888198758	135.512422360248 \\
0.512422360248447	135.512422360248 \\
0.614906832298137	135.512422360248 \\
0.717391304347826	135.512422360248 \\
0.819875776397516	135.512422360248 \\
0.922360248447205	135.512422360248 \\
1.02484472049689	135.512422360248 \\
1.12732919254658	135.512422360248 \\
1.22981366459627	135.512422360248 \\
1.33229813664596	135.512422360248 \\
1.43478260869565	135.512422360248 \\
1.53726708074534	135.512422360248 \\
1.63975155279503	135.512422360248 \\
1.74223602484472	135.512422360248 \\
1.84472049689441	135.512422360248 \\
1.9472049689441	135.512422360248 \\
2.04968944099379	135.512422360248 \\
2.15217391304348	135.512422360248 \\
2.25465838509317	135.512422360248 \\
2.35714285714286	135.512422360248 \\
2.45962732919255	135.512422360248 \\
2.56211180124224	135.512422360248 \\
2.66459627329193	135.512422360248 \\
2.76708074534162	135.512422360248 \\
2.8695652173913	135.512422360248 \\
2.97204968944099	135.512422360248 \\
3.07453416149068	135.512422360248 \\
3.17701863354037	135.512422360248 \\
3.27950310559006	135.512422360248 \\
3.38198757763975	135.512422360248 \\
3.48447204968944	135.512422360248 \\
3.58695652173913	135.512422360248 \\
3.68944099378882	135.512422360248 \\
3.79192546583851	135.512422360248 \\
3.8944099378882	135.512422360248 \\
3.99689440993789	135.512422360248 \\
4.09937888198758	135.512422360248 \\
4.20186335403727	135.512422360248 \\
4.30434782608696	135.512422360248 \\
4.40683229813665	135.512422360248 \\
4.50931677018634	135.512422360248 \\
4.61180124223602	135.512422360248 \\
4.71428571428571	135.512422360248 \\
4.8167701863354	135.512422360248 \\
4.91925465838509	135.512422360248 \\
5.02173913043478	135.512422360248 \\
5.12422360248447	135.512422360248 \\
5.22670807453416	135.512422360248 \\
5.32919254658385	135.512422360248 \\
};
\addplot [semithick, color0, forget plot]
table [row sep=\\]{%
5.32919254658385	135.512422360248 \\
5.4341111297633	135.453346581431 \\
5.53902971294276	135.276119244979 \\
5.64394829612221	134.980740350893 \\
5.74886687930166	134.567209899172 \\
5.85378546248111	134.035527889817 \\
5.95870404566057	133.385694322827 \\
6.06362262884002	132.617709198203 \\
6.16854121201947	131.731572515944 \\
6.27345979519893	130.72728427605 \\
6.37837837837838	129.604844478523 \\
6.48329696155783	128.36425312336 \\
6.58821554473728	127.005510210563 \\
6.69313412791674	125.528615740132 \\
6.79805271109619	123.933569712066 \\
6.90297129427564	122.220372126365 \\
7.00788987745509	120.38902298303 \\
7.11280846063455	118.43952228206 \\
7.217727043814	116.371870023456 \\
7.32264562699345	114.186066207218 \\
7.42756421017291	111.882110833345 \\
7.53248279335236	109.460003901837 \\
7.63740137653181	106.919745412695 \\
7.74231995971126	104.261335365918 \\
7.84723854289072	101.484773761507 \\
7.95215712607017	98.5900605994613 \\
8.05707570924962	95.5771958797811 \\
8.16199429242907	92.4461796024663 \\
8.26691287560853	89.1970117675171 \\
8.37183145878798	85.8296923749333 \\
8.47675004196743	82.344221424715 \\
8.58166862514688	78.7405989168622 \\
8.68658720832634	75.0188248513749 \\
8.79150579150579	71.178899228253 \\
8.89642437468524	67.2208220474966 \\
9.0013429578647	63.1445933091057 \\
9.10626154104415	58.9502130130803 \\
9.2111801242236	54.6376811594204 \\
};
\addplot [semithick, color0, forget plot]
table [row sep=\\]{%
9.2111801242236	54.6376811594203 \\
9.31935514811276	50.1303884973722 \\
9.42753017200191	45.6230958353241 \\
9.53570519589106	41.1158031732761 \\
9.64388021978022	36.6085105112279 \\
9.75205524366937	32.1012178491798 \\
9.86023026755853	27.5939251871317 \\
9.96840529144768	23.0866325250837 \\
10.0765803153368	18.5793398630356 \\
10.184755339226	14.0720472009874 \\
10.2929303631151	9.56475453893933 \\
10.4011053870043	5.0574618768913 \\
10.5092804108935	0.55016921484318 \\
10.6174554347826	-3.95712344720494 \\
};
\end{axis}

\end{tikzpicture}

%% file: OverApprox.tex
\begin{tikzpicture}

\definecolor{color0}{rgb}{0,0.75,0.75}

\begin{axis}[
legend cell align={left},
legend entries={{Original},{Linear}},
legend style={at={(0.98,0.02)}, anchor=south east, draw=white!80.0!black},
tick align=outside,
tick pos=left,
x grid style={white!69.01960784313725!black},
xlabel={$\tau$ (sec)},
xmin=-0.455351498447205, xmax=9.56238146739131,
y grid style={white!69.01960784313725!black},
ylabel={$h$ (ft)},
ymin=-168.137676630435, ymax=112.768460791925
]
\addlegendimage{no markers, black}
\addlegendimage{no markers, color0}
\addplot [semithick, black]
table [row sep=\\]{%
0	-155.369215838509 \\
0.104265619291195	-155.369215838509 \\
0.208531238582389	-155.369215838509 \\
0.312796857873584	-155.369215838509 \\
0.417062477164779	-155.369215838509 \\
0.521328096455974	-155.369215838509 \\
0.625593715747168	-155.369215838509 \\
0.729859335038363	-155.369215838509 \\
0.834124954329558	-155.369215838509 \\
0.938390573620752	-155.369215838509 \\
1.04265619291195	-155.369215838509 \\
1.14692181220314	-155.369215838509 \\
1.25118743149434	-155.369215838509 \\
1.35545305078553	-155.369215838509 \\
1.45971867007673	-155.369215838509 \\
1.56398428936792	-155.369215838509 \\
1.66824990865912	-155.369215838509 \\
1.77251552795031	-155.369215838509 \\
1.8767811472415	-155.369215838509 \\
1.9810467665327	-155.369215838509 \\
2.08531238582389	-155.369215838509 \\
2.18957800511509	-155.369215838509 \\
2.29384362440628	-155.369215838509 \\
2.39810924369748	-155.369215838509 \\
2.50237486298867	-155.369215838509 \\
2.60664048227987	-155.369215838509 \\
2.71090610157106	-155.369215838509 \\
2.81517172086226	-155.369215838509 \\
2.91943734015345	-155.369215838509 \\
3.02370295944465	-155.369215838509 \\
3.12796857873584	-155.369215838509 \\
3.23223419802704	-155.369215838509 \\
3.33649981731823	-155.369215838509 \\
3.44076543660943	-155.369215838509 \\
3.54503105590062	-155.369215838509 \\
};
\addplot [semithick, black, forget plot]
table [row sep=\\]{%
3.54503105590062	-155.369215838509 \\
3.64994963908007	-155.310140059692 \\
3.75486822225953	-155.13291272324 \\
3.85978680543898	-154.837533829154 \\
3.96470538861843	-154.424003377433 \\
4.06962397179788	-153.892321368078 \\
4.17454255497734	-153.242487801088 \\
4.27946113815679	-152.474502676464 \\
4.38437972133624	-151.588365994205 \\
4.4892983045157	-150.584077754311 \\
4.59421688769515	-149.461637956783 \\
4.6991354708746	-148.221046601621 \\
4.80405405405405	-146.862303688824 \\
4.90897263723351	-145.385409218392 \\
5.01389122041296	-143.790363190326 \\
5.11880980359241	-142.077165604626 \\
5.22372838677186	-140.245816461291 \\
5.32864696995132	-138.296315760321 \\
5.43356555313077	-136.228663501717 \\
5.53848413631022	-134.042859685479 \\
5.64340271948967	-131.738904311606 \\
5.74832130266913	-129.316797380098 \\
5.85323988584858	-126.776538890956 \\
5.95815846902803	-124.118128844179 \\
6.06307705220749	-121.341567239768 \\
6.16799563538694	-118.446854077722 \\
6.27291421856639	-115.433989358042 \\
6.37783280174584	-112.302973080727 \\
6.4827513849253	-109.053805245778 \\
6.58766996810475	-105.686485853194 \\
6.6925885512842	-102.201014902976 \\
6.79750713446366	-98.597392395123 \\
6.90242571764311	-94.8756183296357 \\
7.00734430082256	-91.0356927065138 \\
7.11226288400201	-87.0776155257574 \\
7.21718146718147	-83.0013867873665 \\
7.32210005036092	-78.8070064913411 \\
7.42701863354037	-74.4944746376812 \\
};
\addplot [semithick, black, forget plot]
table [row sep=\\]{%
7.42701863354037	-74.4944746376812 \\
7.53901938923395	-69.8277764837819 \\
7.65102014492754	-65.1610783298827 \\
7.76302090062112	-60.4943801759834 \\
7.8750216563147	-55.8276820220842 \\
7.98702241200828	-51.1609838681849 \\
8.09902316770186	-46.4942857142857 \\
8.21102392339545	-41.8275875603864 \\
8.32302467908903	-37.1608894064872 \\
8.43502543478261	-32.494191252588 \\
8.54702619047619	-27.8274930986887 \\
8.65902694616977	-23.1607949447895 \\
8.77102770186336	-18.4940967908902 \\
8.88302845755694	-13.8273986369909 \\
8.99502921325052	-9.16070048309174 \\
9.1070299689441	-4.49400232919248 \\
};
\addplot [semithick, black, forget plot]
table [row sep=\\]{%
0	100 \\
0.120772946859903	96.2985328323493 \\
0.241545893719807	92.4405081409912 \\
0.36231884057971	88.4259259259259 \\
0.483091787439613	84.2547861871533 \\
0.603864734299517	79.9270889246735 \\
0.72463768115942	75.4428341384863 \\
0.845410628019323	70.8020218285919 \\
0.966183574879227	66.0046519949902 \\
1.08695652173913	61.0507246376812 \\
};
\addplot [semithick, black, forget plot]
table [row sep=\\]{%
1.08695652173913	61.0507246376812 \\
1.1899896619919	56.7576771271489 \\
1.29302280224468	52.4646296166166 \\
1.39605594249745	48.1715821060844 \\
1.49908908275023	43.8785345955521 \\
1.602122223003	39.5854870850198 \\
1.70515536325578	35.2924395744875 \\
1.80818850350855	30.9993920639553 \\
1.91122164376133	26.706344553423 \\
2.0142547840141	22.4132970428907 \\
2.11728792426688	18.1202495323584 \\
2.22032106451965	13.8272020218262 \\
2.32335420477242	9.53415451129393 \\
2.4263873450252	5.24110700076165 \\
2.52942048527797	0.948059490229376 \\
2.63245362553075	-3.34498802030291 \\
2.73548676578352	-7.63803553083518 \\
2.8385199060363	-11.9310830413674 \\
2.94155304628907	-16.2241305518997 \\
3.04458618654185	-20.517178062432 \\
3.14761932679462	-24.8102255729643 \\
3.2506524670474	-29.1032730834965 \\
3.35368560730017	-33.3963205940288 \\
3.45671874755294	-37.6893681045611 \\
3.55975188780572	-41.9824156150933 \\
3.66278502805849	-46.2754631256256 \\
3.76581816831127	-50.5685106361579 \\
3.86885130856404	-54.8615581466902 \\
3.97188444881682	-59.1546056572224 \\
4.07491758906959	-63.4476531677547 \\
4.17795072932237	-67.740700678287 \\
4.28098386957514	-72.0337481888192 \\
4.38401700982791	-76.3267956993515 \\
4.48705015008069	-80.6198432098838 \\
4.59008329033346	-84.9128907204161 \\
4.69311643058624	-89.2059382309483 \\
4.79614957083901	-93.4989857414806 \\
4.89918271109179	-97.7920332520129 \\
5.00221585134456	-102.085080762545 \\
5.10524899159734	-106.378128273077 \\
5.20828213185011	-110.67117578361 \\
};
\addplot [semithick, black, forget plot]
table [row sep=\\]{%
0	100 \\
0.0330794847658176	99.0164241881401 \\
0.0661589695316352	98.0504658385093 \\
};
\addplot [semithick, black, forget plot]
table [row sep=\\]{%
0.0661589695316352	98.0504658385093 \\
0.169243635633141	98.0504658385093 \\
0.272328301734646	98.0504658385093 \\
0.375412967836151	98.0504658385093 \\
0.478497633937657	98.0504658385093 \\
0.581582300039162	98.0504658385093 \\
0.684666966140668	98.0504658385093 \\
0.787751632242173	98.0504658385093 \\
0.890836298343679	98.0504658385093 \\
0.993920964445184	98.0504658385093 \\
1.09700563054669	98.0504658385093 \\
1.20009029664819	98.0504658385093 \\
1.3031749627497	98.0504658385093 \\
1.40625962885121	98.0504658385093 \\
1.50934429495271	98.0504658385093 \\
1.61242896105422	98.0504658385093 \\
1.71551362715572	98.0504658385093 \\
1.81859829325723	98.0504658385093 \\
1.92168295935873	98.0504658385093 \\
2.02476762546024	98.0504658385093 \\
2.12785229156174	98.0504658385093 \\
2.23093695766325	98.0504658385093 \\
2.33402162376475	98.0504658385093 \\
2.43710628986626	98.0504658385093 \\
2.54019095596777	98.0504658385093 \\
2.64327562206927	98.0504658385093 \\
2.74636028817078	98.0504658385093 \\
2.84944495427228	98.0504658385093 \\
2.95252962037379	98.0504658385093 \\
3.05561428647529	98.0504658385093 \\
3.1586989525768	98.0504658385093 \\
3.2617836186783	98.0504658385093 \\
3.36486828477981	98.0504658385093 \\
3.46795295088131	98.0504658385093 \\
3.57103761698282	98.0504658385093 \\
3.67412228308432	98.0504658385093 \\
3.77720694918583	98.0504658385093 \\
3.88029161528734	98.0504658385093 \\
3.98337628138884	98.0504658385093 \\
4.08646094749035	98.0504658385093 \\
4.18954561359185	98.0504658385093 \\
4.29263027969336	98.0504658385093 \\
4.39571494579486	98.0504658385093 \\
4.49879961189637	98.0504658385093 \\
4.60188427799787	98.0504658385093 \\
4.70496894409938	98.0504658385093 \\
};
\addplot [semithick, black, forget plot]
table [row sep=\\]{%
4.70496894409938	98.0504658385093 \\
4.80988752727883	97.9913900596921 \\
4.91480611045828	97.8141627232403 \\
5.01972469363774	97.518783829154 \\
5.12464327681719	97.1052533774332 \\
5.22956185999664	96.5735713680778 \\
5.3344804431761	95.923737801088 \\
5.43939902635555	95.1557526764636 \\
5.544317609535	94.2696159942047 \\
5.64923619271445	93.2653277543113 \\
5.75415477589391	92.1428879567834 \\
5.85907335907336	90.9022966016209 \\
5.96399194225281	89.543553688824 \\
6.06891052543226	88.0666592183925 \\
6.17382910861172	86.4716131903265 \\
6.27874769179117	84.758415604626 \\
6.38366627497062	82.9270664612909 \\
6.48858485815007	80.9775657603214 \\
6.59350344132953	78.9099135017173 \\
6.69842202450898	76.7241096854787 \\
6.80334060768843	74.4201543116056 \\
6.90825919086789	71.9980473800979 \\
7.01317777404734	69.4577888909558 \\
7.11809635722679	66.7993788441791 \\
7.22301494040624	64.0228172397679 \\
7.3279335235857	61.1281040777222 \\
7.43285210676515	58.115239358042 \\
7.5377706899446	54.9842230807272 \\
7.64268927312405	51.735055245778 \\
7.74760785630351	48.3677358531942 \\
7.85252643948296	44.8822649029759 \\
7.95744502266241	41.2786423951231 \\
8.06236360584187	37.5568683296357 \\
8.16728218902132	33.7169427065139 \\
8.27220077220077	29.7588655257575 \\
8.37711935538022	25.6826367873666 \\
8.48203793855968	21.4882564913412 \\
8.58695652173913	17.1757246376812 \\
};
\addplot [semithick, black, forget plot]
table [row sep=\\]{%
8.58695652173913	17.1757246376812 \\
8.71697488354037	11.7582928959628 \\
8.84699324534161	6.34086115424431 \\
8.97701160714286	0.923429412525852 \\
9.1070299689441	-4.4940023291926 \\
};
\addplot [semithick, black, forget plot]
table [row sep=\\]{%
0	-141.925465838509 \\
0.107501194457716	-141.925465838509 \\
0.215002388915432	-141.925465838509 \\
0.322503583373149	-141.925465838509 \\
0.430004777830865	-141.925465838509 \\
0.537505972288581	-141.925465838509 \\
0.645007166746297	-141.925465838509 \\
0.752508361204013	-141.925465838509 \\
0.86000955566173	-141.925465838509 \\
0.967510750119446	-141.925465838509 \\
1.07501194457716	-141.925465838509 \\
1.18251313903488	-141.925465838509 \\
1.29001433349259	-141.925465838509 \\
1.39751552795031	-141.925465838509 \\
1.50501672240803	-141.925465838509 \\
1.61251791686574	-141.925465838509 \\
1.72001911132346	-141.925465838509 \\
1.82752030578118	-141.925465838509 \\
1.93502150023889	-141.925465838509 \\
2.04252269469661	-141.925465838509 \\
2.15002388915432	-141.925465838509 \\
2.25752508361204	-141.925465838509 \\
2.36502627806976	-141.925465838509 \\
2.47252747252747	-141.925465838509 \\
2.58002866698519	-141.925465838509 \\
2.6875298614429	-141.925465838509 \\
2.79503105590062	-141.925465838509 \\
};
\addplot [semithick, black, forget plot]
table [row sep=\\]{%
2.79503105590062	-141.925465838509 \\
2.89995501572451	-141.866384004757 \\
3.0048789755484	-141.689138503501 \\
3.10980293537229	-141.39372933474 \\
3.21472689519618	-140.980156498475 \\
3.31965085502008	-140.448419994705 \\
3.42457481484397	-139.798519823431 \\
3.52949877466786	-139.030455984653 \\
3.63442273449175	-138.14422847837 \\
3.73934669431564	-137.139837304583 \\
3.84427065413953	-136.017282463292 \\
3.94919461396342	-134.776563954496 \\
4.05411857378731	-133.417681778196 \\
4.1590425336112	-131.940635934392 \\
4.26396649343509	-130.345426423083 \\
4.36889045325898	-128.63205324427 \\
4.47381441308287	-126.800516397953 \\
4.57873837290677	-124.850815884131 \\
4.68366233273066	-122.782951702805 \\
4.78858629255455	-120.596923853975 \\
4.89351025237844	-118.29273233764 \\
4.99843421220233	-115.870377153801 \\
5.10335817202622	-113.329858302457 \\
5.20828213185011	-110.67117578361 \\
};
\addplot [semithick, color0]
table [row sep=\\]{%
0	-155.369215838509 \\
0.103296160361378	-155.369215838509 \\
0.206592320722755	-155.369215838509 \\
0.309888481084133	-155.369215838509 \\
0.413184641445511	-155.369215838509 \\
0.516480801806889	-155.369215838509 \\
0.619776962168266	-155.369215838509 \\
0.723073122529644	-155.369215838509 \\
0.826369282891022	-155.369215838509 \\
0.9296654432524	-155.369215838509 \\
1.03296160361378	-155.369215838509 \\
1.13625776397516	-155.369215838509 \\
1.23955392433653	-155.369215838509 \\
1.34285008469791	-155.369215838509 \\
1.44614624505929	-155.369215838509 \\
1.54944240542067	-155.369215838509 \\
1.65273856578204	-155.369215838509 \\
1.75603472614342	-155.369215838509 \\
1.8593308865048	-155.369215838509 \\
1.96262704686618	-155.369215838509 \\
2.06592320722755	-155.369215838509 \\
2.16921936758893	-155.369215838509 \\
2.27251552795031	-155.369215838509 \\
2.37581168831169	-155.369215838509 \\
2.47910784867307	-155.369215838509 \\
2.58240400903444	-155.369215838509 \\
2.68570016939582	-155.369215838509 \\
2.7889963297572	-155.369215838509 \\
2.89229249011858	-155.369215838509 \\
2.99558865047995	-155.369215838509 \\
3.09888481084133	-155.369215838509 \\
3.20218097120271	-155.369215838509 \\
3.30547713156409	-155.369215838509 \\
3.40877329192547	-155.369215838509 \\
3.51206945228684	-155.369215838509 \\
3.61536561264822	-155.369215838509 \\
3.7186617730096	-155.369215838509 \\
3.82195793337098	-155.369215838509 \\
3.92525409373235	-155.369215838509 \\
4.02855025409373	-155.369215838509 \\
4.13184641445511	-155.369215838509 \\
4.23514257481649	-155.369215838509 \\
4.33843873517787	-155.369215838509 \\
4.44173489553924	-155.369215838509 \\
4.54503105590062	-155.369215838509 \\
};
\addplot [semithick, color0, forget plot]
table [row sep=\\]{%
4.54503105590062	-155.369215838509 \\
4.65029421379536	-153.109566715702 \\
4.75555737169009	-150.849917592895 \\
4.86082052958483	-148.590268470088 \\
4.96608368747957	-146.330619347281 \\
5.0713468453743	-144.070970224474 \\
5.17661000326904	-141.811321101667 \\
5.28187316116378	-139.55167197886 \\
5.38713631905852	-137.292022856053 \\
5.49239947695325	-135.032373733246 \\
5.59766263484799	-132.772724610439 \\
5.70292579274273	-130.513075487632 \\
5.80818895063746	-128.253426364825 \\
5.9134521085322	-125.993777242018 \\
6.01871526642694	-123.734128119211 \\
6.12397842432167	-121.474478996404 \\
6.22924158221641	-119.214829873597 \\
6.33450474011115	-116.95518075079 \\
6.43976789800588	-114.695531627983 \\
6.54503105590062	-112.435882505176 \\
};
\addplot [semithick, color0, forget plot]
table [row sep=\\]{%
6.54503105590062	-112.435882505176 \\
6.67003105590062	-107.069215838509 \\
6.79503105590062	-101.702549171843 \\
6.92003105590062	-96.335882505176 \\
7.04503105590062	-90.9692158385093 \\
7.17003105590062	-85.6025491718427 \\
7.29503105590062	-80.235882505176 \\
7.42003105590062	-74.8692158385093 \\
};
\addplot [semithick, color0, forget plot]
table [row sep=\\]{%
7.42701863354037	-74.4944746376812 \\
7.53901938923395	-69.8277764837819 \\
7.65102014492754	-65.1610783298827 \\
7.76302090062112	-60.4943801759834 \\
7.8750216563147	-55.8276820220842 \\
7.98702241200828	-51.1609838681849 \\
8.09902316770186	-46.4942857142857 \\
8.21102392339545	-41.8275875603864 \\
8.32302467908903	-37.1608894064872 \\
8.43502543478261	-32.494191252588 \\
8.54702619047619	-27.8274930986887 \\
8.65902694616977	-23.1607949447895 \\
8.77102770186336	-18.4940967908902 \\
8.88302845755694	-13.8273986369909 \\
8.99502921325052	-9.16070048309174 \\
9.1070299689441	-4.49400232919248 \\
};
\addplot [semithick, color0, forget plot]
table [row sep=\\]{%
0	100 \\
0.120772946859903	95.6723027375201 \\
0.241545893719807	91.3446054750403 \\
0.36231884057971	87.0169082125604 \\
0.483091787439613	82.6892109500806 \\
0.603864734299517	78.3615136876007 \\
0.72463768115942	74.0338164251208 \\
0.845410628019323	69.706119162641 \\
0.966183574879227	65.3784219001611 \\
1.08695652173913	61.0507246376812 \\
};
\addplot [semithick, color0, forget plot]
table [row sep=\\]{%
1.08695652173913	61.0507246376812 \\
1.1899896619919	56.7576771271489 \\
1.29302280224468	52.4646296166166 \\
1.39605594249745	48.1715821060844 \\
1.49908908275023	43.8785345955521 \\
1.602122223003	39.5854870850198 \\
1.70515536325578	35.2924395744875 \\
1.80818850350855	30.9993920639553 \\
1.91122164376133	26.706344553423 \\
2.0142547840141	22.4132970428907 \\
2.11728792426688	18.1202495323584 \\
2.22032106451965	13.8272020218262 \\
2.32335420477242	9.53415451129393 \\
2.4263873450252	5.24110700076165 \\
2.52942048527797	0.948059490229376 \\
2.63245362553075	-3.34498802030291 \\
2.73548676578352	-7.63803553083518 \\
2.8385199060363	-11.9310830413674 \\
2.94155304628907	-16.2241305518997 \\
3.04458618654185	-20.517178062432 \\
3.14761932679462	-24.8102255729643 \\
3.2506524670474	-29.1032730834965 \\
3.35368560730017	-33.3963205940288 \\
3.45671874755294	-37.6893681045611 \\
3.55975188780572	-41.9824156150933 \\
3.66278502805849	-46.2754631256256 \\
3.76581816831127	-50.5685106361579 \\
3.86885130856404	-54.8615581466902 \\
3.97188444881682	-59.1546056572224 \\
4.07491758906959	-63.4476531677547 \\
4.17795072932237	-67.740700678287 \\
4.28098386957514	-72.0337481888192 \\
4.38401700982791	-76.3267956993515 \\
4.48705015008069	-80.6198432098838 \\
4.59008329033346	-84.9128907204161 \\
4.69311643058624	-89.2059382309483 \\
4.79614957083901	-93.4989857414806 \\
4.89918271109179	-97.7920332520129 \\
5.00221585134456	-102.085080762545 \\
5.10524899159734	-106.378128273077 \\
5.20828213185011	-110.67117578361 \\
};
\addplot [semithick, color0, forget plot]
table [row sep=\\]{%
0	100 \\
0.0330794847658176	99.0252329192547 \\
0.0661589695316352	98.0504658385093 \\
};
\addplot [semithick, color0, forget plot]
table [row sep=\\]{%
0.0661589695316352	98.0504658385093 \\
0.168682787251049	98.0504658385093 \\
0.271206604970462	98.0504658385093 \\
0.373730422689876	98.0504658385093 \\
0.476254240409289	98.0504658385093 \\
0.578778058128703	98.0504658385093 \\
0.681301875848116	98.0504658385093 \\
0.78382569356753	98.0504658385093 \\
0.886349511286943	98.0504658385093 \\
0.988873329006357	98.0504658385093 \\
1.09139714672577	98.0504658385093 \\
1.19392096444518	98.0504658385093 \\
1.2964447821646	98.0504658385093 \\
1.39896859988401	98.0504658385093 \\
1.50149241760342	98.0504658385093 \\
1.60401623532284	98.0504658385093 \\
1.70654005304225	98.0504658385093 \\
1.80906387076167	98.0504658385093 \\
1.91158768848108	98.0504658385093 \\
2.01411150620049	98.0504658385093 \\
2.11663532391991	98.0504658385093 \\
2.21915914163932	98.0504658385093 \\
2.32168295935873	98.0504658385093 \\
2.42420677707815	98.0504658385093 \\
2.52673059479756	98.0504658385093 \\
2.62925441251697	98.0504658385093 \\
2.73177823023639	98.0504658385093 \\
2.8343020479558	98.0504658385093 \\
2.93682586567521	98.0504658385093 \\
3.03934968339463	98.0504658385093 \\
3.14187350111404	98.0504658385093 \\
3.24439731883345	98.0504658385093 \\
3.34692113655287	98.0504658385093 \\
3.44944495427228	98.0504658385093 \\
3.5519687719917	98.0504658385093 \\
3.65449258971111	98.0504658385093 \\
3.75701640743052	98.0504658385093 \\
3.85954022514994	98.0504658385093 \\
3.96206404286935	98.0504658385093 \\
4.06458786058876	98.0504658385093 \\
4.16711167830818	98.0504658385093 \\
4.26963549602759	98.0504658385093 \\
4.372159313747	98.0504658385093 \\
4.47468313146642	98.0504658385093 \\
4.57720694918583	98.0504658385093 \\
4.67973076690524	98.0504658385093 \\
4.78225458462466	98.0504658385093 \\
4.88477840234407	98.0504658385093 \\
4.98730222006348	98.0504658385093 \\
5.0898260377829	98.0504658385093 \\
5.19234985550231	98.0504658385093 \\
5.29487367322173	98.0504658385093 \\
5.39739749094114	98.0504658385093 \\
5.49992130866055	98.0504658385093 \\
5.60244512637997	98.0504658385093 \\
5.70496894409938	98.0504658385093 \\
};
\addplot [semithick, color0, forget plot]
table [row sep=\\]{%
5.70496894409938	98.0504658385093 \\
5.81023210199412	95.7908167157023 \\
5.91549525988885	93.5311675928953 \\
6.02075841778359	91.2715184700883 \\
6.12602157567833	89.0118693472813 \\
6.23128473357306	86.7522202244742 \\
6.3365478914678	84.4925711016672 \\
6.44181104936254	82.2329219788602 \\
6.54707420725727	79.9732728560532 \\
6.65233736515201	77.7136237332462 \\
6.75760052304675	75.4539746104392 \\
6.86286368094148	73.1943254876321 \\
6.96812683883622	70.9346763648251 \\
7.07338999673096	68.6750272420181 \\
7.1786531546257	66.4153781192111 \\
7.28391631252043	64.1557289964041 \\
7.38917947041517	61.896079873597 \\
7.4944426283099	59.63643075079 \\
7.59970578620464	57.376781627983 \\
7.70496894409938	55.117132505176 \\
};
\addplot [semithick, color0, forget plot]
table [row sep=\\]{%
7.70496894409938	55.117132505176 \\
7.82996894409938	49.7504658385093 \\
7.95496894409938	44.3837991718427 \\
8.07996894409938	39.017132505176 \\
8.20496894409938	33.6504658385093 \\
8.32996894409938	28.2837991718427 \\
8.45496894409938	22.917132505176 \\
8.57996894409938	17.5504658385093 \\
};
\addplot [semithick, color0, forget plot]
table [row sep=\\]{%
8.58695652173913	17.1757246376812 \\
8.71697488354037	11.7582928959628 \\
8.84699324534161	6.3408611542443 \\
8.97701160714286	0.923429412525852 \\
9.1070299689441	-4.4940023291926 \\
};
\addplot [semithick, color0, forget plot]
table [row sep=\\]{%
0	-141.925465838509 \\
0.107501194457716	-141.925465838509 \\
0.215002388915432	-141.925465838509 \\
0.322503583373149	-141.925465838509 \\
0.430004777830865	-141.925465838509 \\
0.537505972288581	-141.925465838509 \\
0.645007166746297	-141.925465838509 \\
0.752508361204013	-141.925465838509 \\
0.86000955566173	-141.925465838509 \\
0.967510750119446	-141.925465838509 \\
1.07501194457716	-141.925465838509 \\
1.18251313903488	-141.925465838509 \\
1.29001433349259	-141.925465838509 \\
1.39751552795031	-141.925465838509 \\
1.50501672240803	-141.925465838509 \\
1.61251791686574	-141.925465838509 \\
1.72001911132346	-141.925465838509 \\
1.82752030578118	-141.925465838509 \\
1.93502150023889	-141.925465838509 \\
2.04252269469661	-141.925465838509 \\
2.15002388915432	-141.925465838509 \\
2.25752508361204	-141.925465838509 \\
2.36502627806976	-141.925465838509 \\
2.47252747252747	-141.925465838509 \\
2.58002866698519	-141.925465838509 \\
2.6875298614429	-141.925465838509 \\
2.79503105590062	-141.925465838509 \\
};
\addplot [semithick, color0, forget plot]
table [row sep=\\]{%
2.79503105590062	-141.925465838509 \\
2.90029421379536	-140.795641277106 \\
3.00555737169009	-139.665816715702 \\
3.11082052958483	-138.535992154299 \\
3.21608368747957	-137.406167592895 \\
3.3213468453743	-136.276343031492 \\
3.42661000326904	-135.146518470088 \\
3.53187316116378	-134.016693908685 \\
3.63713631905852	-132.886869347281 \\
3.74239947695325	-131.757044785878 \\
3.84766263484799	-130.627220224474 \\
3.95292579274273	-129.497395663071 \\
4.05818895063746	-128.367571101667 \\
4.1634521085322	-127.237746540264 \\
4.26871526642694	-126.10792197886 \\
4.37397842432167	-124.978097417457 \\
4.47924158221641	-123.848272856054 \\
4.58450474011115	-122.71844829465 \\
4.68976789800588	-121.588623733247 \\
4.79503105590062	-120.458799171843 \\
};
\addplot [semithick, color0, forget plot]
table [row sep=\\]{%
4.79503105590062	-120.458799171843 \\
4.93278141455045	-117.196258042432 \\
5.07053177320028	-113.933716913021 \\
5.20828213185011	-110.67117578361 \\
};
\end{axis}

\end{tikzpicture}

%% file: Safeable_SAT_COC.tex
\begin{tikzpicture}

\begin{axis}[
tick align=outside,
tick pos=left,
x grid style={lightgray!92.02614379084967!black},
xlabel={$\tau$ (sec)},
xmin=0, xmax=7.52765620295442,
y grid style={lightgray!92.02614379084967!black},
ylabel={$h$ (ft)},
ymin=-50, ymax=407.966006386041,
height=6cm,
width=8cm
]
\addplot [semithick, black]
table [row sep=\\]{%
0	187.869791666667 \\
0.0101190476190476	187.869791666667 \\
};
\addlegendentry{COC Unsafeable Region};

\addplot graphics [includegraphics cmd=\pgfimage,xmin=0, xmax=7.52765620295442, ymin=-50, ymax=407.966006386041] {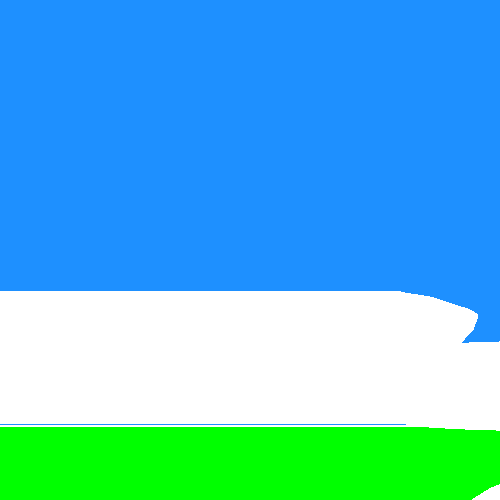};
\addplot [semithick, black, forget plot]
table [row sep=\\]{%
5.84800409243785	129.64716844785 \\
5.98392226083349	135.310425464335 \\
6.11984042922913	140.97368248082 \\
6.25575859762477	146.636939497305 \\
6.39167676602041	152.300196513791 \\
6.52759493441606	157.963453530276 \\
};
\addplot [semithick, black, forget plot]
table [row sep=\\]{%
0	161.344681677019 \\
0.103566723131941	161.344681677019 \\
0.207133446263881	161.344681677019 \\
0.310700169395822	161.344681677019 \\
0.414266892527762	161.344681677019 \\
0.517833615659703	161.344681677019 \\
0.621400338791643	161.344681677019 \\
0.724967061923584	161.344681677019 \\
0.828533785055524	161.344681677019 \\
0.932100508187465	161.344681677019 \\
1.03566723131941	161.344681677019 \\
1.13923395445135	161.344681677019 \\
1.24280067758329	161.344681677019 \\
1.34636740071523	161.344681677019 \\
1.44993412384717	161.344681677019 \\
1.55350084697911	161.344681677019 \\
1.65706757011105	161.344681677019 \\
1.76063429324299	161.344681677019 \\
1.86420101637493	161.344681677019 \\
1.96776773950687	161.344681677019 \\
2.07133446263881	161.344681677019 \\
2.17490118577075	161.344681677019 \\
2.27846790890269	161.344681677019 \\
2.38203463203463	161.344681677019 \\
2.48560135516657	161.344681677019 \\
2.58916807829851	161.344681677019 \\
2.69273480143045	161.344681677019 \\
2.79630152456239	161.344681677019 \\
2.89986824769433	161.344681677019 \\
3.00343497082628	161.344681677019 \\
3.10700169395822	161.344681677019 \\
3.21056841709016	161.344681677019 \\
3.3141351402221	161.344681677019 \\
3.41770186335404	161.344681677019 \\
};
\addplot [semithick, black, forget plot]
table [row sep=\\]{%
3.41770186335404	161.344681677019 \\
3.52336717766203	161.284761992275 \\
3.62903249197002	161.105002938046 \\
3.73469780627801	160.80540451433 \\
3.840363120586	160.385966721127 \\
3.946028434894	159.846689558438 \\
4.05169374920199	159.187573026262 \\
4.15735906350998	158.4086171246 \\
4.26302437781797	157.509821853452 \\
4.36868969212596	156.491187212817 \\
4.47435500643395	155.352713202696 \\
4.58002032074195	154.094399823088 \\
4.68568563504994	152.716247073994 \\
4.79135094935793	151.218254955413 \\
4.89701626366592	149.600423467346 \\
5.00268157797391	147.862752609792 \\
5.10834689228191	146.005242382752 \\
5.2140122065899	144.027892786225 \\
5.31967752089789	141.930703820212 \\
5.42534283520588	139.713675484713 \\
5.53100814951387	137.376807779727 \\
5.63667346382187	134.920100705254 \\
5.74233877812986	132.343554261296 \\
5.84800409243785	129.64716844785 \\
};
\addplot [semithick, black, forget plot]
table [row sep=\\]{%
0	187.850931677019 \\
0.104192546583851	187.850931677019 \\
0.208385093167702	187.850931677019 \\
0.312577639751553	187.850931677019 \\
0.416770186335404	187.850931677019 \\
0.520962732919255	187.850931677019 \\
0.625155279503106	187.850931677019 \\
0.729347826086957	187.850931677019 \\
0.833540372670807	187.850931677019 \\
0.937732919254658	187.850931677019 \\
1.04192546583851	187.850931677019 \\
1.14611801242236	187.850931677019 \\
1.25031055900621	187.850931677019 \\
1.35450310559006	187.850931677019 \\
1.45869565217391	187.850931677019 \\
1.56288819875776	187.850931677019 \\
1.66708074534161	187.850931677019 \\
1.77127329192547	187.850931677019 \\
1.87546583850932	187.850931677019 \\
1.97965838509317	187.850931677019 \\
2.08385093167702	187.850931677019 \\
2.18804347826087	187.850931677019 \\
2.29223602484472	187.850931677019 \\
2.39642857142857	187.850931677019 \\
2.50062111801242	187.850931677019 \\
2.60481366459627	187.850931677019 \\
2.70900621118012	187.850931677019 \\
2.81319875776398	187.850931677019 \\
2.91739130434783	187.850931677019 \\
3.02158385093168	187.850931677019 \\
3.12577639751553	187.850931677019 \\
3.22996894409938	187.850931677019 \\
3.33416149068323	187.850931677019 \\
3.43835403726708	187.850931677019 \\
3.54254658385093	187.850931677019 \\
3.64673913043478	187.850931677019 \\
3.75093167701863	187.850931677019 \\
3.85512422360248	187.850931677019 \\
3.95931677018634	187.850931677019 \\
4.06350931677019	187.850931677019 \\
4.16770186335404	187.850931677019 \\
};
\addplot [semithick, black, forget plot]
table [row sep=\\]{%
4.16770186335404	187.850931677019 \\
4.27496973022049	187.789180689112 \\
4.38223759708695	187.603927725393 \\
4.4895054639534	187.29517278586 \\
4.59677333081986	186.862915870515 \\
4.70404119768631	186.307156979356 \\
4.81130906455277	185.627896112385 \\
4.91857693141923	184.8251332696 \\
5.02584479828568	183.898868451003 \\
5.13311266515214	182.849101656593 \\
5.24038053201859	181.675832886369 \\
5.34764839888505	180.379062140333 \\
5.4549162657515	178.958789418484 \\
5.56218413261796	177.415014720821 \\
5.66945199948441	175.747738047346 \\
5.77671986635087	173.956959398058 \\
5.88398773321732	172.042678772956 \\
5.99125560008378	170.004896172042 \\
6.09852346695023	167.843611595315 \\
6.20579133381669	165.558825042774 \\
6.31305920068314	163.150536514421 \\
6.4203270675496	160.618746010255 \\
6.52759493441606	157.963453530276 \\
};

\node[text=black] at (3,250)  {\footnotesize DES1500};
\node[text=black] at (3,80)  {\footnotesize COC};
\node[text=black] at (3,-10)  {\footnotesize CL1500};

\end{axis}

\end{tikzpicture}

%% file: ResultsTable.tex
\begin{table*}[t]
    \centering
    \caption{Number of counterexamples discovered with Reluplex
    \label{tab:counterexamples}}
    \begin{tabular} {lccccccccc} 
    \toprule
        & \multicolumn{9}{c}{\textbf{Current Advisory}} \\
      \textbf{Previous Advisory} & COC & DNC & DND	& DES1500 & CL1500 & SDES1500 & SCL1500 & SDES2500 & SCL2500 \\
        \midrule
        
       COC & 359 & 28 & 0 & 48 & 21 & \text{N/A} & \text{N/A} & \text{N/A} & \text{N/A} \\
       DNC & 438 & 30 & 0 & 40 & 47 & \text{N/A} & \text{N/A} & \text{N/A} & \text{N/A} \\
       DND & 249 & 0  & 17& 133& 50 & \text{N/A} & \text{N/A} & \text{N/A} & \text{N/A} \\
       DES1500 & 284& 0& 1& 1& 0& 65& 76 & \text{N/A} & \text{N/A} \\
       CL1500  & 223& 0& 0& 0& 0& 117& 21& \text{N/A} & \text{N/A}\\
       SDES1500 & 281& 0& 0& 0& 0& 26& 6& 32& 65 \\
       SCL1500  & 238& 0& 3& 0& 0& 53& 66& 43& 51\\
       SDES2500 & 324& 0&  0& 0& 0& 12&  1& 89& 25\\
       SCL2500  & 209& 0& 12& 0& 0& 52& 15& 48& 58\\
       
      \bottomrule
    \end{tabular}
\end{table*}

%% file: Safeable_SAT_Boundary.tex
\begin{tikzpicture}

\begin{axis}[
tick align=outside,
tick pos=left,
x grid style={lightblack!30!92.02614379084967!black},
xlabel={$\tau$ (sec)},
xmin=0, xmax=6.75097604259095,
y grid style={lightblack!30!92.02614379084967!black},
ylabel={$h$ (ft)},
ymin=-199.338843907128, ymax=186.594469398318,
height=6cm,
width=8cm
]

\addplot [semithick, black!30]
table [row sep=\\]{%
0.0312499999999831	120.125 \\
0.0437499999999932	120.125 \\
0.0562500000000034	120.125 \\
0.0687500000000136	120.125 \\
0.0812500000000237	120.125 \\
0.0937500000000339	120.125 \\
};
\addlegendentry{CL1500 Unsafeable Region}

\addplot [semithick, black]
table [row sep=\\]{%
4.53124999999996	-0.632812500001904 \\
};
\addlegendentry{DES1500 Unsafeable Region};

\addplot graphics [includegraphics cmd=\pgfimage,xmin=0, xmax=6.75097604259095, ymin=-199.338843907128, ymax=186.594469398318] {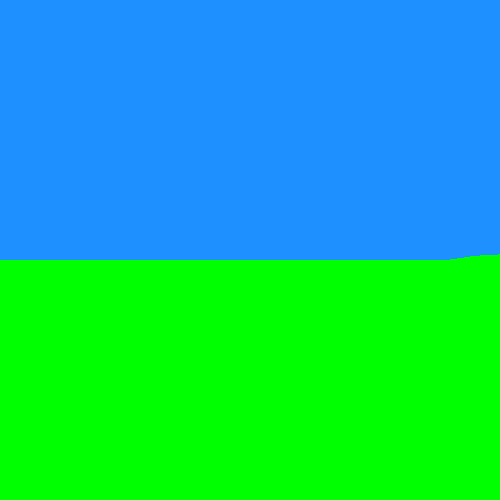};
\addplot [semithick, black!30, forget plot]
table [row sep=\\]{%
4.66824751552795	4.2251552795031 \\
4.77729234082523	8.7686896668897 \\
4.88633716612251	13.3122240542763 \\
4.99538199141979	17.8557584416629 \\
5.10442681671706	22.3992928290495 \\
5.21347164201434	26.9428272164361 \\
5.32251646731162	31.4863616038227 \\
5.4315612926089	36.0298959912094 \\
5.54060611790618	40.573430378596 \\
5.64965094320346	45.1169647659826 \\
5.75869576850074	49.6604991533692 \\
5.86774059379801	54.2040335407558 \\
5.97678541909529	58.7475679281424 \\
6.08583024439257	63.291102315529 \\
};
\addplot [semithick, black!30, forget plot]
table [row sep=\\]{%
0	100.248447204969 \\
0.248447204968944	100.248447204969 \\
};
\addplot [semithick, black!30, forget plot]
table [row sep=\\]{%
0.248447204968944	100.248447204969 \\
0.373706004140787	100.18529589372 \\
0.498964803312629	99.9958419599724 \\
0.624223602484472	99.6800854037267 \\
0.749482401656315	99.2380262249827 \\
0.874741200828157	98.6696644237405 \\
1	97.975 \\
};
\addplot [semithick, black!30, forget plot]
table [row sep=\\]{%
1	97.975 \\
1.1036975931677	97.2899207705543 \\
1.2073951863354	96.4894239595464 \\
1.31109277950311	95.5735095669764 \\
1.41479037267081	94.5421775928442 \\
1.51848796583851	93.3954280371498 \\
1.62218555900621	92.1332608998932 \\
1.72588315217391	90.7556761810745 \\
1.82958074534161	89.2626738806936 \\
1.93327833850932	87.6542539987505 \\
2.03697593167702	85.9304165352452 \\
2.14067352484472	84.0911614901778 \\
2.24437111801242	82.1364888635481 \\
2.34806871118012	80.0663986553563 \\
2.45176630434783	77.8808908656024 \\
2.55546389751553	75.5799654942862 \\
2.65916149068323	73.1636225414079 \\
2.76285908385093	70.6318620069674 \\
2.86655667701863	67.9846838909647 \\
2.97025427018633	65.2220881933998 \\
3.07395186335404	62.3440749142728 \\
3.17764945652174	59.3506440535836 \\
3.28134704968944	56.2417956113322 \\
3.38504464285714	53.0175295875186 \\
3.48874223602484	49.6778459821429 \\
3.59243982919255	46.222744795205 \\
3.69613742236025	42.6522260267049 \\
3.79983501552795	38.9662896766426 \\
3.90353260869565	35.1649357450181 \\
4.00723020186335	31.2481642318315 \\
4.11092779503105	27.2159751370827 \\
4.21462538819876	23.0683684607717 \\
4.31832298136646	18.8053442028986 \\
};
\addplot [semithick, black!30, forget plot]
table [row sep=\\]{%
4.31832298136646	18.8053442028986 \\
4.4932852484472	11.5152497412008 \\
4.66824751552795	4.2251552795031 \\
};
\addplot [semithick, black!30, forget plot]
table [row sep=\\]{%
0	125.311335403727 \\
0.107453416149068	125.311335403727 \\
0.214906832298137	125.311335403727 \\
0.322360248447205	125.311335403727 \\
0.429813664596273	125.311335403727 \\
0.537267080745342	125.311335403727 \\
0.64472049689441	125.311335403727 \\
0.752173913043478	125.311335403727 \\
0.859627329192547	125.311335403727 \\
0.967080745341615	125.311335403727 \\
1.07453416149068	125.311335403727 \\
1.18198757763975	125.311335403727 \\
1.28944099378882	125.311335403727 \\
1.39689440993789	125.311335403727 \\
1.50434782608696	125.311335403727 \\
1.61180124223602	125.311335403727 \\
1.71925465838509	125.311335403727 \\
1.82670807453416	125.311335403727 \\
1.93416149068323	125.311335403727 \\
2.0416149068323	125.311335403727 \\
2.14906832298137	125.311335403727 \\
2.25652173913043	125.311335403727 \\
2.3639751552795	125.311335403727 \\
2.47142857142857	125.311335403727 \\
2.57888198757764	125.311335403727 \\
2.68633540372671	125.311335403727 \\
};
\addplot [semithick, black!30, forget plot]
table [row sep=\\]{%
2.68633540372671	125.311335403727 \\
2.79256961749752	125.250768769852 \\
2.89880383126832	125.069068868226 \\
3.00503804503913	124.76623569885 \\
3.11127225880994	124.342269261724 \\
3.21750647258075	123.797169556847 \\
3.32374068635156	123.13093658422 \\
3.42997490012237	122.343570343842 \\
3.53620911389317	121.435070835714 \\
3.64244332766398	120.405438059836 \\
3.74867754143479	119.254672016207 \\
3.8549117552056	117.982772704828 \\
3.96114596897641	116.589740125699 \\
4.06738018274722	115.075574278819 \\
4.17361439651802	113.440275164189 \\
4.27984861028883	111.683842781808 \\
4.38608282405964	109.806277131677 \\
4.49231703783045	107.807578213796 \\
4.59855125160126	105.687746028164 \\
4.70478546537207	103.446780574782 \\
4.81101967914287	101.08468185365 \\
4.91725389291368	98.6014498647666 \\
5.02348810668449	95.9970846081333 \\
5.1297223204553	93.2715860837496 \\
5.23595653422611	90.4249542916156 \\
5.34219074799691	87.4571892317311 \\
5.44842496176772	84.3682909040963 \\
5.55465917553853	81.1582593087111 \\
5.66089338930934	77.8270944455754 \\
5.76712760308015	74.3747963146894 \\
5.87336181685096	70.8013649160531 \\
5.97959603062176	67.1068002496663 \\
6.08583024439257	63.2911023155291 \\
};

\addplot [semithick, black, forget plot]
table [row sep=\\]{%
0	-120.125 \\
0.104166666666667	-120.125 \\
0.208333333333333	-120.125 \\
0.3125	-120.125 \\
0.416666666666667	-120.125 \\
0.520833333333333	-120.125 \\
0.625	-120.125 \\
0.729166666666667	-120.125 \\
0.833333333333333	-120.125 \\
0.9375	-120.125 \\
1.04166666666667	-120.125 \\
1.14583333333333	-120.125 \\
1.25	-120.125 \\
1.35416666666667	-120.125 \\
1.45833333333333	-120.125 \\
1.5625	-120.125 \\
1.66666666666667	-120.125 \\
1.77083333333333	-120.125 \\
1.875	-120.125 \\
1.97916666666667	-120.125 \\
2.08333333333333	-120.125 \\
2.1875	-120.125 \\
2.29166666666667	-120.125 \\
2.39583333333333	-120.125 \\
2.5	-120.125 \\
};
\addplot [semithick, black, forget plot]
table [row sep=\\]{%
2.5	-120.125 \\
2.60409581928455	-120.066847124187 \\
2.7081916385691	-119.892388496747 \\
2.81228745785365	-119.601624117681 \\
2.91638327713819	-119.194553986989 \\
3.02047909642274	-118.67117810467 \\
3.12457491570729	-118.031496470725 \\
3.22867073499184	-117.275509085153 \\
3.33276655427639	-116.403215947955 \\
3.43686237356094	-115.414617059131 \\
3.54095819284549	-114.30971241868 \\
3.64505401213003	-113.088502026603 \\
3.74914983141458	-111.750985882899 \\
3.85324565069913	-110.29716398757 \\
3.95734146998368	-108.727036340613 \\
4.06143728926823	-107.04060294203 \\
4.16553310855278	-105.237863791821 \\
4.26962892783733	-103.318818889986 \\
4.37372474712187	-101.283468236524 \\
4.47782056640642	-99.1318118314355 \\
4.58191638569097	-96.8638496747207 \\
4.68601220497552	-94.4795817663796 \\
4.79010802426007	-91.9790081064121 \\
4.89420384354462	-89.3621286948182 \\
4.99829966282917	-86.6289435315978 \\
5.10239548211372	-83.7794526167511 \\
5.20649130139826	-80.813655950278 \\
5.31058712068281	-77.7315535321785 \\
5.41468293996736	-74.5331453624526 \\
5.51877875925191	-71.2184314411003 \\
5.62287457853646	-67.7874117681216 \\
5.72697039782101	-64.2400863435165 \\
5.83106621710556	-60.576455167285 \\
5.9351620363901	-56.7965182394272 \\
6.03925785567465	-52.9002755599429 \\
};
\addplot [semithick, black, forget plot]
table [row sep=\\]{%
4.66824751552795	4.2251552795031 \\
4.78249837720684	-0.535297290450731 \\
4.89674923888573	-5.29574986040456 \\
5.01100010056463	-10.0562024303584 \\
5.12525096224352	-14.8166550003122 \\
5.23950182392241	-19.5771075702661 \\
5.3537526856013	-24.3375601402199 \\
5.46800354728019	-29.0980127101737 \\
5.58225440895909	-33.8584652801275 \\
5.69650527063798	-38.6189178500813 \\
5.81075613231687	-43.3793704200352 \\
5.92500699399576	-48.139822989989 \\
6.03925785567465	-52.9002755599428 \\
};
\addplot [semithick, black, forget plot]
table [row sep=\\]{%
0	-100 \\
0.111111111111111	-99.9503086419753 \\
0.222222222222222	-99.8012345679012 \\
0.333333333333333	-99.5527777777778 \\
0.444444444444444	-99.2049382716049 \\
0.555555555555556	-98.7577160493827 \\
0.666666666666667	-98.2111111111111 \\
0.777777777777778	-97.5651234567901 \\
0.888888888888889	-96.8197530864198 \\
1	-95.975 \\
};
\addplot [semithick, black, forget plot]
table [row sep=\\]{%
1	-95.975 \\
1.10439958592132	-95.0760905653324 \\
1.20879917184265	-94.060195594663 \\
1.31319875776397	-92.9273150879917 \\
1.4175983436853	-91.6774490453186 \\
1.52199792960663	-90.3105974666437 \\
1.62639751552795	-88.8267603519669 \\
1.73079710144928	-87.2259377012882 \\
1.8351966873706	-85.5081295146078 \\
1.93959627329193	-83.6733357919255 \\
2.04399585921325	-81.7215565332413 \\
2.14839544513458	-79.6527917385553 \\
2.2527950310559	-77.4670414078675 \\
2.35719461697723	-75.1643055411778 \\
2.46159420289855	-72.7445841384863 \\
2.56599378881988	-70.207877199793 \\
2.6703933747412	-67.5541847250978 \\
2.77479296066253	-64.7835067144007 \\
2.87919254658385	-61.8958431677019 \\
2.98359213250518	-58.8911940850012 \\
3.0879917184265	-55.7695594662986 \\
3.19239130434783	-52.5309393115942 \\
3.29679089026915	-49.175333620888 \\
3.40119047619048	-45.7027423941799 \\
3.5055900621118	-42.11316563147 \\
3.60998964803313	-38.4066033327582 \\
3.71438923395445	-34.5830554980446 \\
3.81878881987578	-30.6425221273292 \\
3.9231884057971	-26.5850032206119 \\
4.02758799171843	-22.4104987778928 \\
4.13198757763975	-18.1190087991719 \\
};
\addplot [semithick, black, forget plot]
table [row sep=\\]{%
4.13198757763975	-18.1190087991719 \\
4.2660525621118	-12.5329677795031 \\
4.40011754658385	-6.9469267598344 \\
4.5341825310559	-1.36088574016565 \\
4.66824751552795	4.2251552795031 \\
};

\node[text=black] at (1,50)  {\footnotesize DES1500};
\node[text=black] at (1,-50)  {\footnotesize CL1500};

\end{axis}

\end{tikzpicture}